\documentclass[3p,preprint]{elsarticle}

\biboptions{sort&compress}

\usepackage{latexsym}
\usepackage{amsfonts}
\usepackage{amsmath}
\usepackage{amssymb}
\usepackage{mathrsfs}
\usepackage{color}
\usepackage{graphicx}
\usepackage{mathptmx}      
\usepackage{bm}
\usepackage{enumerate}
\usepackage{subfigure}
\usepackage{listings}
\usepackage{grffile}
\usepackage{multirow}
\usepackage{qcircuit}
\usepackage{hyperref}
\usepackage{mathtools}
\usepackage[normalem]{ulem}
\usepackage{xcolor}
\usepackage{hyperref}
\hypersetup{
  colorlinks=true,
}

\definecolor{amethyst}{rgb}{0.6, 0.1, 0.6}
\definecolor{DarkGreen}{RGB}{0,100,0}
\newcommand{\nint}[1]{\ensuremath\left\lfloor#1\right\rceil}

\newcommand{\onlinecite}{\cite}

\DeclareMathOperator\sign{sign}
\newcommand{\ket}[1]{| #1 \rangle} 

\newcommand\ba{\mathbf{a}}
\newcommand\bb{\mathbf{b}}
\newcommand\bc{\mathbf{c}}
\newcommand\bd{\mathbf{d}}
\newcommand\be{\mathbf{e}}

\newcommand\bv{\mathbf{v}}
\newcommand\bw{\mathbf{w}}
\newcommand\bx{\mathbf{x}}

\newcommand\bA{\mathbf{A}}
\newcommand\bB{\mathbf{B}}
\newcommand\bC{\mathbf{C}}
\newcommand\bD{\mathbf{D}}

\newcommand\bF{\mathbf{F}}
\newcommand\bM{\mathbf{M}}
\newcommand\bP{\mathbf{P}}

\newcommand\bS{\mathbf{S}}
\newcommand\bX{\mathbf{X}}
\newcommand\bY{\mathbf{Y}}

\newcommand*\widebar[1]{%
   \hbox{%
     \vbox{%
       \hrule height 0.5pt 
       \kern0.5ex
       \hbox{%
         \kern-0.1em
         \ensuremath{#1}%
         \kern-0.1em
       }%
     }%
   }%
}

\newcommand\Cac{1}
\newcommand\Cad{2}
\newcommand\Cbc{3}
\newcommand\Cbd{4}

\newcommand\QUAD{{\Delta}}
\newcommand{\LHS}{\mathrm{LHS}}
\newcommand{\RHS}{\mathrm{RHS}}

\journal{Annals of physics}

\begin{document}

\definecolor{urlorange}{HTML}{ff8000}
\hypersetup{
    colorlinks=true,
    citecolor=red,
    linkcolor=blue,
    filecolor=darkgreen,
    urlcolor=urlorange,
    pdftoolbar=false,
    pdfmenubar=false,
    pdftitle={Foreign exchange rates can violate Bell inequalities},
    }

\begin{frontmatter}

\title{Einstein-Podolsky-Rosen-Bohm experiments: a discrete data driven approach\footnote{With corrections}}

\cortext[cor1]{Corresponding author: deraedthans@gmail.com\\Published in Ann. Phys.: \url{https://doi.org/10.1016/j.aop.2023.169314}}
\address[FZJ]{J\"ulich Supercomputing Centre, Institute for Advanced Simulation, Forschungzentrum J\"ulich, D-52425 J\"ulich, Germany}
\address[RUG]{Zernike Institute for Advanced Materials, University of Groningen, Nijenborgh 4, NL-9747 AG Groningen, Netherlands}
\address[RAD]{Radboud University, Institute for Molecules and Materials, Heyendaalseweg 135, 6525AJ Nijmegen,  Netherlands}
\address[FRA]{Modular Supercomputing and Quantum Computing, Goethe University Frankfurt, Kettenhofweg 139, 60325 Frankfurt am Main, Germany}
\address[RWTH]{RWTH Aachen University, D-52056 Aachen, Germany}

\author[FZJ,RUG]{Hans De Raedt\corref{cor1}}
\author[RAD]{Mikhail I. Katsnelson}
\author[FZJ,FRA]{Manpreet S. Jattana}
\author[FZJ,RWTH]{Vrinda Mehta}
\author[FZJ]{Madita Willsch}
\author[FZJ]{Dennis Willsch}
\author[FZJ,RWTH]{Kristel Michielsen}
\author[FZJ]{Fengping Jin}

\begin{abstract}
We take the point of view that
building a one-way bridge from experimental data to
mathematical models instead of the other way around
avoids running into controversies resulting from attaching meaning to the symbols used in the latter.
In particular, we show that adopting this view
offers new perspectives for constructing mathematical models for and interpreting the results of
Einstein-Podolsky-Rosen-Bohm experiments.
We first prove new Bell-type inequalities constraining the values
of the four correlations obtained by performing Einstein-Podolsky-Rosen-Bohm experiments
under four different conditions.
The proof is ``model-free'' in the sense that it does not refer to any mathematical model
that one imagines to have produced the data.
The constraints only depend on the number of quadruples obtained by reshuffling the data in the four data sets
without changing the values of the correlations.
These new inequalities reduce to model-free versions of the well-known Bell-type inequalities if
the maximum fraction of quadruples is equal to one.
Being model-free, a violation of the latter by experimental data implies
that not all the data in the four data sets can be reshuffled to form quadruples.
Furthermore, being model-free inequalities, a violation of the latter by experimental data
only implies that any mathematical model assumed to produce this data does not apply.
Starting from the data obtained by performing Einstein-Podolsky-Rosen-Bohm experiments,
we construct instead of postulate mathematical models that describe the main features of these data.
The mathematical framework of plausible reasoning is applied
to reproducible and robust data, yielding without using
any concept of quantum theory, the expression of the correlation
for a system of two spin-1/2 objects in the singlet state.
Next, we apply Bell's theorem to the Stern-Gerlach experiment
and demonstrate how the requirement of separability leads to the quantum-theoretical description of
the averages and correlations obtained from an Einstein-Podolsky-Rosen-Bohm experiment.
We analyze the data of an Einstein-Podolsky-Rosen-Bohm experiment
and debunk the popular statement that Einstein-Podolsky-Rosen-Bohm experiments have vindicated quantum theory.
We argue that it is not quantum theory but the processing of data from EPRB experiments that should be questioned.
We perform Einstein-Podolsky-Rosen-Bohm experiments on a superconducting quantum information processor
to show that the event-by-event generation of discrete data
can yield results that are in good agreement with the quantum-theoretical description
of the Einstein-Podolsky-Rosen-Bohm thought experiment.
We demonstrate that a stochastic and a subquantum model can also produce data that are in excellent agreement with the quantum-theoretical description of the Einstein-Podolsky-Rosen-Bohm thought experiment.
\end{abstract}

\date{\today}

\end{frontmatter}
\clearpage
\tableofcontents

\section{Introduction}\label{section1}

\begin{figure*}[!htp]
\centering
\includegraphics[width=0.90\hsize]{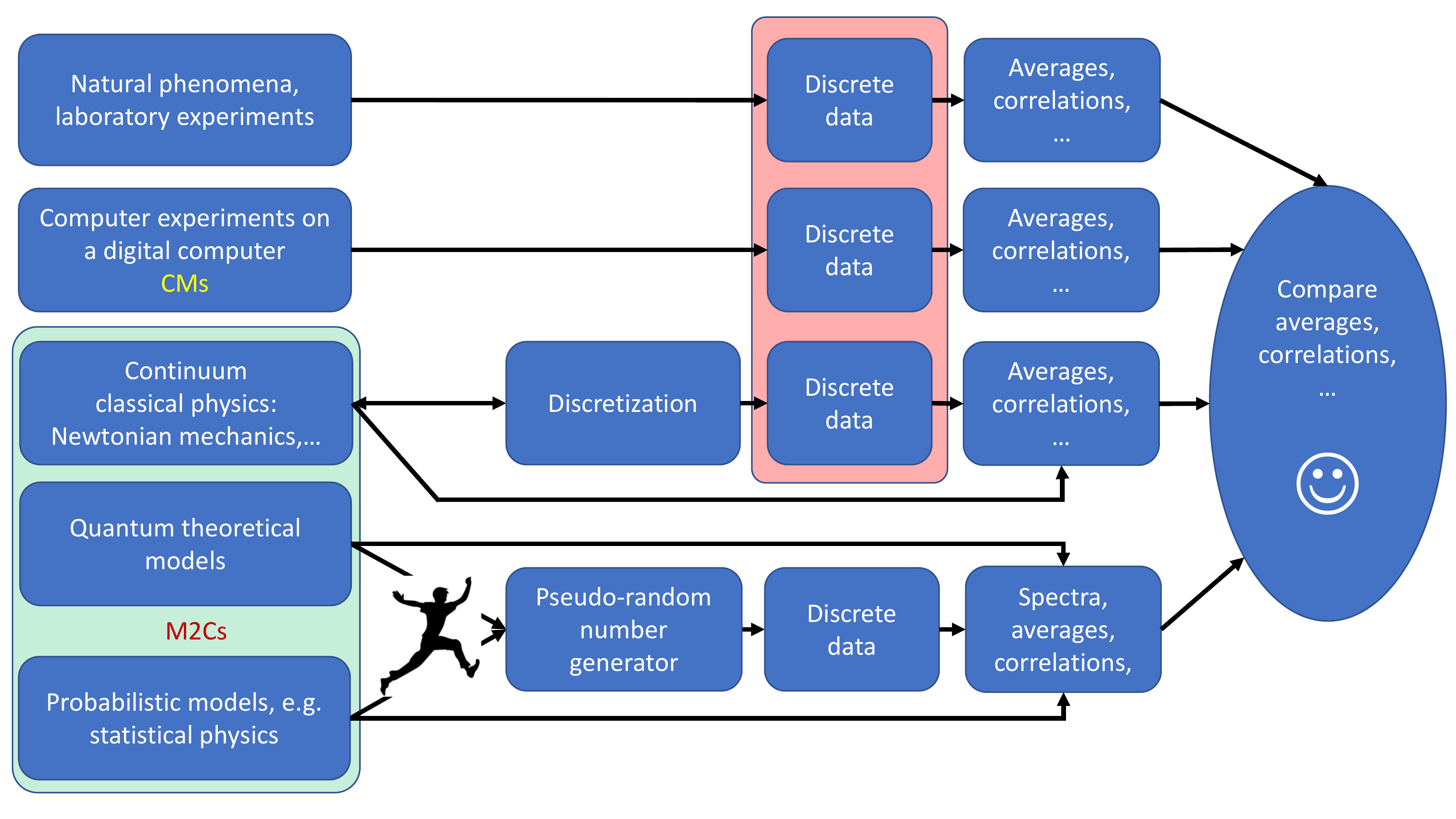}
\caption{(color online)
Graphical representation of the view adopted in this paper.
}
\label{wall}
\end{figure*}

All experiments which yield results
in numerical form generate a finite amount of discrete data represented by (ratios of) finite integers.
Obviously, also algorithms running on digital computers generate discrete data
(a finite number of bits).
The same can be said of analog simulations by means of e.g., electronic circuits.
In practice, the data gathered from these experiments
also comes in the form of a finite amount of sampled, discrete data, even though
we often imagine them as continuous.

In this paper, discrete data are considered to be immutable facts, free of personal judgment.
Of course, there first has to be a consensus among individuals that the discrete data are indeed immutable facts.
Once this consensus has been established these immutable facts constitute the ``reality'',
the ``real world'' that we refer to in this paper.
By adopting this very narrow definition of ``reality'', there is little room left for philosophical
arguments about the nature of reality, realism etc.~\cite{FINE96}.
In brief, experimental or computer generated data are considered as immutable facts, constituting the ``reality''.
At the risk of overemphasizing the importance of taking this narrow view of ``reality'',
it is necessary to carefully distinguish the definition of reality as immutable facts
adopted in this paper for the aim of analyzing specific scientific questions
from the question ``{\bf what is reality really?}'', which goes far beyond the scope of this paper.

We take the common view that a mathematical model (MM),
that is a model formulated in the language of mathematics,
of (the process that generates) the data should provide
a description of the discrete data that is more concise than simply tabulating all the data.
The MM should describe the data or the relevant features thereof, either in terms
of discrete data itself or, as is more common in physics, by
providing a function of one or more variables that fits well to the data.
If possible, a MM should also describe relations between features extracted from the data.

We distinguish between two classes of MMs.
The first class (M1C) contains all MMs which generate discrete data in a finite number of steps.
MMs of this class can be represented by a terminating algorithm running on a digital computer.
Any such algorithm is an instance of a computer model (CM),
the acronym that will be used to refer to the first class of MMs.
On the other hand, as digital computers are physical devices on which numerical experiments
are being carried out, CMs can also be viewed as metaphors for real experiments
in which {\sl all} conditions are known and under control (assuming the digital computer
is operating flawlessly which, in practice, is easily verified by repeating the numerical experiment)~\cite{RAED16c}.
Furthermore, the logical operation of the electronic digital computers we are all used to can equally well be
realized by a mechanical machine, albeit at great cost and great loss of efficiency.
Thus, any CM executed on a digital computer has, at least in principle, a macroscopic, mechanical equivalent.
The second class (M2C), symbolically represented in Fig.~\ref{wall}
by the green rectangle with rounded edges contains all MMs that do not belong to M1C.

Most of the fundamental models in theoretical physics are based on the notions of the space-time continuum and real numbers.
After suitable discretization,
the equations of classical physics for
Newtonian mechanics, Maxwell's electrodynamics, special relativity, etc., produce discrete data when
these equations are solved numerically on a digital computer.
Although the discretization procedure is not unique, different procedures all share
the property that they yield the same continuum model.
Thus, as indicated in Fig.~\ref{wall}, the relation between the MM and CM is bidirectional.

The transition from any probabilistic or quantum-theoretical model to discrete data
requires the use of an algorithm that is external to both these models.
This transition is unidirectional.
Conceptually, these models are separated from the discrete data by a gap
that takes the proportion of an abyss.
A probabilistic model is defined by its real-valued probability (density) measure on a probability space~\cite{KOLM56,GRIM01}.
It describes the probability (density) distribution of events~\cite{KOLM56,GRIM01}.
Probability theory does not contain a recipe/algorithm to generate the events.
Adding such an algorithm, which to make contact to the realm of discrete data
is necessarily finite and terminating (e.g., a pseudo-random number generator),
fundamentally changes the mathematical structure of the probabilistic model,
turning it into an event-by-event simulation on a digital computer, a CM.

In quantum theory, the state of the system  is described by a vector (or density matrix) in an abstract Hilbert space~\cite{NEUM55,BALL03}.
Quantum theory plus any of the interpretations allegedly explaining the existence
of the discrete, definite events encountered in real life faces a similar abyss.
Also, this M2C does not contain a recipe/algorithm to generate the events but, exactly as in the case
of probabilistic models, can be turned into CM by appealing to Born's rule, a key postulate of quantum theory.
The existence of the named abyss proves itself through the fact that
more than 100 years after their conception, there seem to be
irreconcilable differences in opinion about the interpretation
of probability~\cite{JAYN03} and quantum theory~\cite{HOME97,Leggett2005}.
The conundrum of not being able to deduce within the context of the latter
theory that, in general, each measurement yields a definite outcome~\cite{HOME97,Leggett2005}
has, for a Curie-Weiss model of the measurement device, been shown
to be amenable to detailed analysis without invoking elusive concepts such
as the wave function collapse~\cite{NIEU13,ALLA17,Nieuwenhuizen2022}.

From the foregoing, it is clear that discretizing differential equations or using of
pseudo-random number generators, maps M2Cs onto CMs.
This category of CMs is ``inspired'' by M2Cs.
In contrast, there are CMs that are not based, also do not have any relation to,
one of the M2Cs that are used to describe physical phenomena.
They are defined by specifying a set of rules, an algorithm.
A prominent example of such a CM is a pseudo-random number generator.
Its algorithm consists of a set of arithmetic operations, designed to create
the illusion that the numbers being generated are unpredictable.
More generally, discrete event simulations belong to this category of CMs.

In many but not all cases, the discrete data generated by laboratory experiments,
computer experiments on a digital computer and classical physics model
may directly be confronted with each other, as indicated by the red rectangle containing
the three discrete data boxes in Fig.~\ref{wall}.
For instance, we can compare the observed trajectory of a satellite with the
numerical solution of the classical equation of motion for that object.

In general, the comparison between experimental data and discrete data
obtained from model calculations is through averages, correlations, etc.,
that is through quantities that capture the salient features of the discrete data.
The applicability of the models is established a posteriori by comparing
their features with those of the laboratory experiment or, if the latter is not
available, by comparing features among models.

If the comparison is considered to be successful (by some necessarily subjective criterion),
as indicated by the smiley, the MM has been validated.
If the MM does not describe the discrete data, it is not ``wrong'' (assuming it is mathematically sound).
Then, we have two options.
First, following common practice of all subfields of physics,
we should try to include into the MM elements that are of relevance to the real experiment
but have been left out in the construction of the MM.
Second, like in the case of classical mechanics failing to describe relativistic mechanics,
one has to come up with a new MM, a task that is much more daunting than the first one.

With one exception, all the arrows in Fig.~\ref{wall} are unidirectional.
Therefore, if the comparison between experimental data and a MM (e.g., a probabilistic or a quantum model)
is found to be unsatisfactory (by whatever criterion), it is a logical fallacy to conclude that
one of the premises underlying the MM must be ``wrong''.
The logically correct conclusion is that the predictions of the MM in terms of averages, correlations, etc.
do not agree. Of course, logically unjustified conclusions may sometimes provide inspiration
to construct other MMs that eliminate some or all of the differences in their predictions.

Nevertheless, once the model has been validated, the assumption that
the phenomenon, which was the subject of the laboratory experiment, shares the same
properties as the model is expressing a belief, a logical fallacy of false analogy.

A recurring theme of this paper is that there is no direct relation between
any of the properties of CMs, MMs (all contained in the green area of Fig.~\ref{wall}),
and those of natural phenomena or laboratory experiments.
In this view, there exists an impenetrable barrier
between discrete data produced by (computer) experiments
and MMs designed to capture the salient features of these data, see Fig.~\ref{wall}.
Indeed, there is no reason why valid ``theorems'' derived from a MM
should have a bearing on ``reality'' represented by discrete data.
The idea that they would have a bearing reminds us of the ``mind projection fallacy'',
the assertion that one's own thoughts and sensations are realities
existing in the world in which we live~\cite[p. 22]{JAYN03}.

\subsection{Some further thoughts on relations between ``model'' (theory) and ``reality''}

To the best of our knowledge, a view similar to the ``mind projection fallacy'' was first clearly expressed in Heinrich Hertz' last work
``Die Prinzipien der Mechanik in neuem Zusammenhange dargestellt (1894)''~\cite{HERTZ2014}.
In the introduction, Hertz discussed the relation between object and observer, subject and
object, nature and culture, theory and practice.
In the introduction he wrote~\cite{HAGE17}
``We form for ourselves mental pictures or symbols of
external objects; and the form which we give them is such that the necessary
consequences of the pictures in thought are always the pictures of
the necessary consequences in nature of the things pictured.''
Hertz's position represents a significant departure from Galileo's
view that the ``book of nature is written in geometric symbols'',
a position which does not assume that the mathematical symbols used in physical theories
have meaning outside these theories.
The first lines of Ref.~\cite{EPR35} read ``Any serious consideration of a physical theory must take into account the distinction
between the objective reality, which is independent of any theory, and the physical
concepts with which the theory operates. These concepts are intended to correspond with the
objective reality, and by means of these concepts we picture this reality to ourselves'',
which seem to be in concert with Hertz' view.
An in-depth discussion of the relevance of Hertz' view to the foundations
of quantum theory can be found in Ref.~\cite{Khrennikov2019}.

More generally, the essential part of the whole philosophy, from Ancient Greece
to modern times, is related to the problem of the adequacy of our worldview and
its relation to ``reality'' (whatever that ``reality'' means). Of course, it
is far beyond both the scope of the paper and the expertise of the authors to
discuss this issue in its generality, but several remarks seem to be not only
useful but even necessary.

\begin{itemize}
\item
There is a strong tendency to identify our description of reality, represented
in a mathematical way, with reality. Usually, this tendency is associated with
the philosophy of Plato and his followers but one can go even farther in the
past, e.g., to Pythagoras. Importantly, this view is still alive and quite
popular among physicists and mathematicians, the philosophical views of
Heisenberg probably being the most striking example~\cite{Heisenberg1959}. There
are many varieties of this worldview but, roughly speaking, mathematics is
identified with the deepest level of ``reality''. Our physical world is
supposed to be a shadow of this true, or supreme, reality.
\item
One advantage of this approach is obvious: ``the unreasonable effectiveness of
mathematics in natural sciences''~\cite{Wigner1960} is no longer a problem.
Strangely enough, counterexamples to this statement such as the famous
Banach-Tarski paradox~\cite{Banach1924} are routinely ignored. In our physical
world, we cannot cut a ball into a finite number of pieces and reconnect them
into a ball twice the size of the original ball. This observation alone should
force us to reconsider the idea that the connection between the world of
mathematical concepts and the physical world are related in trivial way.
\item
Importantly enough, even accepting Plato's main concept does not imply, in any
way, that this divine mathematics, this supreme reality, should coincide with
our human mathematics. The latter may be merely a projection, and the procedure
of projecting could change dramatically its character. There is some analogy
with Bohr's complementarity principle: an electron is neither wave nor
particle but these concepts naturally arise with our attempts to describe the
results of interaction of invisible micro objects like the electron with
macroscopic measuring devices~\cite{Irkhin2003}. We cannot go deeper into this
analogy here but should mention that there is also the complementarity of
continuity and discreteness which seems to be an unavoidable property of such a
projection~\cite{Irkhin2003}. This is directly related to the fact that, as
mentioned in the beginning of the introduction, the results of physical
experiments have finite precision and are represented by discrete numbers.
\item
The previous observations imply that even if we take the idealistic positions in
spirit of Plato or Hegel philosophies, one needs to distinguish carefully
between superior reality, whatever rational and even mathematical it may be by
itself, and its reflection in one's mind, unavoidably restricted by our own
everyday experience, by our language reflecting this experience (``The limits
of my language mean the limits of my world''~\cite{Witgenstein1922}) and by
physiology of our bodies and brains. \item The Hertzian view on scientific
theories as images of reality seems to be careful enough in this respect. The
rest depends on our general world view. For example, if we believe in evolution
and in the origin of our mind as a result of this evolution, these images should
be correct, at least to some significant degree. Indeed, our survival, as well
as the survival of our ancestors, heavily depends on them. This world view
already enforces some quite strong restrictions on the structure of both our
mind and physical reality~\cite{Vanchurin2022}. \item However, the previous
statement should not be misunderstood. What is required from the internal image
(world view) is its ability to make reasonably accurate predictions of the
events in the external world. The power of this ability is directly related to
the requirement of the robustness of the description. The latter lies at the
base of our logical inference approach to the foundations of quantum
theory~\cite{RAED14b}. Also, the compactness of the representation of the
information about the external world is crucially important to limit resources
necessary to operate with this information.  In our separation of conditions
principle we use this idea to construct the formal framework of quantum
theory~\cite{RAED19b}.
\end{itemize}

\section{Structure of the paper}\label{STRUC}

In this paper, we demonstrate by means of the application to Einstein-Podolsky-Rosen-Bohm (EPRB)
experiments and appeal to Bell's theorem that building a one-way bridge {\bf from} discrete data {\bf to}
a MM eliminates all the problems of interpreting the results of EPRB experiments. The reason for this is simple.
Starting from the discrete data, immutable facts,  instead of from imaginary MMs which usually
support very rich mathematical structures, there is no room for going astray in interpretations.

We start by describing the discrete data obtained by both EPRB thought and laboratory experiments,
see sections~\ref{EXPI},~\ref{LABE} and~\ref{COUNT},
and in Section~\ref{INEQDATA}, we present a new inequality for the correlations computed from these data.
The proof of this inequality does not depend on the existence of a MM
for the process that (one imagines having) produced the data.
This inequality is ``model free'', involving discrete data only.
Correlations obtained from EPRB experiments can never violate this inequality.
This inequality contains the Bell-CHSH inequality, valid for discrete data, as a special case.
A violation of the Bell-CHSH inequality by discrete data only indicates that not all
the data contributing to the correlations can be reshuffled to form quadruples, see~\ref{PAIRS} for
the definition of ``quadruples''.
It follows that any interpretation of a violation of the latter in terms of an imagined physical process
generating the data is a logical fallacy of the kind mentioned earlier.
A short note focusing on a derivation of this inequality for the simplest case can be found in Ref.~\onlinecite{DeRaedt2023}.

The model-free inequality puts a constraint on correlations
computed from the discrete data but does not contribute to
the theoretical modeling of the EPRB experiment.
In this paper we systematically scrutinize various alternatives for constructing such models
starting from the assumed or imagined features of the discrete data of an EPRB experiment.

We start by constructing, as opposed to postulating, the quantum-theoretical description of the EPRB experiment.
First, in Section~\ref{LI} we apply the elementary theory of plausible reasoning to data obtained
from reproducible and robust EPRB experiments and derive, without using
any concept of quantum theory, the expressions for the averages and the correlation
for a system of two spin-1/2 objects in the singlet state.
This approach builds a bridge between the discrete data
produced by experiment and a theoretical model but does not provide insight into the process that led to the data.

Second, in Section~\ref{SOC}, we demonstrate how the requirement
of separability of the condition under which the EPRB experiment
is carried out leads to the quantum-theoretical description of
the averages and correlations obtained from an EPRB experiment.
Remarkably, a crucial step in this construction is the application of Bell's theorem
to the Stern-Gerlach experiment.
Section~\ref{QADV} contains a discussion about the efficiency of quantum theory
in terms of compressing data and Section~\ref{NOGO}
presents a proof of a no-go theorem for quantum theory of two spins-1/2 objects.
We also identify the point at which interpretations of mathematical symbols that appear in MMs lose contact with the reality, represented by discrete data.

In Section~\ref{FAIL} we confront the data of an EPRB experiment~\cite{WEIH98} with
the quantum-theoretical predictions for two spin-1/2 objects in the singlet state,
debunking the popular statement that EPRB experiments confirm
these predictions~\cite{ADEN09,RAED12,RAED13a,Bednorz2017,Adenier2017}.
We also argue that it is not quantum theory but rather the practical realization of the EPRB experiments that
should be questioned.

Section~\ref{QCE} presents results of EPRB-like experiments performed
by means of a superconducting quantum information processor
and show that this event-by-event generation of discrete data
yields results that are in good agreement with the quantum-theoretical description
of the EPRB thought experiment.

Section~\ref{NQM} reviews non-quantum models (NQMs) that cannot (sections~\ref{BELL} and~\ref{EXTE})
and can (sections~\ref{TPM} and~\ref{EVENT}) reproduce
the averages and correlations of two spin-1/2 objects in the singlet or product state. Additional examples can be found in~\ref{RECO}.
We also provide an alternative proof of Fine's theorem~\cite{FINE82a,FINE82b} and discuss its implications in the light of the central theme of this paper.

In Section~\ref{CONC}, we summarize the key ideas and results of our work.

In the main text, we only address the main points.
Technicalities of proofs and examples illustrating the main points are given in the appendices.

Finally, a remark about the notation used throughout the paper:
discrete data generated by real (computer) experiments, e.g.,  $A_{C,n}$,
are labeled by a subscript $C$ specifying the condition under which the data has been generated
and a subscript to label the instance ($n$) of the data item.
The data most likely change if the condition $C$ changes or the experiment is repeated. Note that this notation does not refer to any particular MM.
Objects that belong to the domain of MMs are regarded as functions
of the arguments that appear in parentheses, e.g., $A(\ba,\lambda)$.

\section{Einstein-Podolsky-Rosen-Bohm thought experiment}\label{EXPI}

\begin{figure}[!htp]
\centering
\includegraphics[width=0.90\hsize]{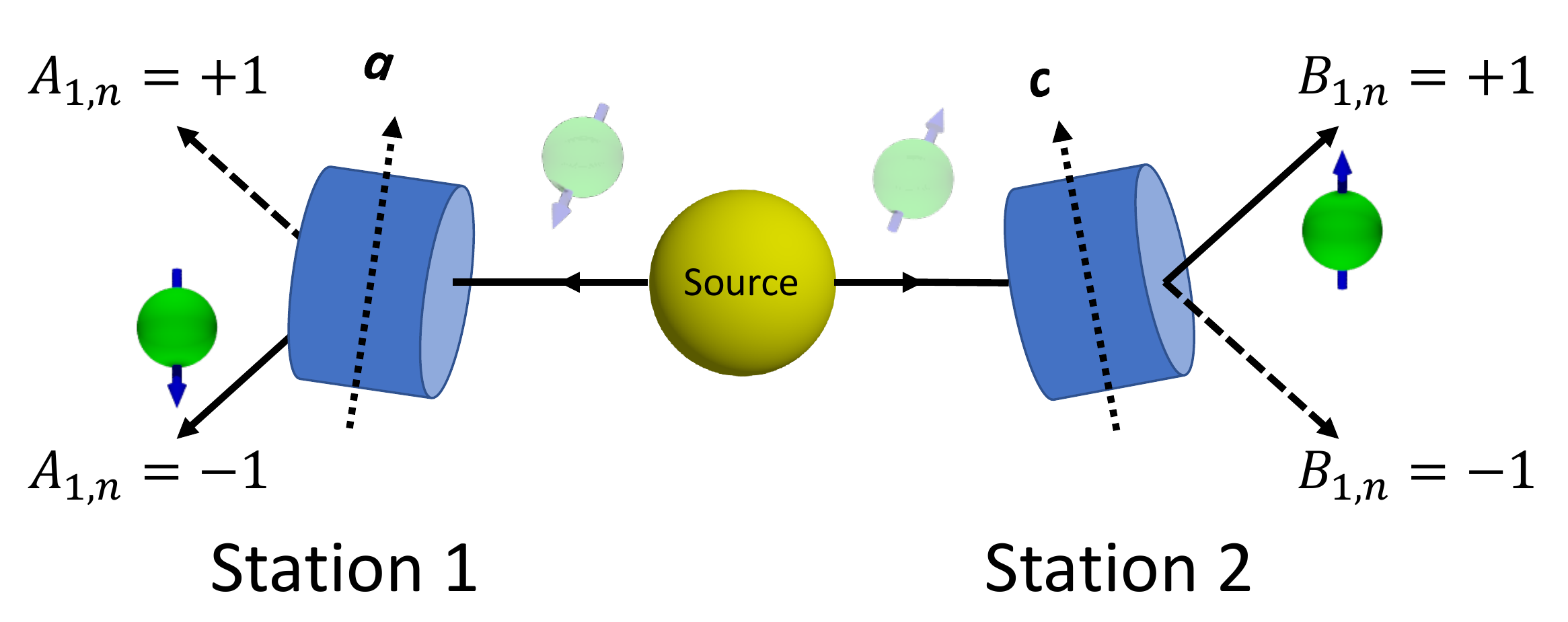}
\caption{(color online)
Conceptual representation of the Einstein-Podolsky-Rosen thought experiment~\cite{EPR35}
in the modified form proposed by Bohm~\cite{BOHM51}.
A source produces pairs of particles.
The particles of each pair carry opposite magnetic moments implying that
there is a correlation between the two magnetic moments of each pair leaving the source.
The magnetic field gradients of the Stern-Gerlach magnets (cylinders)
with their uniform magnetic field component
along the directions of the unit vectors $\ba$ and $\bc$ divert each incoming
particle into one of the two, spatially separated directions labeled by $+1$ and $-1$.
The pair $(\ba,\bc)$ determines the conditions, to be denoted by the subscript ``$1$'',
under which the data $(A_{\Cac,n},B_{\Cac,n})$ is collected.
The values of $A_{\Cac,n}$ and $B_{\Cac,n}$ correspond to the labels of the
directions in which the particles have been diverted.
The result of this experiment is the set of data pairs
${\cal D}_{\Cac}=\{ (A_{\Cac,1},B_{\Cac,1}),\ldots, (A_{\Cac,N},B_{\Cac,N})\}$
where $N$ denotes the total number of pairs emitted by the source.
Note that according to our notational convention, the subscript ``$\Cac$'' stands for
the condition under which the experiment has been performed.
}
\label{eprbidea}
\end{figure}

The Einstein-Podolsky-Rosen thought experiment was introduced to question the completeness
of quantum theory~\cite{EPR35}, ``completeness'' being defined in Ref.~\cite{EPR35}. 
Bohm proposed a modified version that employs the spins-1/2 objects instead of coordinates and momenta of
a two-particle system~\cite{BOHM51}.
This modified version, which we refer to as the Einstein-Podolsky-Rosen-Bohm (EPRB) experiment,
has been the subject of many experiments~\cite{KOCH67,CLAU78,ASPE82b,Kiess1993,WEIH98,CHRI13,HENS15,GIUS15,SHAL15} and
theoretical studies~\cite{BELL64,PEAR70,PENA72,FINE74,FINE82,FINE82a,FINE82b,MUYN86,KUPC86,BRAN87,JAYN89,BROD89,BROD93,PITO94,FINE96,KHRE09z,
SICA99,BAER99,HESS01a,HESS01b,HESS05,ACCA05,KRAC05,SANT05,
KUPC05,MORG06,KHRE07,ADEN07,Khrennikov2008,NIEU09,MATZ09,KARL09,KHRE09,GRAF09,KHRE11,NIEU11,Brunner2014,HESS15,KUPC16z,KUPC17,HESS17a,NIEU17,
Adenier2017,Khrennikov2018,Drummond2019,Lad2020,Blasiak2021,Cetto2021,Lad2022}.

The essence of the EPRB thought experiment is shown and described in Fig.~\ref{eprbidea}.
Performing the EPRB thought experiment under the first condition
defined by the directions $(\mathbf{a},\mathbf{c})$ yields the data set of pairs
\begin{eqnarray}
{\cal D}_{\Cac}&=&\{(A_{\Cac,n},B_{\Cac,n})\,|\,A_{\Cac,n},B_{\Cac,n}=\pm1\,;\,n=1,\ldots,N\}
\;,
\label{DATA0}
\end{eqnarray}
where $N$ is the number of pairs emitted by the source.
In this paper, we reserve the symbol $A$ ($B$) for representing discrete data (if it carries subscripts denoting
conditions or a model function thereof if it has arguments enclosed in parentheses)
originating from stations 1 (2) of the EPRB experiment shown in Fig.~\ref{eprbidea}.

Repeating the EPRB thought experiment under the three conditions $(\ba,\bd)$, $(\bb,\bc)$, and $(\bb,\bd)$
yields the corresponding data sets ${\cal D}_{\Cad}$, ${\cal D}_{\Cbc}$ and ${\cal D}_{\Cbd}$.
We repeat once more that according to our notational convention, the subscripts represent
the condition under which the experiment has been performed.

The data sets ${\cal D}_{s}$ for $s=1,2,3,4$ are {\sl imagined} exhibiting the following features:
\begin{itemize}
\item[1.]
There is no relation between the numerical values of $A_{s,n}$ and $A_{s,n'}$ if $n\not=n'$
and similarly for $B_{s,n}$ and $B_{s,n'}$.
This implies that, based on the knowledge of all $A_{s,n\not=m}$ and all $B_{s,n\not=m}$,
it is impossible to predict $A_{s,m}$ or $B_{s,m}$ with certainty.
Similarly, it is impossible to predict with certainty $A_{s,m}$ or $B_{s,m}$ knowing all $A_{s',n}$ or $B_{s',n}$ for all $s'\not=s$ and all $n$.
\end{itemize}

Inspired by the quantum-theoretical description of the EPRB thought experiment (see~\ref{QTDE}),
the data sets ${\cal D}_{s}$ for $s=1,2,3,4$ are {\sl imagined} exhibiting the following additional features:
\begin{itemize}
\item[2.]
The averages and the correlation defined by
\begin{eqnarray}
E^{(1)}_{s} =\frac{1}{N}\sum_{n=1}^N A_{s,n}
\quad,\quad
E^{(2)}_{s} =\frac{1}{N}\sum_{n=1}^N B_{s,n}
\quad,\quad
E^{(12)}_{s}= \frac{1}{N}\sum_{n=1}^N A_{s,n}B_{s,n}
\;,
\label{CORR}
\end{eqnarray}
are invariant under simultaneous rotation of the Stern-Gerlach magnets,
that is they can only depend on the directions of Stern-Gerlach magnets through the scalar product
of their respective direction vectors.
\item[3.]
The averages $E^{(1)}_{s}\approx0$, $E^{(2)}_{s}\approx0$, and
correlations $E^{(12)}_{1}\approx-\ba\cdot\bc$,
$E^{(12)}_{2}\approx-\ba\cdot\bd$,
$E^{(12)}_{3}\approx-\bb\cdot\bc$, and
$E^{(12)}_{4}\approx-\bb\cdot\bd$.
\item[4.]
From 3, it follows that the data shows perfect anticorrelation (correlation), that is $A_{s,n}=-B_{s,n}$ ($A_{s,n}=+B_{s,n}$)
for all $n=1,\ldots,N$,
if the directions of the Stern-Gerlach magnets are the same (opposite).
\end{itemize}
\begin{figure*}[!htp]
\begin{center}
\includegraphics[width=14cm]{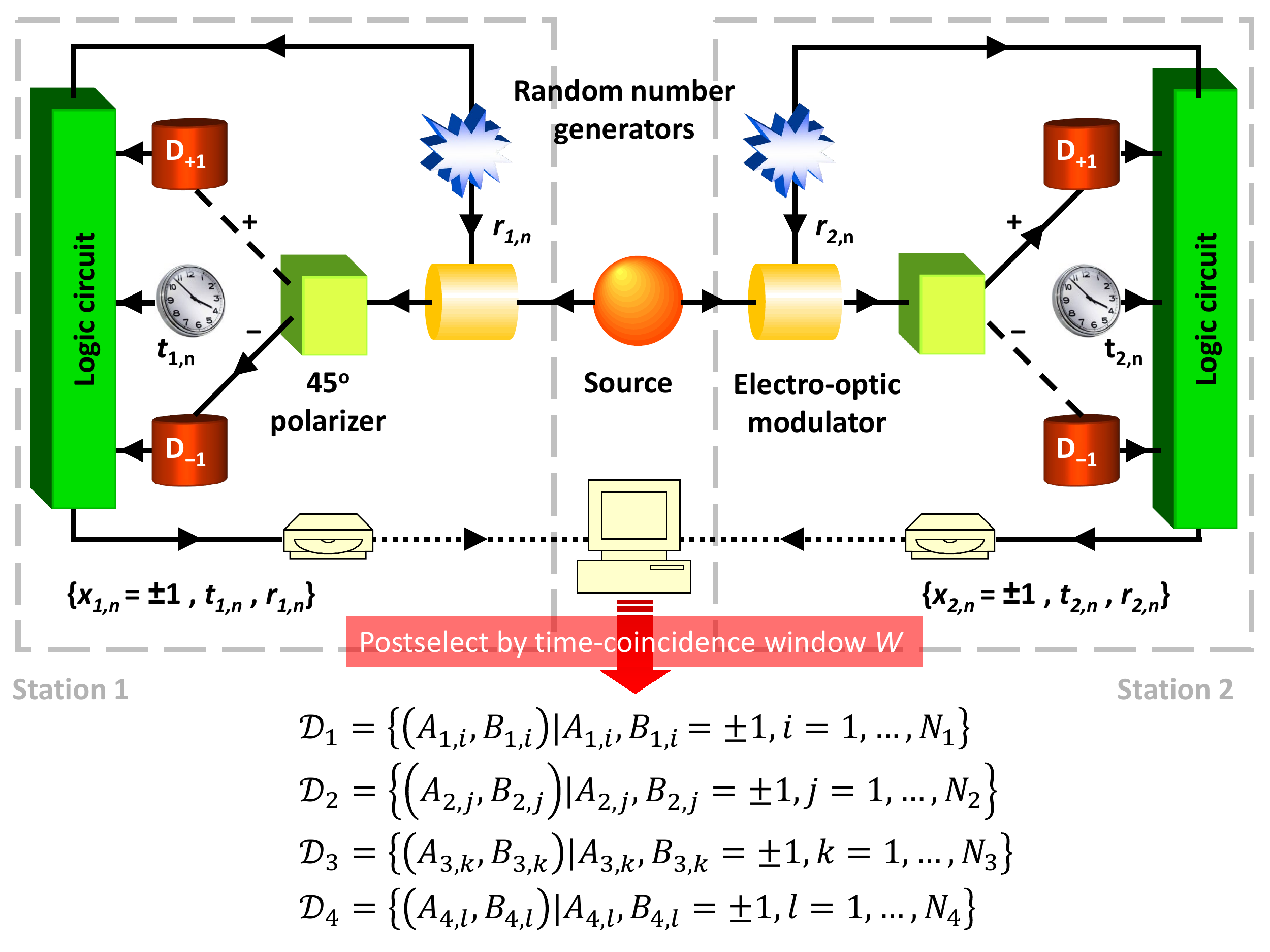}
\caption{Schematic diagram of an EPRB experiment with photons performed by Weihs {\sl et al.}~\cite{WEIH98,WEIH00}.
A source emits pairs of photons in spatially separated directions.
Photons arriving at station $1$ pass through an electro-optic
modulator (EOM) which rotates the polarization of the photon that passes through it
by an angle corresponding to the voltage applied to that EOM.
The latter is controlled by a binary variable $r_{1,n}$, which is chosen at random~\cite{WEIH98,WEIH00}.
The two different angles to choose from in station 1 (2) are
denoted by two-dimensional vectors $\ba$ ($\bb$) and $\bc$ ($\bd$).
As the photon leaves the EOM, a polarizing beam splitter
directs the photon to either detector $D_{+1}$ or detector $D_{-1}$.
Depending of the detection efficiency (about 5\%~\cite{WEIH98}),
one of these detectors fires, producing either a signal $x_{1,n}=+1$
or a signal $x_{1,n}=-1$ and a time stamp $t_{1,n}$.
Each triple $(x_{1,n},t_{1,n},r_{1,n})$ is written to a file.
The same holds for photons arriving at station $2$.
After all data has been written to the two files, that is when the experiment has finished,
a time window $W$ is used to remove all data that does not satisfy a time-coincidence criterion~\cite{WEIH98,WEIH00}.
The remaining data is organized in four data sets,
corresponding to the four different values of the pairs of random numbers $(r_{1,n},r_{2,n})$.
Note that the values of the $A$'s ($B$'s) that appear in say ${\cal D}_{\Cac}$ (${\cal D}_{\Cbc}$),
and those that appears in say ${\cal D}_{\Cad}$ may be different,
even though they may have been recorded for the same angle $\ba$ ($\bb$).
The four sets of discrete data ${\cal D}_{\Cac}$, ${\cal D}_{\Cad}$, ${\cal D}_{\Cbc}$, and ${\cal D}_{\Cbd}$
are the result of the experiment for the particular value of the time-coincidence window $W$.
}
\label{weihsexp}
\end{center}
\end{figure*}

\section{Einstein-Podolsky-Rosen-Bohm laboratory experiment}\label{LABE}

An EPRB laboratory experiment is, of course, more complicated than the thought experiment.
In Fig.~\ref{weihsexp}, we show a schematic of the EPRB experiment with photons performed by
Weihs et al.~\cite{WEIH98,WEIH00}.
In experiments with photons, the photon polarization plays the role of the spin-1/2 object
in the EPRB thought experiment depicted in Fig.~\ref{eprbidea} (see Section~\ref{INSTEAD}).
Prominent features of this particular experiment are that for each event that triggers
the emission of photons by the source,
binary random numbers are used to select one of the two EOM settings
and that all the time tags and detector clicks are stored in files
which can be analyzed long after the experiment has finished (as we do in this paper, see Section~\ref{FAIL}).

A key element, not present in the thought experiment, is a procedure to identify pairs of particles.
Many EPRB experiments~\cite{KOCH67,CLAU78,ASPE82b,WEIH98,WEIH00,ADEN12,AGUE09},
use the $t$'s, the time tags, for coincidence counting.
More recent experiments use thresholds on the voltages generated by the transition edge detectors
to identify pairs~\cite{GIUS15,SHAL15}.
In essence, in all cases, the identification procedure removes (a lot) of data from the
raw data sets~\cite{WEIH98,WEIH00,ADEN12,AGUE09,GIUS15,SHAL15}.
As indicated in Fig.~\ref{weihsexp}, this procedure yields data sets of different size
$N_{\Cac}$, $N_{\Cad}$, $N_{\Cbc}$, and $N_{\Cbd}$.
In the following, to simplify the discussion and notation somewhat,
we truncate the four data sets by keeping only the first
$N=\min (N_{\Cac}, N_{\Cad}, N_{\Cbc}, N_{\Cbd})$ data pairs.
The four data sets then read (see also Fig.~\ref{weihsexp})
\begin{eqnarray}
{\cal D}_{s}&=&\{(A_{s,n},B_{s,n})\,|\,
A_{s,n},B_{s,n}=\pm1\,;\,n=1,\ldots,N\}
\;,
\label{DATAz}
\end{eqnarray}
where $s=1,2,3,4$ labels the different runs of the experiment.

The identification process is irrelevant for the material presented in this paper, except for Section~\ref{FAIL} and also for sections~\ref{TPM} and~\ref{EVENT} in which we briefly discuss
a probabilistic M2C and an event-by-event, cause-and-effect CM which describe, respectively
generate the raw data $\{x_{i,n},t_{i,n},r_{i,n}\}$ for $i=1,2$ (see Fig.~\ref{weihsexp})
and can reproduce the averages and the correlation obtained from the quantum-theoretical description of the EPRB experiment.
As indicated in Fig.~\ref{weihsexp}, the data that are subject to further analysis
are only those that remain after the identification process has played its part.

\subsection{What is the main issue?}\label{ISSUE}

The following is an attempt to explain the main issue without taking recourse to mathematics.
Therefore, some aspects which are important for a precise formulation of the issue have been
left out. They are mentioned in the sections that follow.

Referring to Fig.~\ref{eprbidea}, imagine that the particle of a pair
traveling to the left (right) is very close to the leftmost (rightmost) magnet
but has not yet interacted with it (we assume that there is no faster-than-light communication between the particles).
Also imagine that the distance between the two Stern-Gerlach magnets
is so large that light emitted by one particle will not arrive at the other particle
before both particles complete their journeys by arriving at one of the detectors.
Under these conditions, changing the direction of the left (right) magnet
cannot have an effect on the particle passing through the right (left) magnet.
Thus, under these circumstances, knowledge of $A_{1,n}$'s ($B_{1,n}$'s) value
cannot affect $B_{1,n}$'s ($A_{1,n}$'s) value.
The picture of what happens to the particle going left
is completely separated from the picture of what happens to the particle going right.

Next, consider the case in which
the directions of the two magnets are either parallel or antiparallel, that is $\bc=\pm\ba$.
Then, according to the features of the data listed above,
\begin{enumerate}
\item
The values of the $A_{1,n}$ and $B_{1,n}$ for the $n$th pair,
are unpredictable, randomly taking values $\pm1$.
\item
The value of the product $A_{1,n}B_{1,n}=\mp1$ for all $n=1,\ldots,N$ pairs,
depending on the direction $\bc=\pm\ba$ of the magnets.
\end{enumerate}
Thus, even though the value of, say $A_{1,n}$ is random,
once it is known, because of the assumed perfect (anti)correlation, see the above feature 2,
the value of $B_{1,n}$ is known too, even before it is actually recorded.

In 1964, Bell~\cite{BELL64} presented a simple model that (i) describes the two-particle system
in terms of two separated one-particle systems and (ii) provides a picture of the observations that we have just described.
The two one-particle systems are separated in the sense that what happens to
a particle only depends on the direction of the Stern-Gerlach magnet with which it interacts and on
some variables that it shares with the other particle, the initial values of which are
determined at the time the particles leave the source.

In the same paper~\cite{BELL64}, Bell also proved his theorem (see Section~\ref{BELL} for a precise statement
of the theorem) stating that there does not exist
a description in terms of two separated one-particle systems that yields
the correlation ($-\ba\cdot\bc$) of two spin-1/2 objects in a singlet state.
Thus, although Bell's simple model can reproduce the main features listed in points 1 and 2 above,
it fails to agree with the quantum-theoretical description of the EPRB thought experiment.

The key question is then ``what is the outcome of an EPRB laboratory experiment?''
Instead of generating data for many directions of the Stern-Gerlach magnets
and comparing the correlation with the quantum-theoretical result $-\ba\cdot\bc$,
it is easier to demonstrate a violation of the so-called Bell-CHSH inequality (see Section~\ref{INEQDATA}),
an inequality that Bell used to prove his theorem.
The argument is that any two-particle system which can be separated into two one-particle systems
as envisaged by Bell cannot violate a Bell-CHSH inequality.
Therefore, the argument goes, if the experimental data violates the Bell-CHSH inequality,
the ``separability principle'',
which asserts that any two spatially separated systems possess their own separate real states~\cite{Howard1985},
or in Bell's words, that ``mutually distant systems are independent of one another''~\cite{Bell1966},
must be abandoned.

The main issue is that the logic of this argument is seriously flawed.
The Bell-CHSH inequality holds for some MMs but certainly not
for experimental data (see Section~\ref{INEQDATA}).
Thus, a violation of the Bell-CHSH inequality by experimental data can only imply that this particular MM does not apply to the experiment.
Moreover, as shown in Section~\ref{FAIL},
the analysis of data obtained from an EPRB laboratory experiment shows
that (i) under suitable conditions, this experiment yields $-\ba\cdot\bc$ to good approximation
and (ii) by changing these conditions a little, that correlation becomes compatible with
the results of a separable, Bell-like model.
From the viewpoint portrayed by Fig.~\ref{wall}, none of the apparent conflicts is surprising.
They all result from the idea that ``theorems'' derived in the context of a MM
have a bearing on the ``reality'' represented by experimental data.
As mentioned in Section~\ref{section1}, if a MM leads to the conclusion that
there is a conflict with the ``reality'' represented by experimental data,
the appropriate course of action is to revise/extend/abandon the MM, not immediately call into
question the elementary concepts on which our picture-building of natural phenomena is based.

\section{Description of discrete data: statistics}\label{COUNT}

As the $A$'s and $B$'s with different indices $n$ and different indices $s$
are assumed to be unrelated, the order in which data items appear
is irrelevant for the characterization of the data set.
In this case, the (relative) frequencies by which a pair of values $(x,y)$ ($x,y=\pm1$)
appears in the data set Eq.~(\ref{DATA0})
are sufficient to characterize this data set.
Formally, for the data collected under condition $s=1,2,3,4$, these frequencies are defined by
\begin{eqnarray}
f_{s}(x,y)=\frac{1}{N}\sum_{n=1}^N \delta_{x,A_{s,n}}\delta_{y,B_{s,n}}
=\frac{1}{4N}\sum_{n=1}^N (1+xA_{s,n})(1+yB_{s,n})
\quad,\quad x,y=\pm1
\;,
\label{DATA1}
\end{eqnarray}
where the Kronecker delta $\delta_{i,j}$ takes the value one if the variables
$i$ and $j$ are equal and is zero otherwise.
By construction $0\le f_{s}(x,y)\le 1$ and $\sum_{x,y=\pm1} f_{s}(x,y)=1$.
Frequencies, such as the one  defined by Eq.~(\ref{DATA1}), are discrete-valued functions
of the variables $x$ and $y$ which take discrete values and are denoted as such,
using the subscript ``$s$'' to indicate that the data has been collected
under the conditions represented by the symbol ``$s$''.
Frequencies belong to the domain of discrete data, not to M2C.

As it is clear from Eq.~(\ref{DATA1}), computing a frequency is a form
of data compression. In this particular case,
computing the frequencies Eq.~(\ref{DATA1}) compresses the whole data set Eq.~(\ref{DATA0})
to three, discrete-valued numbers (three instead of four because of the normalization).

From Eq.~(\ref{DATA1}) it follows immediately that the averages and the correlation
and their relation to the frequencies Eq.~(\ref{DATA1}) are given by
\begin{subequations}
\label{DATA2}
\begin{eqnarray}
E^{(1)}_{s}&=&
\frac{1}{N}\sum_{n=1}^N A_{s,n}
=\sum_{x,y=\pm1} x f_{s}(x,y)
\;,
\label{DATA2a}
\\
E^{(2)}_{s}&=&
\frac{1}{N}\sum_{n=1}^N B_{s,n}
=\sum_{x,y=\pm1} y f_{s}(x,y)
\;,
\label{DATA2b}
\\
E^{(12)}_{s}&=&
\frac{1}{N}\sum_{n=1}^N A_{s,n}B_{s,n}
=\sum_{x,y=\pm1} xy f_{s}(x,y)
\label{DATA2c}
\;,
\\
f_{s}(x,y)&=&\frac{1+x\,E^{(1)}_{s}+y\,E^{(2)}_{s}+xy\,E^{(12)}_{s}}{4}
=\frac{1+x\,E^{(1)}_{s}}{2}\frac{1+y\,E^{(2)}_{s}}{2}
+\frac{xy\left(E^{(12)}_{s}-E^{(1)}_{s}E^{(2)}_{s}\right)}{4}
\label{DATA2d}
\label{DATA3}
\;
\end{eqnarray}
\end{subequations}
showing that the frequencies Eq.~(\ref{DATA1}) or the expectations $E^{(1)}_{s}$,
$E^{(2)}_{s}$, and $E^{(12)}_{s}$ are equivalent characterizations of the data
in the set Eq.~(\ref{DATAz}).

Later, we need the notion of ``independence'' and this is a good place to discuss this notion.
The two variables $x$ and $y$ are said to be independent if the frequency $f_{s}(x,y)$
can be written in the factorized form $f_{s}(x,y)=f_{s}^{(1)}(x|\ba,\bc)f_{s}^{(2)}(y|\ba,\bc)$
where $f_{s}^{(i)}(z)=({1+z\,E^{(i)}_{s}})/{2}$ for $i=1,2$ are the
marginal frequency distributions of $f_{s}(x,y)$.
Equation~(\ref{DATA3}) shows that $x$ and $y$ are independent if and only if
the correlation $E^{(12)}_{s}-E^{(1)}_{s}E^{(2)}_{s}=0$.
In general, independence implies vanishing correlation~\cite{GRIM01}
but for two-valued variables the vanishing of correlations also implies independence.

Keeping the condition fixed,
different data sets ${\cal D}_s$ that yield the same frequencies $f_{s}(x,y)$
are equivalent in the sense that they yield the same values for the averages and the correlation.
Repeating an EPRB laboratory experiment with the same conditions
is expected (see Section~\ref{LI}) to yield numerical values of the frequencies $f_{s}(x,y)$ that are subject
to statistical fluctuations which decrease as the number of pairs $N$ increases.

\section{Model-free inequality for correlations computed from discrete data}\label{INEQDATA}

Motivated by the work of Bell~\cite{BELL93} and Clauser et al.~\cite{CLAU69,CLAU74}, many EPRB experiments~\cite{CLAU78,ASPE82b,WEIH98,CHRI13,HENS15,GIUS15,SHAL15}
focus on demonstrating a violation of the Bell-CHSH inequality~\cite{CLAU69,BELL93}.
To this end, one collects experimental data for four, well-chosen conditions denoted by
$\Cac$, $\Cad$, $\Cbc$ and $\Cbd$, and computes the corresponding
correlations according to Eq.~(\ref{DATA2c}).

In ~\ref{DISD}, we present a rigorous proof that
for any (real or computer or thought) experiment producing discrete data in the range $[-1,+1]$,
the correlations computed from the four data sets Eq.~(\ref{DATAz}) must satisfy the model-free inequalities
\begin{eqnarray}
\left|E^{(12)}_{\Cac\phantom{\bd}}\mp E^{(12)}_{\Cad}\right|+\left|E^{(12)}_{\Cbc}\pm E^{(12)}_{\Cbd}\right|
&\le& 4-2\QUAD
\;,
\label{DATA4}
\end{eqnarray}
where $0\le\QUAD\le 1$ is the maximum fraction of quadruples that can be found by rearranging/reshuffling the
data in
${\cal D}_{\Cac}$, ${\cal D}_{\Cad}$, ${\cal D}_{\Cbc}$, and ${\cal D}_{\Cbd}$
without affecting the value of the correlations
$E^{(12)}_{\Cac\phantom{\bd}}$, $E^{(12)}_{\Cad}$, $E^{(12)}_{\Cbc}$, and $E^{(12)}_{\Cbd}$.
For a detailed description of the reshuffling procedure and the definition of quadruples, see ~\ref{DISD}.
We emphasize that inequality Eq.~(\ref{DATA4}) holds for discrete data in the range $[-1,+1]$,
independent of how the data was generated and/or processed and, most importantly independent of any MM.
In essence, the upper bound $4-2\QUAD$ in Eq.~(\ref{DATA4}) results from the fact that for every quadruple
that we can create by reshuffling data pairs in each of the four data sets, the contribution to the expression on the left hand side of
Eq.~(\ref{DATA4}) is limited in magnitude by two, not by four.

The proof of Eq.~(\ref{DATA4}) requires that the $A$'s and $B$'s that appear in the expressions of the correlations
take discrete values (ratios of finite integers) in the interval $[-1,+1]$.
However, as mentioned in Fig.~\ref{eprbidea}, the data produced by an EPRB experiment is most
conveniently represented by variables $A$ or $B$ taking values $+1$ or $-1$ only.
In other words, Eq.~(\ref{DATA4}) covers both the case of data produced by EPRB experiments and general discrete data
in the range $[-1,+1]$.

It is expedient to introduce the Bell-CHSH function
\begin{eqnarray}
S_{\mathrm{CHSH}}=\max_{(i,j,k,l)\in\bm\pi_4}
\left|E^{(12)}_{i}-E^{(12)}_{j}+E^{(12)}_{k}+E^{(12)}_{l}\right|
\;,
\label{CHSHdef}
\end{eqnarray}
where $\bm\pi_4$ denotes the set of all permutations of $(\Cac,\Cad,\Cbc,\Cbd)$.
By application of the triangle inequality, it directly follows from
Eq.~(\ref{DATA4}) that for any (real or computer or thought) experiment
producing discrete data in the range $[-1,+1]$,
\begin{eqnarray}
S_{\mathrm{CHSH}}\le 4 - 2\Delta
\;.
\label{CHSHineq}
\end{eqnarray}

The symbol $\QUAD$ in Eqs.~(\ref{DATA4})--(\ref{CHSHineq}) quantifies the structure in terms of quadruples exhibited by
data ${\cal D}_{\Cac}$, ${\cal D}_{\Cad}$, ${\cal D}_{\Cbc}$, and ${\cal D}_{\Cbd}$.
If $\QUAD=0$, it is impossible to find a reshuffling that yields even one quadruple.
If $\QUAD=1$, the four sets can be reshuffled such that they can be viewed as being
generated from $N$ quadruples. Then we recover the ``model-free'' Bell-CHSH inequality for discrete data
\begin{eqnarray}
S_{\mathrm{CHSH}}\le 2
\;,
\label{CHSH}
\end{eqnarray}
usually derived within the context of a MM (see~\ref{ERGO})~\cite{CLAU69,BELL71,CLAU74,BELL93}.

In~\ref{EEPRB}, we briefly discuss the extended EPRB experiment (EEPRB)~\cite{SICA99,RAED20a} which 
always generates data for which $\QUAD=1$.
Perhaps somewhat counterintuitive is that $\QUAD\lesssim1$
if all the $A$'s and $B$'s take independent random values $\pm1$, see~\ref{DISD}.
In the case of interest, namely $E^{(12)}_{\Cac}=-\ba\cdot\bc$, $E^{(12)}_{\Cad}=-\ba\cdot\bd$,
$E^{(12)}_{\Cbc}=-\bb\cdot\bc$, and $E^{(12)}_{\Cbd}=-\bb\cdot\bd$,
the maximum value over all directions $\ba,\bb,\bc,\bd$
of the left-hand side of Eq.~(\ref{DATA4}) is $2\sqrt{2}\approx2.83$~\cite{CIRE80} (see~\ref{EXQT}),
implying that the maximum fraction of quadruples which can be found in
the data ${\cal D}_{\Cac}$, ${\cal D}_{\Cad}$, ${\cal D}_{\Cbc}$, and ${\cal D}_{\Cbd}$
must satisfy $\QUAD\le 2-\sqrt{2}\approx0.59$.
In~\ref{DISD}, we present simulation results obtained by generating four times one million independent pairs
according to the quantum-theoretical distribution of two spin-1/2 objects in the singlet state
and obtain $|E^{(12)}_{\Cac}-E^{(12)}_{\Cad}|+|E^{(12)}_{\Cbc}+E^{(12)}_{\Cbd}|\approx2.83$ and $4-2\QUAD\approx2.83$,
strongly suggesting that the value of quantum-theoretical upper bound $2\sqrt{2}$
is reflected in the fraction of quadruples that one can find by reshuffling the data.

{\bf
Suppose that the (post processed) data of an EPRB experiment
yields $S_{\mathrm{CHSH}}>2$, that is the data violates
the Bell-CHSH inequality Eq.~(\ref{CHSH}).
From Eq.~(\ref{CHSHineq}), it follows that $\QUAD\le 2-S_{\mathrm{CHSH}}/2<1$ if $S_{\mathrm{CHSH}}>2$.
Therefore, if $S_{\mathrm{CHSH}}>2$ not all
the data in ${\cal D}_{\Cac}$, ${\cal D}_{\Cad}$, ${\cal D}_{\Cbc}$,
and ${\cal D}_{\Cbd}$ can be reshuffled such that they originate from quadruples only.
Indeed, the data produced by these experiments have to comply with Eq.~(\ref{DATA4}),
and certainly not with the Bell-CHSH inequality. The reason is that the Bell-CHSH inequality is obtained from
Eq.~(\ref{DATA4}) in the exceptional case $\QUAD=1$ in which all data can be extracted from $N$ quadruples.

In other words, all EPRB experiments which have been performed and may be performed in the future
and which only focus on demonstrating a violation of Eq.~(\ref{CHSH}) merely provide evidence
that not all contributions to the correlations can be reshuffled to form quadruples (yielding $\QUAD<1$).
These violations do not provide a clue about the nature of the physical processes that produce the data.}

More specifically, Eq.~(\ref{DATA4}) holds for discrete data, rational numbers in the range $[-1,+1]$, irrespective of how
the data sets Eq.~(\ref{DATAz}) were obtained. Inequality~(\ref{DATA4}) shows that correlations of discrete data violate the
Bell-CHSH inequality Eq.~(\ref{CHSH}) only if not all
the pairs of data in Eq.~(\ref{DATAz}) can be reshuffled to create quadruples.
The proofs of Eq.~(\ref{DATA4}) and Eq.~(\ref{CHSH}) reflect a certain structure in the data.
They do not refer to notions such as ``locality'', ``realism'', ``non-invasive measurements'', ``action at a distance'',
``free will'', ``superdeterminism'', ``complementarity'', etc.
Logically speaking, a violation of Eq.~(\ref{CHSH}) by experimental data cannot be used to argue about the relevance of one or more of these notions used as motivation to formulate mathematical models of the process that generated the experimental data.

A violation of the original (non model-free) Bell-CHSH inequality $S\le2$ may lead to a variety of
conclusions about certain properties of a MM for which this inequality has been derived.
However, projecting these logically correct conclusions about the MM, obtained within the context of that MM,
to the domain of EPRB laboratory experiments requires some care, as we now discuss.

The first step in this projection is to feed real-world, discrete data (rational numbers in the range $[-1,+1]$)
into the original Bell-CHSH inequality $S\le2$ derived, not for discrete data as we did by considering the case $\QUAD=1$
in Eq.~(\ref{CHSHineq}), but rather in the context of some mathematical model, and to conclude that this inequality is violated.
Considering the discrete data for the correlations as given, it may indeed be tempting
to plug these rational numbers into an expression obtained from some mathematical model. However, then it is no longer clear
what a violation actually means in terms of the mathematical model because the latter (possibly
by the help of pseudo-random number generators) may not be able to produce these experimental data at all.
The second step is to conclude from this violation that the mathematical model
cannot produce the numerical values
of the correlations, implying that the mathematical model simply does not apply and has to be replaced by a more adequate one or
that one or more premises underlying the mathematical model must be wrong.
In the latter case, the final step is to project at least one of these wrong premises to properties of the world around us.

The key question is then to what extent the premises or properties of a mathematical model
can be transferred to those of the world around us.
{\bf Based on the rigorous analysis presented in this paper, the authors' point of view is that in the case of laboratory EPRB experiments, they cannot.}

Using only Eq.~(\ref{DATA4}),
the logically and mathematically
correct conclusion one can draw on the basis of the correlations computed from
these data obtained under conditions ($\Cac$, $\Cad$, $\Cbc$, $\Cbd$)
(listed in Section~\ref{EXPI}) is that a fraction of the experimental data Eq.~(\ref{DATAz})
can be reshuffled to create quadruples.
However, Eq.~(\ref{DATA4}) and the conclusions drawn from it do
not significantly contribute to the modeling of the EPRB experiment as such.
To this end, we need to develop MMs that describe the change of the data as the conditions change.
In the sections that follow, we explore various ways of constructing MMs which reproduce
the features of the experimental data mentioned in Section~\ref{EXPI}.

\section{Modeling data: logical inference}\label{LI}

In the exact sciences, an elementary requirement for the outcomes of
an experiment to be considered meaningful is that they are reproducible.
Obviously, in the case of the EPRB experiment, the (hypothesized) unpredictable nature
of the individual events renders the data set Eq.~(\ref{DATA0}) itself irreproducible.
However, repeating the experiment and analyzing the resulting data sets,
there is the possibility that the frequencies Eq.~(\ref{DATA1}) computed
from the different sets are reproducible (within reasonable statistical errors).
Or, if it is difficult to repeat the experiment, dividing the data set into
subsets and comparing the frequencies obtained from the different subsets
may also lead to the conclusion that the data is reproducible.

In this paper, we assume that the frequencies Eq.~(\ref{DATA1})
produced by the EPRB experiment are reproducible.
But even if the frequencies Eq.~(\ref{DATA1}) are reproducible,
they may still show an erratic dependence on $\ba$ or $\bc$
which can only be captured in tabular form.
The latter has very little descriptive power.
Thus, in order for an experiment to yield frequencies that allow
for a description that goes beyond simply tabulating all values,
the frequencies not only have to be reproducible but also have to be
robust, meaning that the frequencies should smoothly change
if the conditions under which the data was taken changes a little~\cite{RAED14b}.
This is the key idea of the logical inference approach for deriving, not postulating,
several of the basic equations of quantum physics~\cite{RAED14b,RAED15b,DONK16,RAED16b,RAED18a}.
We briefly recall the main elements of the logical inference approach as it has been applied to the EPRB experiment~\cite{RAED14b}.

The first step of the logical inference approach is to assign a plausibility~\cite{POLY54}
$0\le p(x,y|\ba,\bc)\le1$ for observing a data pair $(x=\pm1,y=\pm1)$ under the conditions $(\ba,\bc)$.
As explained in~\ref{PLAUS}, plausibility and (mathematical) probability are distinct concepts but
for the present, practical purposes, the difference is not important.
Recall that frequencies are discrete data whereas the concept of probability belongs to M2C.

The second step is to use the Cox-Jaynes approach~\cite{COX46,COX61,JAYN03}, the notion
of robust, reproducible discrete data and symmetry arguments
to derive, not postulate, $p(x,y|\ba,\bc)$.
The most general form of a function $p(x,y|\ba,\bc)$ of variables that only take values $\pm1$ reads
\begin{eqnarray}
p(x,y|\ba,\bc)&=&\frac{1+x\,E_{1}(\ba,\bc)+y\,E_{2}(\ba,\bc)+xy\,E_{12}(\ba,\bc)}{4}
\;,
\label{LI00}
\end{eqnarray}
see also Eq.~(\ref{DATA2d}).
Specializing to the case $E_1(\ba,\bc)=E_2(\ba,\bc)=0$
and accounting for rotational invariance by imposing $E_{12}(\ba,\bc)=E_{12}(\ba\cdot\bc)$, we have~\cite{RAED14b}
\begin{eqnarray}
p(x,y|\ba,\bc)&=&
\frac{1 + x\,y\,E_{12}(\ba\cdot\bc)}{4}=
\frac{1 + x\,y\,E_{12}(\theta)}{4}=p(x,y|\theta)
\;,
\label{LI0}
\end{eqnarray}
where $0\le\theta=\arccos(\ba\cdot\bc)\le\pi$ is the angle between the unit vectors $\ba$ and $\bc$.
Using the assumption that $(x_n,y_n)$ and $(x_m,y_m)$ are independent if $n\not=m$, we can
express the plausibility to observe several pairs in terms of
the plausibility Eq.~(\ref{LI0}) for a single pair~\cite{RAED14b}.

Expressing the notions of reproducible and robust statistical experiments mathematically leads
to the requirements (i) that frequencies should be used
to assign values to the plausibilities, thereby eliminating their subjective character~\cite{RAED14b}
and (ii) that, in the case at hand, the Fisher information
\begin{eqnarray}
I_\mathrm{F}(\theta)&=&\sum_{x,y=\pm1}\frac{1}{p(x,y|\theta)}\left(\frac{\partial p(x,y|\theta)}{\partial \theta}\right)^2
=\frac{1}{1-E_{12}^2(\theta)}\left(\frac{\partial E_{12}(\theta)}{\partial \theta}\right)^2>0
\;,
\label{LI1}
\end{eqnarray}
should be independent of $\theta$, positive and minimal~\cite{RAED14b}.
In~\ref{LIapp}, we solve this optimization problem. From Eq.~(\ref{LI3}) it is obvious that, discarding the irrelevant solution $n=0$,
the Fisher information is minimal if $n=1$, yielding
\begin{eqnarray}
E_{12}(\theta)&=&\cos (\theta+\varphi)\quad,\quad I_F(\theta)=1
\;.
\label{LI4}
\end{eqnarray}
where $\varphi$ is a constant of integration.

Requiring perfect anticorrelation (correlation coefficient $-1$)
in the case that $\ba=\bc$ ($\theta=0$), the phase $\varphi$ must be equal to $\pi$ and we obtain
\begin{eqnarray}
E_{12}(\theta)&=&-\ba\cdot\bc=-\cos\theta
\;,
\label{LI5}
\end{eqnarray}
and
\begin{eqnarray}
p(x,y|\ba,\bc)&=&\frac{1 - x\,y\,\ba\cdot\bc}{4}
\;.
\label{LI5a}
\end{eqnarray}
The correlation Eq.~(\ref{LI5}) with the minus sign known from
the quantum theory of two spin-1/2 objects in the singlet state (see~\ref{QTDE}),
plays a crucial role in Bell's theorem (see Section~\ref{BELL}).
Remarkably, the solution Eq.~(\ref{LI4}) with $\varphi=0$, that is
$E_{12}(\theta)=+\ba\cdot\bc$, cannot be obtained from
the quantum theory of two spin-1/2 objects~\cite{RAED19b}, see Section~\ref{NOGO}.

It is worth noting that the derivation that led to Eqs.~(\ref{LI5}) and~(\ref{LI5a})
does not make {\bf any} reference to concepts of quantum theory.
Apparently, the requirement that an EPRB experiment yields unpredictable
individual outcomes but reproducible and robust results for the averages (which are zero)
and perfect anticorrelation if $\ba=\bc$ suffices to show
that the correlation must be of the form Eq.~(\ref{LI5}), with the plausibility
to observe a pair $(x,y)$ given by Eq.~(\ref{LI5a}).
In short, any ``good'' EPRB experiment is expected to yield Eq.~(\ref{LI5a}).

Whether the theoretical description embodied in Eq.~(\ref{LI5a}) survives
the confrontation with the discrete data obtained from experiments
can only be established a posteriori.
For instance, in Section~\ref{FAIL} we plot Eq.~(\ref{LI5}) and the data
given by Eq.~(\ref{DATA2c})
and find satisfactory agreement in one case (Fig.~\ref{fig.A5}a) and significant disagreement in the other (Fig.~\ref{fig.A5}b).

Not surprisingly, the same logical inference
reasoning also yields a description of an experiment with an idealized
Stern-Gerlach magnet~\cite{RAED14b}.
To see this, imagine that the particles leaving the rightmost Stern-Gerlach magnet (see Fig.~\ref{eprbidea})
in the direction labeled $y=+1$ (or $y=-1$)
pass through another identical Stern-Gerlach magnet (not shown in Fig.~\ref{eprbidea}
but see Fig.~\ref{doublesg} in~\ref{SOCapp})
with its uniform magnetic field component in the direction $\bd$ and outputs labeled by $z=\pm1$.

Note that if an ideal Stern-Gerlach magnet is to function as an ideal filter device, we must require
that $z=y$ if $\bd=\bc$. Otherwise, assigning the attribute/label $y$ to a particle is meaningless.
Furthermore, we assume that the average of the $z$'s does not change
if we apply the same rotation to $\bc$ and $\bd$.
Denoting $\bc\cdot\bd=\cos\xi$, expressing the
notions of reproducible and robust statistical experiments as before
and accounting for rotational invariance,
it follows that the corresponding Fisher information
\begin{eqnarray}
I_\mathrm{F}(\xi)&=&\sum_{z=\pm1}\frac{1}{p(z|\xi)}
\left(\frac{\partial p(z|\xi)}{\partial \xi}\right)^2>0
\;,
\label{LI1b}
\end{eqnarray}
for the plausibility $p(z|\xi)$ to observe the event $z=\pm1$ under the condition $\xi$
must be independent of $\xi$, positive and minimal~\cite{RAED14b}.
Solving the optimization problem~\cite{RAED14b} yields $I_\mathrm{F}(\xi)=1$ and
\begin{eqnarray}
p(z|\xi)&=&\frac{1\pm z\cos\xi}{2}=\frac{1\pm z\, \bc\cdot\bd}{2}
\;,
\label{LI3a}
\end{eqnarray}
where the $\pm$ sign reflects the ambiguity in assigning $+1$ or $-1$ to one of the directions. In the following, we remove this ambiguity by opting for the solution with the ``+'' sign.

Note that quantum theory postulates Eq.~(\ref{LI3a}) (through the Born rule) whereas
the logical inference approach applied to ``good'' experiments
yield Eq.~(\ref{LI3a}) without making reference to any concept of quantum theory.

\subsection{Polarization instead of magnetic moments}\label{INSTEAD}

Most EPRB laboratory experiments employ photons instead of massive, electrically neutral magnetic particles.
The Stern-Gerlach magnets in Fig.~\ref{eprbidea} are then replaced by polarizers
with their plane of incidence perpendicular to the propagation direction of the photons (taken as
the $z$-direction in the following).
The unit vectors $\bc$ and $\bd$, describing the orientations
of the axes of the polarizers can be written as $\bc=(\cos c,\sin c,0)$ and $\bd=(\cos d,\sin d,0)$.

Let us first model a thought experiment aimed at demonstrating perfect filtering of the polarization.
We imagine placing two identical, ideal polarizers in a row (see also Fig.~\ref{eeprb1}).
Repeating the steps that led to Eq.~(\ref{LI3a})
and requiring that a polarizer acts as perfect filtering device for all $c=d$,
we find that reproducible and robust statistical experiments are described by the plausibility (see~\ref{LIapp})
\begin{eqnarray}
p(z|c-d)&=&\frac{1\pm z\cos\, n(c-d)}{2}=
\left\{\begin{array}{ccc}
\cos^2\frac{n(c-d)}{2}&,&z=\pm1\\ \\
\sin^2\frac{n(c-d)}{2}&,&z=\mp1
\end{array}\right.
\;.
\label{LI3b}
\end{eqnarray}
Comparing Eq.~(\ref{LI3b}) with Malus' law for the intensity
of polarized light (= many photons) passing through a polarizer,
we conclude that the solution with $n=1$ is incompatible with experimental facts
and should therefore be discarded.
The solution with $n=2$ yields Malus' law.
As discussed below, quantum theory attributes this empirical finding to the fact that photons are spin-one particles.

Repeating the derivation for the EPRB setup, with polarizers and photons
instead of Stern-Gerlach magnets and magnetic moments, we find
\begin{eqnarray}
E_{12}(\theta)&=&\pm\cos2\theta
\;.
\label{LI4p}
\end{eqnarray}
where $\theta=c-d =\arccos\bc\cdot\bd$ expresses the difference in the axis angles
of the polarizers, replacing the Stern-Gerlach magnets in Fig.~\ref{eprbidea}.
Requiring perfect anticorrelation (correlation coefficient $-1$)
in the case that $\theta=0$, only one of the two solutions in Eq.~\ref{LI4p}
survives and we have $E_{12}(\theta)=-\cos2\theta$.

Equations~(\ref{LI3b}) and~(\ref{LI4p}) differ from
Eqs.~(\ref{LI3a}) and~(\ref{LI5}) by the appearance of the extra factor of two in the argument of the cosine.
Within quantum theory, this can be explained as follows.
A photon is thought of as a massless, electrically neutral particle with spin $S=1$,
with a quantized polarization that can only take two values $\pm\hbar$.
In contrast, a neutron for instance is a massive spin $S=1/2$ particle, with a quantized magnetic moment
that can only take two values $\pm\hbar/2$.
The extra factor of two stems from the difference between $S=1$ and $S=1/2$ particles.

Logical inference provides a mathematically well-defined framework
to model empirical, statistical data acquired by reproducible and robust, that is ``good'' experiments.
It does not provide pictures of the individual objects and processes involved in actually producing the
discrete data. LI models belong to M2C.

\section{Modeling data: separation of conditions}\label{SOC}

Instead of postulating the axioms of quantum theory and reproducing textbook material (see also~\ref{QTDE}),
we use the EPRB experiment to show that its quantum-theoretical description
directly follows from another representation of the frequencies Eq.~(\ref{DATA1}).
By doing so, we do not need to call on Born's rule, for instance.
We proceed directly from discrete data and the logical inference description of Section~\ref{LI}
to the quantum-theoretical description.
In doing so, we avoid all metaphysical problems resulting from the various interpretations of quantum theory.
Thus, we construct a mapping {\bf from} the data (the frequencies) {\bf to} a M2C (quantum theory).

The basic idea is rather simple.
Instead of using the standard expression
\begin{eqnarray}
\langle x(c_1,c_2) \rangle=\sum_{k\in K} x(k) f(k|c_1,c_2)
\;.
\label{SOC0}
\end{eqnarray}
for the average of a function $x(k)$ over the events $k\in K$  that appear with
a (relative) frequency $f(k|c_1,c_2)$, depending on conditions represented by the symbols $c_1$ and $c_2$,
we search for functions ${\widehat x}(j,i|c_1)$ and ${\widehat f}(i,j|c_2)$ satisfying
$|{\widehat x}(j,i|c_1)|\le 1$ and $0\le|{\widehat f}(i,j|c_2)|\le 1$ and for which
\begin{eqnarray}
\langle x(c_1,c_2) \rangle&=&\sum_{k,k'\in K} {\widehat x}(k',k|c_1) {\widehat f}(k,k'|c_2)
\;,
\label{SOC1}
\end{eqnarray}
yields the same numerical value as given by Eq.~(\ref{SOC0}), for all $\langle x(c_1,c_2) \rangle$'s of interest.
Note that the condition $c_1$ appears in ${\widehat x}(k',k|c_1)$ and not in ${\widehat f}(k,k'|c_2)$.

{\bf
Recalling the recurring theme of this paper,
the transition from Eq.~(\ref{SOC0}) to Eq.~(\ref{SOC1})
is a jump over the impenetrable barrier between data (facts) and models thereof.
Although both representations yield the same averages,
any interpretation of the symbols ${\widehat x}(k',k|c_1)$
and ${\widehat f}(k,k'|c_2)$ in terms of discrete data, in terms of ``reality'', is problematic.
Indeed, there is no relation between
${\widehat x}(k',k|c_1)$ and ${\widehat f}(k,k'|c_2)$ and actual discrete data, other than that the sum
in Eq.~(\ref{SOC1}) yields these same numerical value for the average Eq.~(\ref{SOC0}).
The symbols ${\widehat x}(k',k|c_1)$ and ${\widehat f}(k,k'|c_2)$  do not belong to the realm of these data (facts),
they live in the domain of models only.
In Hertz's terminology, the connection with the original picture is completely lost.
Having lost the connection with ``reality'', there is complete freedom regarding
the interpretation one would like to attach to the symbols ${\widehat x}(k',k|c_1)$
and ${\widehat f}(k,k'|c_2)$.
In this paper, we adopt a pragmatic approach.
We refrain from giving an interpretation
to mathematical symbols except for those that represent the original discrete data which we aim
to describe.}

In matrix notation ${\widehat\bY}_{k,k'}(c)={\widehat y}(k,k'|c)$, Eq.~(\ref{SOC1}) reads
\begin{eqnarray}
\langle x(c_1,c_2) \rangle&=&
\mathbf{Tr\;} {\widehat\bX}^{\mathrm{T}}(c_1)\,{\widehat\bF}(c_2)
=\mathbf{Tr\;}{\widehat\bF}(c_2)\, {\widehat\bX}^{\mathrm{T}}(c_1)
\;.
\label{SOC2}
\end{eqnarray}
As Eq.~(\ref{SOC2}) indicates, we will be searching for representations
that allows us to separate the conditions $c_1$ and $c_2$,
very much like solving differential equations by separating variables~\cite{RAED19b}.
Although the appearance of matrices played a key role in Heisenberg's matrix mechanics,
the latter and the approach pursued here are only distantly related~\cite{RAED19b}.
A much more elaborate discussion of how the separation of conditions
leads to the framework of quantum theory can be found in Ref.~\cite{RAED19b}.

The separation of conditions, applied to data produced by experiments performed
under several sets of conditions $(c_1, c_2), (c'_1, c'_2),\ldots$,
is regarded as successful if it yields a decomposition into models which depend on mutually exclusive, proper
subsets $c_1,c'_1,\ldots$ and $c_2,c'_2,\ldots$ of the conditions only,
thereby reducing the complexity of describing the whole~\cite{RAED19b}.
In the following, to keep the presentation short, we limit ourselves to a cursory discussion of the separation-of-conditions approach. A much more detailed treatment
can be found in Ref.~\cite{RAED19b}.

We illustrate the basic idea by application to the idealized Stern-Gerlach magnet,
assuming that the frequencies of counting $z=\pm1$ particles are given by
\begin{eqnarray}
f(z|\bc,\bd)&=&\frac{1+ z\, \bc\cdot\bd}{2}
\quad,\quad
\langle z \rangle =\sum_{z=\pm1}z f(z|\bc,\bd)=\bc\cdot\bd
\;,
\label{LI3aa}
\end{eqnarray}
that is, by the logical inference treatment of the same experiment, see Section~\ref{LI}.

The key idea is to exploit the fact that {\bf any} choice of the (in this case $2\times2$)
matrices ${\widehat\bF}(\bc)$ and ${\widehat\bP}(z,\bd)$ for which
\begin{eqnarray}
\mathbf{Tr\;}{\widehat\bF}(\bc)=1
\quad,\quad
\mathbf{Tr\;}{\widehat\bP}(z,\bd)\,{\widehat\bF}(\bc)=f(z|\bc\cdot\bd)
\;,
\label{SOC5}
\end{eqnarray}
yields an equivalent description of the data {\bf and} realizes the desired separation of conditions.

Before embarking on the search for a representation of Eq.~(\ref{SOC5}) in terms of matrices,
it is worthwhile to ask ``why not try to find a separation in terms of scalar functions
${\widetilde z}(k,\bc)$ and ${\widetilde f}(k,\bd)$ such that
$\langle x(c_1,c_2) \rangle=\sum_{k\in K} {\widetilde z}(k|\bc) {\widetilde f}(k|\bd)$~?''
In~\ref{SOCapp}, we show that Bell's theorem, applied to the Stern-Gerlach experiment,
prohibits a separation in terms of scalar functions.
Then, the next step is to search for a representation in terms of functions of matrices. Conceptually, taking this step is very similar to introducing complex numbers for solving equations such as $x^2=-1$, or using Dirac's gamma matrices to linearize the relativistic wave equation. Indeed, introducing the matrix structure implicit in Eq.~(\ref{SOC1}) will permit us to carry out the desired separation.

Anticipating for the transition to quantum theory but without loss of generality,
we may take as a basis for the vector space of $2\times2$ matrices,
the unit matrix $\be_0=\left(\begin{array}{cc}1&0\\0&1\end{array}\right)$ and the three Pauli matrices
$\be_1=\bm\sigma^x=\left(\begin{array}{cc}0&1\\1&0\end{array}\right)$,
$\be_2=\bm\sigma^y=\left(\begin{array}{cc}0&-i\\+i&0\end{array}\right)$, and
$\be_3=\bm\sigma^z=\left(\begin{array}{cc}+1&0\\0&-1\end{array}\right)$.
The four hermitian matrices ${\be_0,\be_1,\be_2,\be_3}$ are mutually orthonormal
with respect to the inner product $\langle \bv|\bw\rangle=(1/2)\mathbf{Tr\;} \bv^\dagger \bw$.
Furthermore, $\mathbf{Tr\;}\be_0=2$, $\be_n^2=\be_0$, and
$\mathbf{Tr\;}\be_n=\mathbf{Tr\;}\be_0^\dagger\be_n=0$ for $n=1,2,3$.

In terms of these basis vectors we have, in general
\begin{eqnarray}
{\widehat\bF}(\bc)=\frac{f_0\be_0 +f_1(\bc)\be_1+f_2(\bc)\be_2+f_3(\bc)\be_3}{2}
\quad,\quad
{\widehat\bP}(z,\bd)=\frac{p_0(z,\bd)\be_0+p_1(z,\bd)\be_1+p_2(z,\bd)\be_2+p_3(z,\bd)\be_3}{2}
\;,
\label{SOC6}
\end{eqnarray}
where the $f$'s and $p$'s can be complex-valued and the factor 2 was introduced to compensate for the fact that $\mathbf{Tr\;}\be_0=2$.
The constraints expressed by Eq.~(\ref{SOC5}) imply that
\begin{eqnarray}
f_0=1
\quad,\quad
p_0(z,\bd)+\frac{1}{2}\sum_{n=1}^3 f_n(\bc)p_n(z,\bd)=\frac{1+z\bc\cdot\bd}{2}
\;.
\label{SOC7}
\end{eqnarray}
Obviously, Eq.~(\ref{SOC7}) is trivially satisfied by the choice $f_n(\bc)=c_n$ for $n=1,2,3$ and $p_0(z,\bd)=1$, $p_n(z,\bd)=zd_n$ for $n=1,2,3$.
It is easy to verify that this choice yields eigenvalues of ${\widehat\bF}(\bc)$
and ${\widehat\bP}(z,\bd)$ in the range $[0,1]$ (recall that $\bc$ and $\bd$ are unit vectors).
From the arguments given above, it is not clear that the choice of $f$'s and $p$'s is unique, but this does not matter for the present discussion (see Ref.~\cite{RAED19b} for more information).
Our goal was to find at least {\bf one} representation which describes the data and for which
the conditions $\bc$ and $\bd$ are separated.

By construction, the matrix ${\widehat\bF}(\bc)$ has all
the properties of the density matrix ${{\bm\rho}}(\bc)$ (see~\cite{RAED19b}).
In quantum-theory notation and with $f_n(\bc)=c_n$ and $p_n(z,\bd)=zd_n$, we have
\begin{eqnarray}
{\widehat\bF}(\bc)={{\bm\rho}}(\bc)=\frac{1 +\bc\cdot\bm\sigma}{2}
\quad,\quad
{\widehat\bP}(z,\bd)=\frac{1+z\bd\cdot\bm\sigma}{2}
\quad\Longrightarrow\quad
\sum_{z=\pm1}z
\mathbf{Tr\;}{\widehat\bP}(z,\bd)\,{\widehat\bF}(\bc)=
\bc\cdot\bd
\;.
\label{SOC8}
\end{eqnarray}
We emphasize that the quantum-theoretical description Eq.~(\ref{SOC8})
has been constructed, not postulated as in quantum theory textbooks,
by searching for another representation of the same discrete data.

As mentioned before, Stern-Gerlach magnets act as filtering devices.
This follows from
${\widehat\bP}(z,\bd)={\widehat\bP}^2(z,\bd)$, showing
that ${\widehat\bP}(z,\bd)$ is a projection operator.
However, although ${\widehat\bP}(z,\bd)$ appears in the course of constructing a description of the data,
we should not think of ${\widehat\bP}(z,\bd)$ as an object that affects particles
but only as a description of how the frequency distribution of the particles
changes when the particles pass through a Stern-Gerlach magnet.

The frequencies Eq.~(\ref{DATA2d}) describing the outcomes of the EPRB experiment
depend on two, two-valued variables and two conditions $\ba$ and $\bc$.
Therefore, instead of $2\times2$ matrices, we now have to use
$4\times4$ matrices.
Repeating the steps that led to Eq.~(\ref{SOC8}) and making use
of the separated description of the ideal Stern-Gerlach magnet Eq.~(\ref{SOC8}), we
can construct, {\sl not postulate}, the quantum-theoretical description of the
EPRB experiment~\cite{RAED15c,RAED18a}.
Specializing to the case $\widehat{E}_1(\ba,\bc)=\widehat{E}_2(\ba,\bc)=0$
and $\widehat{E}_{12}(\ba,\bc)=-\ba\cdot\bc$
(see Eq.~(\ref{LI5})) we obtain
\begin{eqnarray}
\bm\rho&=&
\frac{1-\bm\sigma_1\cdot\bm\sigma_2}{4}=
\left( \frac{|{\uparrow}{\downarrow}\rangle-|{\downarrow}{\uparrow}\rangle}{\sqrt{2}}\right)
\left( \frac{\langle{\uparrow}{\downarrow}|-\langle{\downarrow}{\uparrow}|}{\sqrt{2}}\right)
\;,
\label{SOC10}
\end{eqnarray}
that is, the quantum-theoretical description of the singlet state of two spin-1/2 objects.

Furthermore, for any $\bm\rho$ (such as Eq.~(\ref{SOC10})) which does not explicitly depend on $\ba$ or $\bc$
we have
\begin{subequations}
\label{SOC11}
\begin{eqnarray}
\widehat{E}_1(\ba,\bc)&=&\langle \bm\sigma_1\cdot\ba\rangle=
\hbox{\bf Tr\;}\bm\rho\,\bm\sigma_1\cdot\ba
=
\langle \bm\sigma_1\rangle\cdot\ba
\label{SOC11a}
\;,\\
\widehat{E}_2(\ba,\bc)&=&\langle \bm\sigma_2\cdot\bc\rangle
=\hbox{\bf Tr\;}\bm\rho\,\bm\sigma_2\cdot\bc
=\langle \bm\sigma_2\rangle\cdot\bc
\label{SOC11b}
\;,\\
\widehat{E}_{12}(\ba,\bc)&=&
\langle \bm\sigma_1\cdot\ba\; \bm\sigma_2\cdot\bc \rangle
=\hbox{\bf Tr\;}\bm\rho\,\bm\sigma_1\cdot\ba\,\bm\sigma_2\cdot\bc
=\sum_{\alpha,\beta} a_\alpha \Gamma_{\alpha,\beta} c_\beta
\label{SOC11c}
\;,
\end{eqnarray}
\end{subequations}
where the $3\times3$ matrix $\Gamma_{\alpha,\beta}=
\langle \bm\sigma_1^\alpha \bm\sigma_2^\beta\rangle$.
Equation Eq.~(\ref{SOC11}) epitomizes the power of the quantum-theoretical description.
It separates the description of the state of the two-spin system
in terms of the expectation values $\langle \bm\sigma_1\rangle$, $\langle \bm\sigma_2\rangle$,
and $\Gamma_{\alpha,\beta}=\langle \bm\sigma_1^\alpha \bm\sigma_2^\beta\rangle$,
from the description of the conditions (or context) $\ba$ and $\bc$ under which the data was collected.

Noteworthy is also that in general,
the single-spin averages Eqs.~(\ref{SOC11a}) and~(\ref{SOC11b})
do not depend on $\bc$ and $\ba$, respectively.
In this sense, quantum theory exhibits a kind of ``locality'', ``separation'', or perhaps better ``independence'', in that
averages pertaining to particle 1 (2) only depend on $\ba$ ($\bc$).
Of course, the correlation Eq.~(\ref{SOC11c}) involves both $\ba$ and $\bc$.

In short, separating conditions led to the construction of the quantum-theoretical description of the EPRB experiment
containing the following elements~\cite{BALL03}
\begin{itemize}
\item
The two-particle system emitted by the source has total spin zero
and after leaving the source, the particles do not interact.
\item
The statistics of the magnetizations, obtained by observing many pairs,
is described by the singlet state
$|\psi\rangle=(|{\uparrow}{\downarrow}\rangle-|{\downarrow}{\uparrow}\rangle)/\sqrt{2}$,
or equivalently, by the density matrix Eq.~(\ref{SOC10}).
\item
The single-particle averages $\langle \bm\sigma_1\rangle=\langle \bm\sigma_2\rangle=0$
and the correlation $\Gamma_{\alpha,\beta}=-\delta_{\alpha,\beta}$,
implying
$\widehat{E}_{1}(\ba,\bc)=\langle \bm\sigma_1\cdot\ba\rangle=0$,
$\widehat{E}_{2}(\ba,\bc)=\langle \bm\sigma_2\cdot\bc\rangle=0$,
and
$\widehat{E}_{12}(\ba,\bc)=\langle \bm\sigma_1\cdot\ba\; \bm\sigma_2\cdot\bc \rangle=-\ba\cdot\bc$.
\end{itemize}

\subsection{Advantages and limitations of using quantum theory}\label{QADV}

The main advantage of using quantum theory as a model for the data is
in the amount of compression that can be achieved.
This can be seen as follows.

For a fixed pair of settings $(\ba,\bc)$, the frequencies $f(x,y|\ba,\bc)$
and the density matrix $\bm\rho$, together with the
projectors $\bM(x|\ba)=(1+x\ba\cdot\bm\sigma_1)/2$ and $\bM(y|\bc)=(1+y\bc\cdot\bm\sigma_2)/2$,
describe the statistics of ${\cal D}$ equally well.

If we repeat the EPRB experiment with $M$ different pairs of settings and characterize
the results by frequencies, we need $3M$ numbers to represent the statistics (not $4M$
because $f(-1,-1|\ba,\bc)=1-f(1,1|\ba,\bc)-f(1,-1|\ba,\bc)-f(-1,1|\ba,\bc)$).

On the other hand, quantum theory describes
the statistics of {\bf all} EPRB experiments for ${\bf all}$ possible settings through the fifteen
real numbers that completely determine the density matrix $\bm\rho$.
To see this, we write the density matrix as a linear combination
of a basis of the Hilbert space of $4\times4$ matrices.
One way to construct these basis vectors is to form the direct product
of each matrix from the set $\{1,\sigma^x_1,\sigma^y_1,\sigma^z_1\}$
with each of the matrices from the set $\{1,\sigma^x_2,\sigma^y_2,\sigma^z_2\}$.
There are sixteen such $4\times4$ matrices with their corresponding coefficients.
Because $\mathbf{Tr\;}\bm\rho=1$, there are only fifteen independent coefficients,
which are real-valued because $\bm\rho$ is a hermitian, non-negative definite matrix~\cite{BALL03}
and all elements of the basis are hermitian matrices too.

It is easy to show that these coefficients
are completely determined by the six single spin averages
$\langle \sigma^\alpha_1\rangle$,
$\langle \sigma^\alpha_2\rangle$,
and nine two-spin averages
$\langle \sigma^\alpha_1\sigma^\beta_2\rangle$,
with $\alpha,\beta=x,y,z$.
Thus, in theory, we need to perform only fifteen experiments to determine these
expectation values and can then use these numbers to
compute
$\widehat{E}_1(\ba,\bc)=\langle \bm\sigma_1\cdot\ba\rangle$,
$\widehat{E}_2(\ba,\bc)=\langle \bm\sigma_2\cdot\bc\rangle$,
and
$\widehat{E}_2(\ba,\bc)=\langle \bm\sigma_1\cdot\ba\,\bm\sigma_2\cdot\bc\rangle$
for {\bf any} pair of settings $(\ba,\bc)$,  see also Eq.~(\ref{SOC11}).

In conclusion, compared to the representation in terms of frequencies
which requires $3M$ numbers to describe the statistics of $M$ EPRB experiments,
the quantum-theoretical description requires only fifteen numbers to describe {\bf all}
possible EPRB experiments, a tremendous compression of the data if $M$ is large.

Regarding limitations of the quantum formalism, the no-go theorem presented in Section~\ref{NOGO}
gives a first indication that there exist data which can be modeled probabilistically
but does not fit into the quantum formalism.
It is also not difficult to see that, as a direct consequence of the linear structure of
the density matrix $\rho=(1+\ba\cdot\bm\sigma_1)/2$
and the projector $\bM(x|\bc)=(1+x\bc\cdot\bm\sigma_1)/2$,
quantum theory can never yield a frequency $f(x|\ba,\bc)=(1+x (\ba\cdot\bc)^2)/2$, for instance.
These limitations are, of course, outweighed by the power of quantum theory to
compress the statistics of the data in a way that probabilistic models cannot.

Finally, it is important to recall that the quantum formalism also applies
to cases where the experimental data does not come in the form of single items of discrete data, i.e., individual events.
For instance, if an experiment measures the specific heat of some material,
there are no ``individual events'' to compute the statistics of,
only a record of numbers for the specific heat as a function of e.g., the temperature.
Still, a quantum-theoretical model calculation of the specific heat is
based on Eq.~(\ref{SOC2}), with suitable matrices ${\widehat\bF}(c_1)$ and
${\widehat\bX}(c_2)$, of course.

\subsection{Quantum theory: a no-go theorem for a system of two spin-1/2 objects}\label{NOGO}

Replacing the requirement of perfect anticorrelation
by complete correlation, the solution of the logical inference
problem reads $\widehat{E}_{12}(\ba,\bc)=+\ba\cdot\bc$.
To prove that such a correlation is incompatible with quantum-theoretical description of two spin-1/2 objects~\cite{RAED20a},
we consider the more general case for which
\begin{eqnarray}
\widehat{E}_1(\ba,\bc)=\widehat{E}_2(\ba,\bc)=0 \;,\quad \widehat{E}_{12}(\ba,\bc)&=&-q\ba\cdot\bc
\;,
\label{NOGO1}
\end{eqnarray}
where $q$ is a real number.
Starting from the most general expression of the density matrix, Eq.~(\ref{NOGO1}) implies that~\cite{RAED20a}
\begin{eqnarray}
\bm\rho&=&
\frac{1-q\bm\sigma_1\cdot\bm\sigma_2}{4}
\;.
\label{NOGO0}
\end{eqnarray}
However, as the eigenvalues of $\bm\sigma_1\cdot\bm\sigma_2$ are $-3,+1,+1,+1$,
the matrix Eq.~(\ref{NOGO0}) is only non-negative definite if $q\ge-1/3$.
In other words, Eq.~(\ref{NOGO0}) is not a valid density matrix if $q<-1/3$.
This then allows us to state the following no-go theorem:
\begin{center}
\framebox{
\parbox[t]{0.9\hsize}{%
There does not exist a quantum model for a system of two spin-1/2 objects
that yields $p(x,y|{\ba},{\bc})=(1 - q\, x\,y\,\ba\cdot\bc)/4$
or, equivalently, the averages $\widehat{E}_1(\ba,\bc)=\widehat{E}_2(\ba,\bc)=0$  and correlation $E_{12}({\ba},{\bc})=-q\,\ba\cdot\bc$ unless  $-1/3\le q\le1$.
}}%
\end{center}
An immediate consequence is that there does not exist a quantum-theoretical
description of a system of two spin-1/2 objects if
\begin{eqnarray}
\widehat{E}_1(\ba,\bc)=\widehat{E}_2(\ba,\bc)=0 \;,\quad \widehat{E}_{12}(\ba,\bc)&=&+\ba\cdot\bc
\;.
\label{NOGO2}
\end{eqnarray}
This ``no-go theorem'' may be viewed as a kind of ``Bell theorem'', although it is not a theorem
about the limited applicability of the separable model introduced by Bell (see Section~\ref{BELL})
but rather a theorem about the limited applicability of quantum theory.
In contrast, a probabilistic model can yield Eq.~(\ref{NOGO2}).
Indeed, $P(x,y|{\ba},{\bc})=(1 + x\,y\,\ba\cdot\bc)/{4}$ does exactly that.

\subsection{Discussion}\label{QDISC}

Recapitulating, starting from sets of two-valued data, we have constructed instead of postulated
the quantum-theoretical description of the ideal Stern-Gerlach and EPRB experiments by
\begin{enumerate}[(i)]
\item
Assuming that the data sets were obtained by reproducible and robust, that is by ``good'',
experiments.
\item
Requiring that the description of the data in terms of the relative frequencies
can be separated in a part that describes the preparation of particles with particular
properties and other parts that describe the process of measuring these properties.
\end{enumerate}
At no point use was made of concepts that are quantum-theoretical in nature.
The data being the immutable facts and quantum theory being a very convenient, minimalistic M2C
describing the data, there is no need to bring in Born's rule to ``explain''
how quantum theory ``produces'' data.
By itself, a quantum-theoretical model simply cannot ``produce'' data and, as a description, also does not need to.

Moreover, as the data is discrete, represented by two-valued variables,
there is no need to postulate the quantization of the spin.
The transition from the statistical description in terms of frequencies of events to a representation
in terms of $2\times2$ matrices made it possible to decompose the description
of the whole in simpler descriptions of the parts.
It is this decomposition that renders quantum theory a powerful mathematical apparatus to describe discrete data.

Starting from the model-free description of any four sets of data pairs presented in sections~\ref{COUNT}
and~\ref{INEQDATA}, sections~\ref{LI} and~\ref{SOC} have shown how
a combination of plausible reasoning and the basic requirement
that a description of the whole can be decomposed in simpler, separated parts
leads to the construction of a MM, that is quantum theory, which can capture the main features of these data.

In the two sections that follow, we scrutinize laboratory experiments
which have been designed to produce averages and correlations of data that,
inspired by the quantum-theoretical description of the EPRB thought experiment (see~\ref{QTDE}),
are expected to exhibit the features of the ``imagined'' data, listed in Section~\ref{EXPI}.

\section{Data analysis of an EPRB laboratory experiment with polarized photons}\label{FAIL}

Most EPRB laboratory experiments focus on demonstrating that the measured correlation cannot
be described by Bell's model, see Eq.~(\ref{IN0}), by showing that the Bell-CHSH inequality (see~\ref{BELLINQ})
is violated.
Perhaps more interesting and ultimately more relevant is that
the analysis of the data of several EPRB experiments points to a conflict with the quantum-theoretical description
of the EPRB experiment~\cite{ADEN09,RAED12,RAED13a,Bednorz2017,Adenier2017}.

As the data analysis of very different experimental
setups~\cite{WEIH98,WEIH00,ADEN12,AGUE09,HENS15,GIUS15,SHAL15}
all point to similar conflicts~\cite{ADEN09,RAED12,RAED13a,Bednorz2017,Adenier2017},
we only show and analyze the data acquired in one of the more complete and sophisticated EPRB experiments,
the one performed by Weihs {\sl et al.}~\cite{WEIH98,WEIH00}, see Fig.~\ref{weihsexp}.
The protocol that we use to analyze these data is
identical to the one employed by the experimenters~\cite{WEIH98,WEIH00},
see Refs.~\cite{ZHAO08,RAED12,RAED13a} for more details.

\begin{figure}[!htp]
\centering
\includegraphics[width=0.45\hsize]{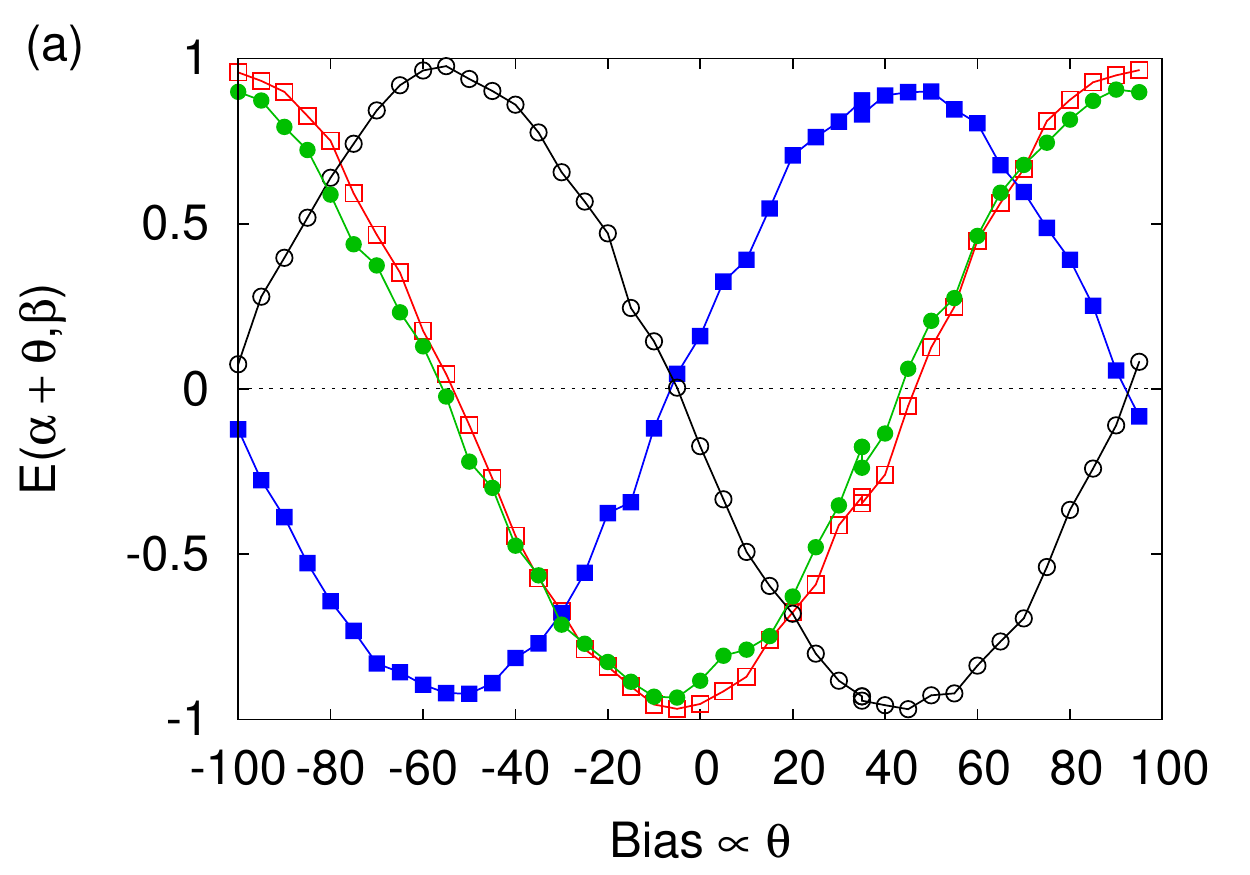} 
\includegraphics[width=0.45\hsize]{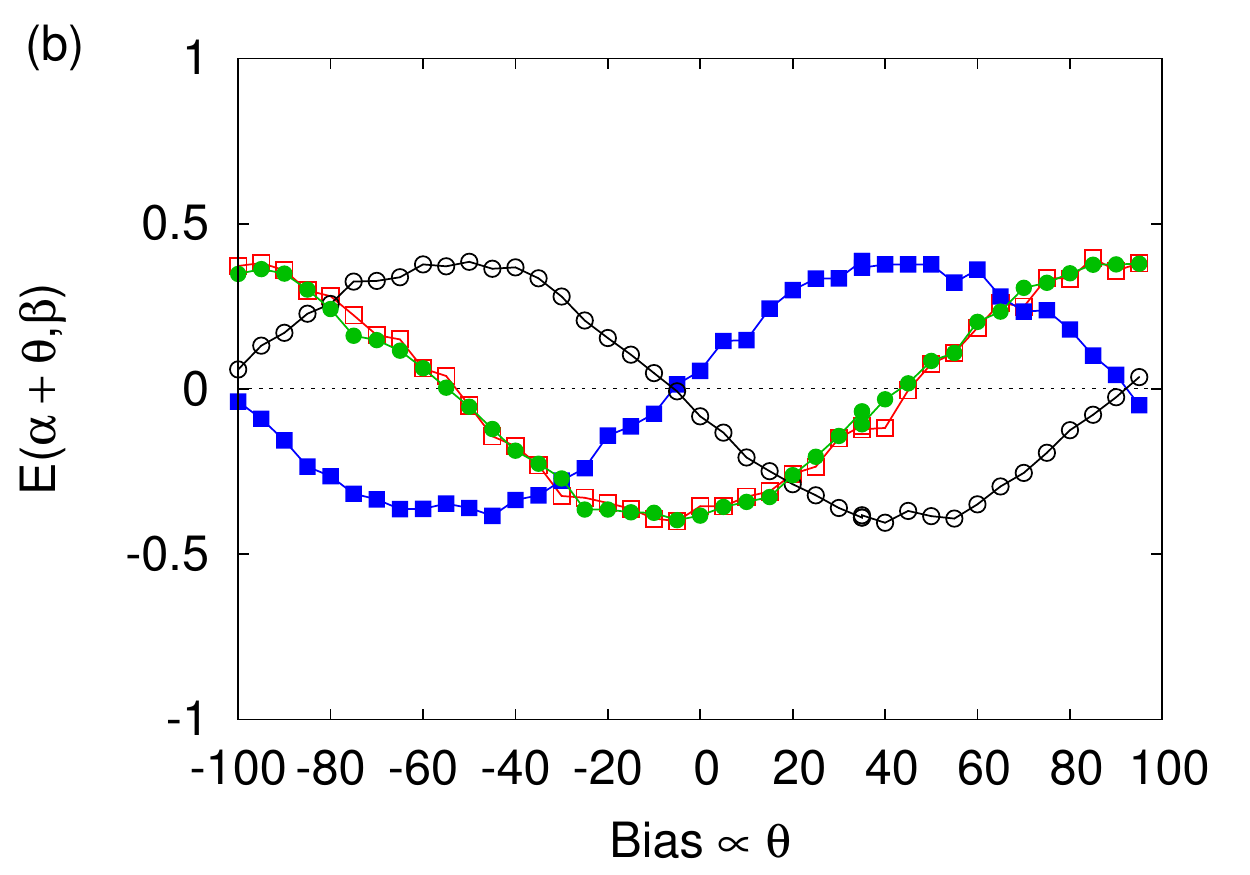}
\caption{(color online)
Analysis of experimental data (data set called {\bf scanblue1})
recorded in the EPRB experiments performed by Weihs {\sl et al.}~\cite{WEIH00}.
(a): the correlation $E(\alpha+\theta,\beta)$ as a function of the bias applied to Alice's EOM
for a time coincidence window $W=2\,\mathrm{ns}$.
In the experiment, the angle $\theta$, which is proportional to the bias, changes every 5 seconds~\cite{WEIH00}.
The number of coincidences during each 5 second period is at least 7500.
The minimum of the average times between detection events is $\langle \delta t \rangle\approx 24000\,\mathrm{ns}$,
an order of ten thousand larger than the time window $W=2\,\mathrm{ns}$ used to compute the correlations.
The pair of setting $(\alpha,\beta)$ is chosen randomly out of four possibilities~\cite{WEIH98,WEIH00}.
Open squares: $(\alpha,\beta)=(a,c)$;                
solid squares: $(\alpha,\beta)=(a,d)$;               
open circles: $(\alpha,\beta)=(b,c)$;                
solid circles: $(\alpha,\beta)=(b,d)$,               
where $a=0$, $b=\pi/4$, $c=\pi/8$ and $d=3\pi/8$.
These correlations cannot be obtained from Bell's model (see Eq.~(\ref{IN0}) below).
(b): same as (a) except that the time coincidence window $W=1000\,\mathrm{ns}$, much smaller
than $\langle \delta t \rangle\approx 24000\,\mathrm{ns}$.
These correlations are compatible with Bell's model (see Eq.~(\ref{IN0}) below).
}
\label{fig.A5}
\end{figure}

\begin{figure}[!htp]
\centering
\includegraphics[width=0.45\hsize]{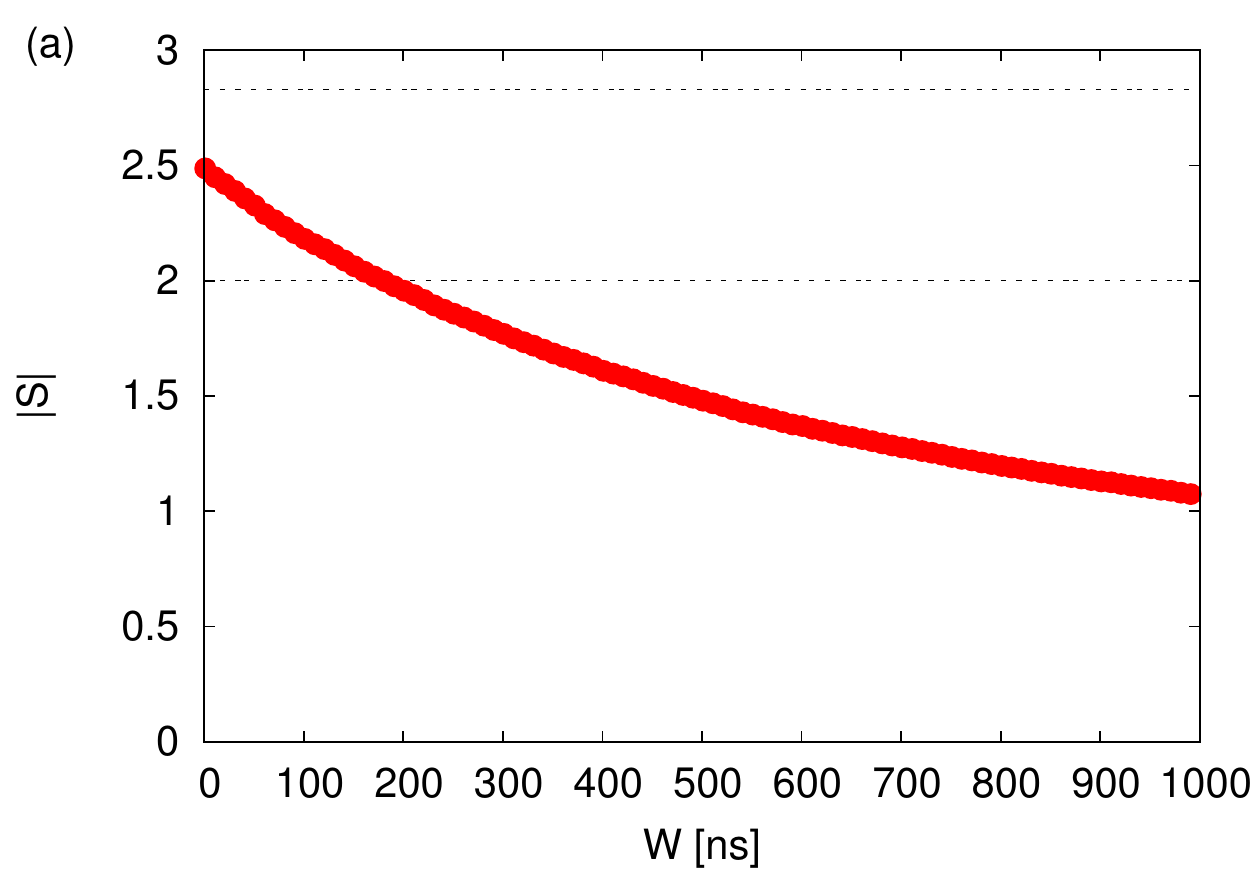}
\includegraphics[width=0.45\hsize]{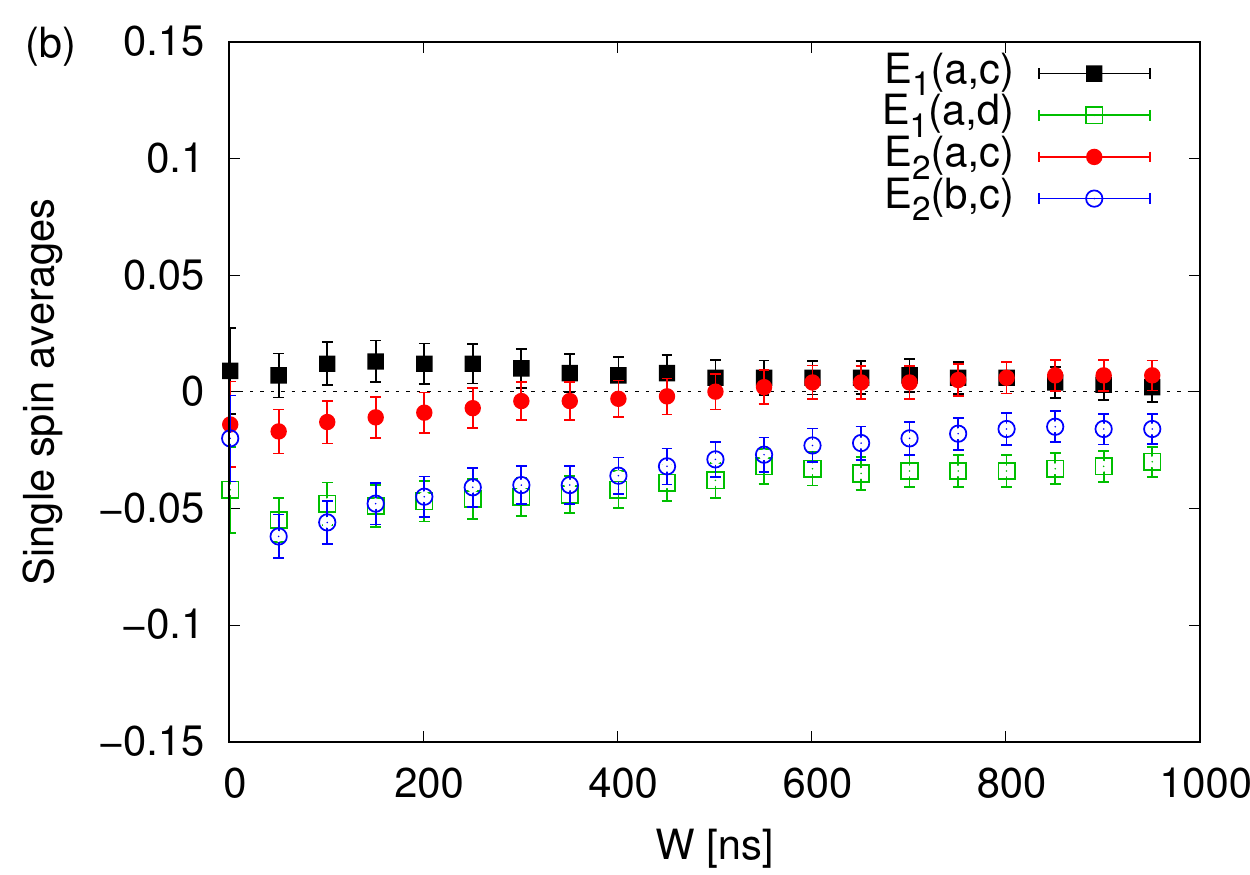} 
\caption{(color online)
Analysis of experimental data (data set called {\bf newlongtime2})
recorded in the EPRB experiments performed by Weihs {\sl et al.}~\cite{WEIH00}.
(a): the Bell-CHSH function $|S|=|E(a,c)-E(a,d)+E(b,c)+E(b,d)|$
as a function of the time window $W$ for $a=0$, $b=\pi/4$, $c=\pi/8$ and $d=3\pi/8$.
The dashed line at $|S|=2$ is the maximum value for Bell's model Eq.~(\ref{IN0}),
or a quantum system of two $S=1/2$ objects in a separable (product) state.
The dashed line at $|S|=2\sqrt{2}$ is the maximum value for
a quantum system of two $S=1/2$ objects in a singlet state.
The total number of photons detected on the left and right side during the experiment which lasted 60 s is
1733902 and 1621229, respectively.
The total number for coincidences with an absolute value of the time-tag difference less than $W=2\,\mathrm{ns}$
is about 35000.
The minimum of average times between detection events is $\langle \delta t \rangle=35000\,\mathrm{ns}$,
more than {\bf 30} times the maximum value $W=1000$ of the time coincidence window used to plot $|S|$.
(b): selected single-particle averages as a function of the time coincidence window $W$,
contradicting the quantum-theoretical description on a very elementary level.
Error bars correspond to $2.5$ standard deviations.
}
\label{fig.A6}
\end{figure}

In Figs.~\ref{fig.A5} and~\ref{fig.A6}, we present some results of our analysis of the experimental data~\cite{WEIH98}.
A first observation is that the correlations $E(\alpha+\theta,\beta)$
shown in Fig.~\ref{fig.A5}(a) are in excellent agreement with the correlation $\widehat{E}_{12}(\ba,\bc)=-\cos2(a-c)$
of a system of two photon-polarizations, described by the singlet state.
The correlations shown in Fig.~\ref{fig.A5}(a) have been obtained by
analysing the raw data with a time-coincidence window $W=2\,\mathrm{ns}$,
{\bf four orders of magnitude} smaller than $\langle \delta t \rangle\approx 24000\,\mathrm{ns}$, the average time
between the registration of a pair of detection events.
The correlations shown in  Fig.~\ref{fig.A5}(a) cannot, not even approximately,  be reproduced by Bell's model, see Eq.~(\ref{IN0}) below.

Figure~\ref{fig.A5}(b) demonstrates that by enlarging the
time-coincidence window to $W=1000\,\mathrm{ns}\ll\langle \delta t \rangle$, the maximum amplitude of the correlations
$E(\alpha+\theta,\beta)$ drops from approximately one to about one half.
Note that a time-coincidence window of $W=1000\,\mathrm{ns}$ is a factor of twenty-four smaller than
$\langle \delta t \rangle\approx 24000\,\mathrm{ns}$, that is the frequency of identifying ``wrong'' pairs is small.
The minor modification that lets Bell's toy model comply with Malus' law (see~\ref{CBELM})
yields $C(\ba,\bc)=-(1/2)\cos2(a-c)$, rather close to $E(\alpha+\theta,\beta)$ shown in Fig.~\ref{fig.A5}(b).

The analysis of the experimental data clearly demonstrates that the maximum amplitude of the
correlations $E(\alpha+\theta,\beta)$ decreases as the time-coincidence window $W\ll\langle \delta t \rangle$ increases.
Within the range $2\,\mathrm{ns}\le W\ll\langle \delta t \rangle$, $W$ can be used to ``tune''
the correlations $E(\alpha+\theta,\beta)$ such that they are
compatible with (i)
Bell's modified toy model (see~\ref{CBELM}) or
two spin-1/2 objects described by a separable state (see~\ref{EXQT}) or with
(ii) two spin-1/2 objects described by a non-separable pure state, e.g., a singlet state (see~\ref{QTDE}).

{\bf Whether the polarization state is ``entangled'' depends
on the choice of the coincidence window $W$.
Therefore, in this case, entanglement is not an intrinsic property of the pairs of photons.
It is a property of the whole experimental setup.}

The obvious conclusion that one has to draw from the observation
that the numerical values of the correlations depend on the value of
the time-coincidence window $W$ is that any CM or MM that
aims to describe this particular EPRB experiment
should account for the time-coincidence window (or another mechanism) that
is required to identify pairs of events.
Phrased differently, the time-coincidence window is essential to the way
the data is processed and the conclusions that are drawn from them therefore
have to be a part of the model description.
Note that quantum theory may be able to describe the experimental data for
one particular $W$ but lacks the capability to also describe the $W$-dependence,
simply because in orthodox quantum theory, time is not an observable.

The importance of including the time-coincidence window $W$ in a model for this particular
laboratory EPRB experiment is further illustrated in Fig.~\ref{fig.A6}(a).
Figure~\ref{fig.A6}(a) demonstrates that the experimental data
might be represented by Eq.~(\ref{IN0})
if the time coincidence window $W$ is chosen properly, that is if the Bell-CHSH function $|S|\le2$.

For $W<100\,\mathrm{ns}$, the value of $|S|>2$ and the correlations are compatible with a quantum-theoretical
description in terms of a non-separable density matrix, but not with Bell's model Eq.~(\ref{IN0})
because the latter implies that $|S|\le2$.
From model-free inequality Eq.~(\ref{DATA4}), it follows that if $|S|\ge2$, the fraction
of quadruples $\QUAD$ that one might be able to identify in the data that led to
Fig.~\ref{fig.A6}(a) must be smaller than $2-|S|/2$.

For $200\,\mathrm{ns}<W<1000\,\mathrm{ns}$, {\bf much less than the average time interval $\langle \delta t \rangle=35000\,\mathrm{ns}$ between two detection events},
the value of $|S|\le2$ admits a description in terms of a
separable density matrix or, possibly, by Bell's model Eq.~(\ref{IN0}).

Clearly, the time-coincidence window $W$ can be ``adjusted'' in a wide range, such that either $|S|\le2$ or $|S|>2$.
In other words, the ``evidence'' for a singlet state description appears or disappears depending
on very reasonable  choices of $W$
(that is $W\ll\langle \delta t \rangle=35000\,\mathrm{ns}$).
These results corroborate our conclusions drawn on the basis of the data depicted in Fig.~\ref{fig.A5}.

According to quantum theory, see Eq.~(\ref{SOC11}),
$\widehat{E}_1(\ba,\bc)=\langle \bm\sigma_1\cdot\ba\rangle$ and
$\widehat{E}_2(\ba,\bc)=\langle \bm\sigma_2\cdot\bc\rangle$ should not depend on $\bc$ and $\ba$, respectively.
For small $W$, the total number of coincidences is too small to yield statistically meaningful results.
For $W>20\,\mathrm{ns}$ the change in some of these single-spin averages
observed on the left (right) when the settings of the right (left) are changed (randomly)
systematically exceeds five standard deviations~\cite{RAED12,RAED13a}.
Indeed, $E_1(a,c)$ and $E_1(a,d)$ (squares in Fig.~\ref{fig.A6}(b))
should, according to the quantum theory of a system in the singlet state, be independent of $c$ and $d$
but in fact, they differ by at least five standard deviations.
The same holds true for $E_2(a,c)$ and $E_2(b,c)$ (circles in Fig.~\ref{fig.A6}(b)).

In conclusion, although most EPRB experiments that have been performed so far
have produced data that violate the Bell-CHSH inequalities, these experiments do not
produce data that agree with the quantum-theoretical description
of the EPRB experiment~\cite{ADEN09,RAED12,RAED13a,Bednorz2017,Adenier2017}.

To head off possible confusion, it is not quantum theory which fails
to describe the data.
Indeed, given experimental data for the averages
$u_{\alpha}=\langle \bm\sigma_1\cdot\be_\alpha\rangle$, $v_{\alpha}=\langle \bm\sigma_2\cdot\be_\alpha\rangle$,
$w_{\alpha,\beta}=\langle \bm\sigma_1\cdot\be_\alpha\,\bm\sigma_2\cdot\be_\beta\rangle$
with $\be_x=(1,0,0)$, $\be_y=(0,1,0)$, $\be_z=(0,0,1)$, the density matrix
\begin{eqnarray}
\bm\rho&=&\frac{1}{4}\left[1+ \sum_{\alpha=x,y,z} \left( u_\alpha\sigma_1^\alpha
+ v_\alpha\sigma_2^\alpha
\right) +\sum_{\alpha,\beta=x,y,z} \sigma_1^\alpha w_{\alpha,\beta} \sigma_2^\beta\right]
\;,
\label{FAIL1}
\end{eqnarray}
always yields a perfect fit to this data if we allow
the expansion coefficients $u_\alpha$, $v_\alpha$, and $w_{\alpha,\beta}$ to depend on $\ba$ and $\bc$.
The conflicts between quantum theory of the EPRB experiment
and experimental data can only be resolved by new, much more accurate experiments.

In Section~\ref{TPM}, we review a probabilistic model which complies with the separation principle and
describes the raw, unprocessed data of the EPRB experiment.
In the limit of a vanishing time-coincidence window,
this M2C yields the quantum-theoretical results for the EPRB experiment~\cite{RAED06c,ZHAO08}.
The latter implies that there is no fundamental problem to have a MM
produce (e.g., with the help of a pseudo-random number generator)
data of the kind gathered in an EPRB experiment and recover the quantum-theoretical results.
Indeed, that is exactly what the subquantum model described in Section~\ref{EVENT} shows.

\section{Quantum computing experiments}\label{QCE}

Instead of performing EPRB experiments with photons,
one can also carry out their own EPRB experiments with publicly accessible quantum computer (QC) hardware~\cite{IBMQE}.
In the case of photons, disregarding technicalities, it is not difficult
to realize situations in which there is no interaction between the photons of a pair at the time
of their detection because photons need a material medium to ``interact''.
In contrast, in superconducting or ion trap quantum devices, the qubits are very close to each other~\cite{NIEL10}.
The qubits of a physical QC device are part of a complicated many-body system, the behavior of which
cannot be described in terms of noninteracting entities.
Therefore, experiments with QC hardware are of little relevance to the EPR argument~\cite{EPR35} as such.
However, they can be used to test to what extent a QC~\cite{NIEL10}
can generate data sets ${\cal D}$ that comply with the quantum-theoretical description
in terms of the singlet state, that is ${\widehat E}_{1}(\ba,\bc)=0$, ${\widehat E}_{2}(\ba,\bc)=0$,
and ${\widehat E}_{12}(\ba,\bc)=-\ba\cdot\bc$.
In contrast to EPRB experiments with photons for which it is essential to
have an external procedure (such as counting time coincidences or using voltage thresholds) to identify photons,
QC experiments can generate bitstrings for all qubits simultaneously and identification is not an issue.

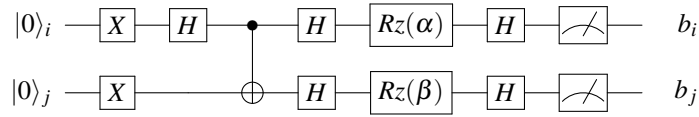
\begin{figure}[!htp]
  \[
  \begin{array}{ccc}
   \mbox{\Qcircuit @C=1.3em @R=.95em {
      &\lstick{\ket{0}_i}&\gate{X} &\gate{H} &\ctrl{1}&\gate{H}&\gate{Rz(\alpha)}&\gate{H}&\meter&\qw& &\lstick{b_i}\\
      &\lstick{\ket{0}_j}&\gate{X} &\qw      &\targ   &\gate{H}&\gate{Rz(\beta)}&\gate{H}&\meter  &\qw& &\lstick{b_j}\\
    }
    }
  \end{array}
  \]
  \caption{Circuit to generate the singlet state $\ket{\Phi}=\left(\ket{01}-\ket{10}\right)/\sqrt{2}$
  and perform measurements of the state of the qubits $(i,j)$, projected on directions
  specified by the angles $\alpha$ and $\beta$, respectively.
  The symbol connecting the two qubit lines denotes the controlled-NOT (CNOT) operation. For the meaning of the other symbols, see~\cite{NIEL10}.
  \label{CIRCfig2a}
  }
\end{figure}

In short, a QC is a physical device that is subject
to a sequence of electromagnetic pulses and changes its state accordingly.
The often complicated physical processes induced by these pulses
are assumed to implement a sequence of unitary operations
that change the wavefunction describing the state of the ideal QC~\cite{NIEL10}.
The sequences of unitary operations, constituting the algorithm,
are conveniently represented by quantum circuits~\cite{NIEL10}.
Actually executing a quantum circuit such as Fig.~\ref{CIRCfig2a} on QC hardware requires
an intermediate step to translate the circuit into pulses. This step is taken
care of by the software of the QC hardware provider~\cite{IBMQE}.

\begin{figure}[!htp]
{\footnotesize 
  \[
  \begin{array}{ccc}
   \mbox{\Qcircuit @C=1.3em @R=.95em {
      &\lstick{\ket{0}_i}&\gate{X} &\gate{Rz(\pi/2)} &\gate{Sx} &\gate{Rz(\pi/2)}&\ctrl{1}&\gate{Rz(\pi/2)} &\gate{Sx}&\gate{Rz(\pi/2)}&\gate{Rz(\alpha)}    &\gate{Rz(\pi/2)}&\gate{Sx}&\gate{Rz(\pi/2)} &\meter&\qw& &\lstick{b_i}\\
      &\lstick{\ket{0}_j}&\gate{X} &\qw              &\qw             &\qw             &\targ   &\gate{Rz(\pi/2)} &\gate{Sx}&\gate{Rz(\pi/2)}&\gate{Rz(\beta)}&\gate{Rz(\pi/2)}&\gate{Sx}&\gate{Rz(\pi/2)} &\meter&\qw& &\lstick{b_j}\\
    }
    }
  \end{array}
  \]
  }
  \caption{Transpiled version of Fig.~\ref{CIRCfig2a} in terms of the native gates ``X'', ``Controlled X'', ``Rz'', and ``Sx'',
  used by the IBM-QE Manila device~\cite{IBMQ}.
  \label{CIRCfig2b}
  }
\end{figure}
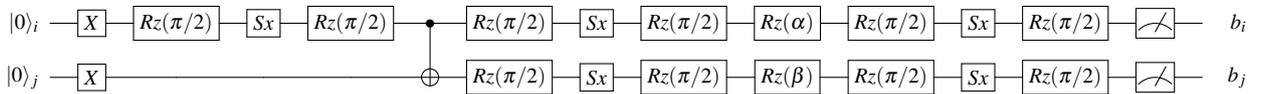

The quantum circuit shown in Fig.~\ref{CIRCfig2a}
changes the initial state $|\psi\rangle=|00\rangle=|\uparrow\uparrow\rangle$, corresponding to both
spins up, to the singlet state $|\Phi\rangle=(|01\rangle - |10\rangle)/\sqrt{2}=
(|\uparrow\downarrow\rangle - |\downarrow\uparrow\rangle)/\sqrt{2}$
and performs measurements on qubits $i$ and $j$, corresponding to rotation angles $\alpha$ and $\beta$ about the vector $(1,1,0)/\sqrt{2}$, respectively.

The analytical calculation of the result of executing the quantum circuit shown in Fig.~\ref{CIRCfig2a}
on the MM of an ideal QC yields for the single- and two-qubit expectation values
${\widehat E}_1(\alpha,\beta)={\widehat E}_2(\alpha,\beta)=0$
and ${\widehat E}_{12}(\alpha,\beta)=-\cos(\alpha-\beta)$, respectively,
as it should be for a quantum-theoretical description of the EPRB experiment.

The software controlling the operation of an IBM-QE device transpiles
the circuit Fig.~\ref{CIRCfig2a} into the mathematically equivalent circuit shown in Fig.~\ref{CIRCfig2b}
in which only so-called ``native'' gates appear~\cite{IBMQ}.
It is the latter circuit that is executed on the QC hardware.

In practice, a QC EPRB experiment consists of repeating the following three steps
\begin{enumerate}
\item
Reset the device to the state with both qubits set to zero.
\item
Apply the pulse sequences as specified by the circuit in Fig.~\ref{CIRCfig2b}.
\item
Read out the state, yielding a pair of bits $(b_i,b_j)$ taking one
out of the four possibilities $(0,0)$, $(1,0)$, $(0,1)$, or $(1,1)$.
\end{enumerate}
The number of repetitions $N$ is typically in the range $1000$ -- $10000$.
Each readout yields a pair $(x=1-2b_i=\pm1,y=1-2b_j)$
which is added to the data set ${\cal D}_1$.
From the $N$ pairs of data items in ${\cal D}_1$, we compute the averages and the correlation according to Eq.~(\ref{DATA2}).

\begin{figure}[!htp]
\begin{center}
\includegraphics[width=0.32\hsize]{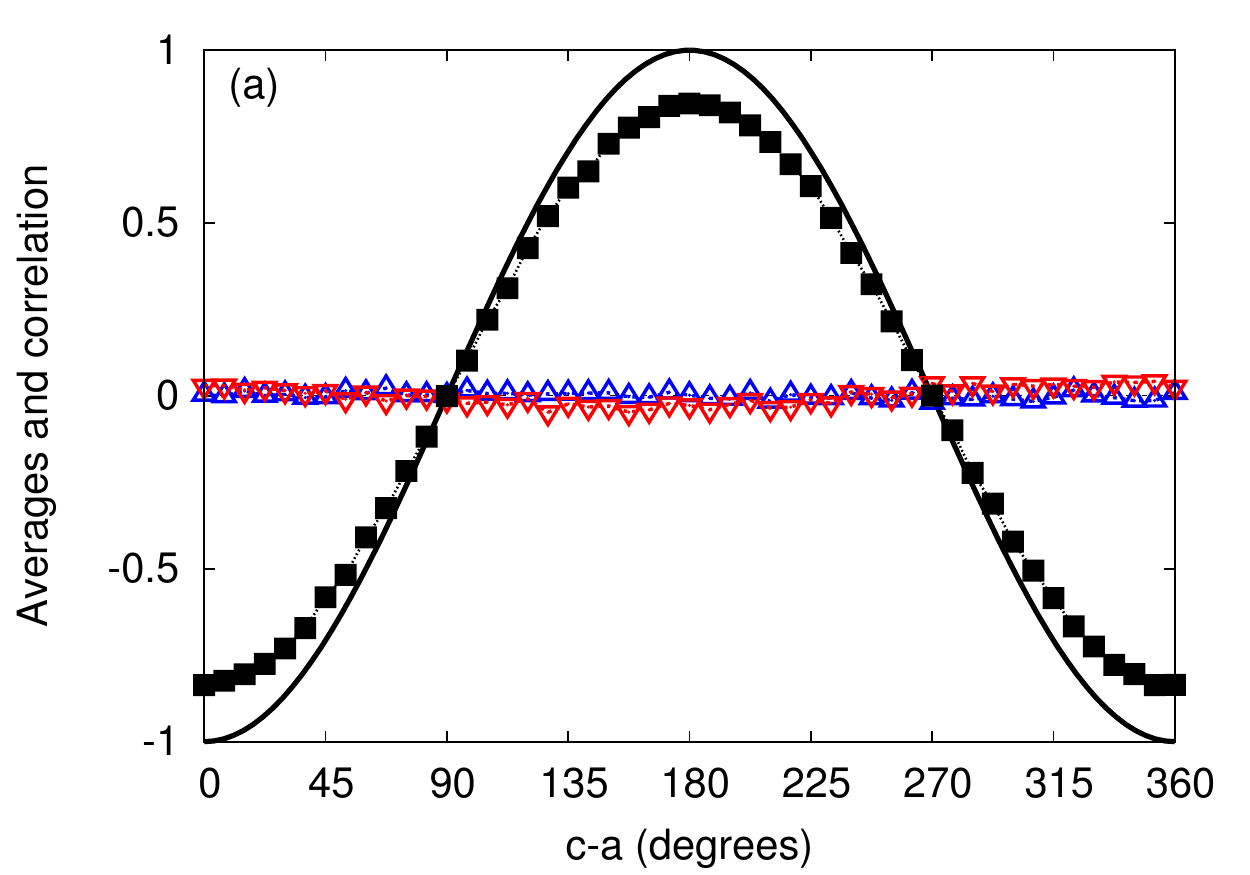}
\includegraphics[width=0.32\hsize]{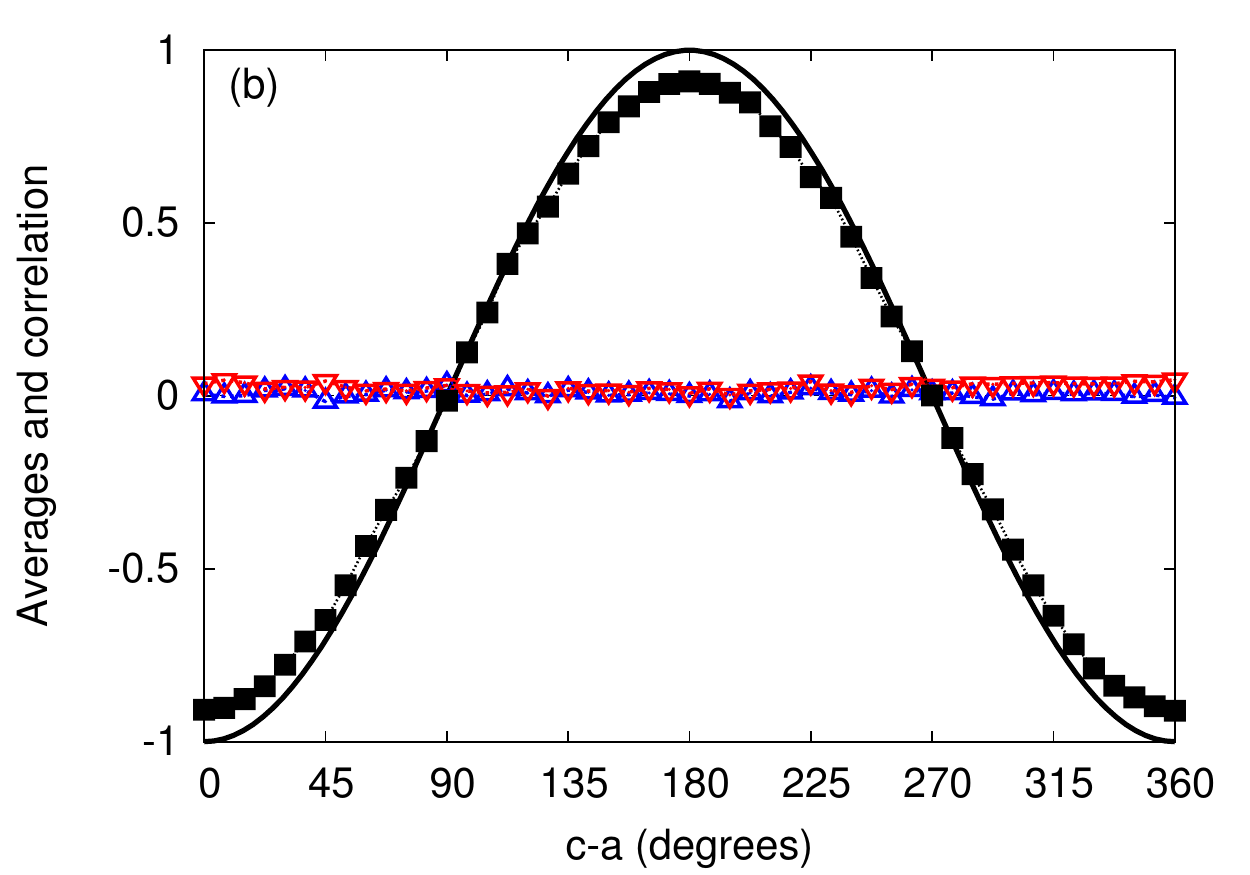}
\includegraphics[width=0.32\hsize]{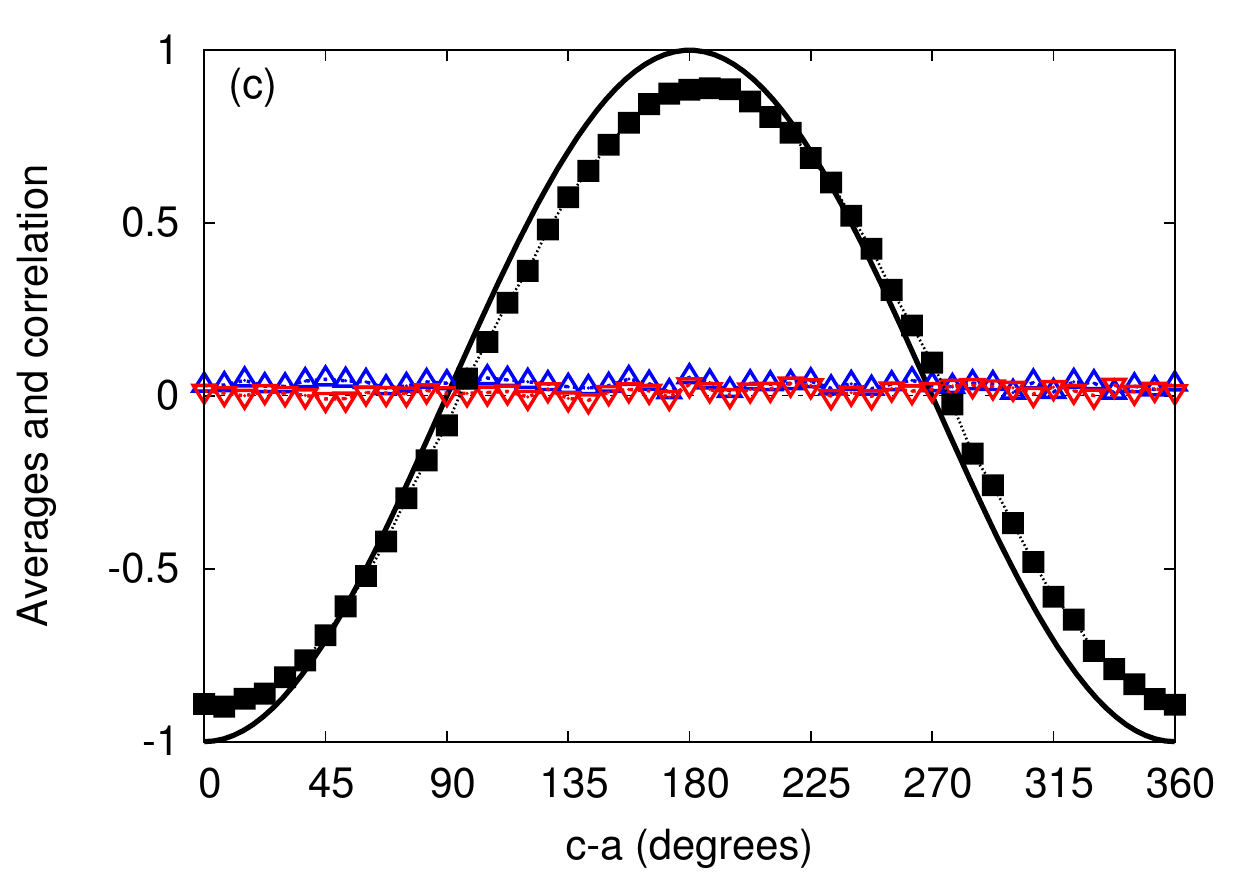}
\caption{(color online) %
Comparison between experimental data produced by the IBM-QE Manila QC device in September 2022
and the quantum-theoretical description in terms of two spin-1/2 objects in the singlet state.
For fixed angles of rotation $a$ and $c$, the circuit shown in Fig.~\ref{CIRCfig2b}  is executed on the IBM-QE Manila QC device ten thousand times.
For each pair of angles $(\alpha,\beta)=(a,c)$, the ten thousand pairs of bits that have been generated determine the ten thousand values of
$(A_{1,n},B_{1,n})$ which are then used to compute
the single spin averages and correlation according to Eq.~(\ref{DATA2}).
In this set of experiments, $a=0$ and $c=0,7.5,\ldots,360$ degrees.
Solid line: correlation ${\widehat E}_{12}(\ba,\bc)=-\cos(a-c)$ of two spin-1/2 objects in the singlet state;
open triangles: experimental results for the single-spin averages ${E}_{\Cac}^{(1)}$ ($\triangle$) and
${E}_{\Cac}^{(2)}$ ($\bigtriangledown$);
solid squares: experimental results for the correlation ${E}_{\Cac}^{(12)}$.
(a): 
data obtained by using qubits $(i,j)=(1,2)$  of the Manila device yields $\max|{\widehat E}_{12}(\ba,\bc)|\approx0.83$;
(b): 
using qubits $(i,j)=(2,3)$ yields $\max|{\widehat E}_{12}(\ba,\bc)|\approx0.90$ with no visible asymmetry;
(c): 
using qubits $(i,j)=(3,4)$ yields $\max|{\widehat E}_{12}(\ba,\bc)|\approx0.90$
and a curve which is slightly shifted or asymmetric around $c-a=180$ degrees.
}
\label{fig4}
\end{center}
\end{figure}

Figure~\ref{fig4} shows the results of performing EPRB experiments on the IBM-QE Manila device,
using three different pairs $(i,j)=(1,2),(2,3),(3,4)$ of the 5-qubit device.
The best results are obtained if we use qubits $(2,3)$, in which case
the correlation ${E}_{\Cac}^{(12)}$ agrees with the quantum-theoretical result within 10\%,
a considerable improvement from the 20\% accuracy obtained with the
IBM-QE devices available in 2016~\cite{MICH17b}.
The averages ${E}_{\Cac}^{(1)}$ and ${E}_{\Cac}^{(2)}$ are close to zero,
as they should be for two spin-1/2 objects in the singlet state.

Next, we perform QC experiments (employing qubits $(i,j)=(2,3)$) to compute the Bell-CHSH function $|S|=|E(a,c)-E(a,d)+E(b,c)+E(b,d)|$.
As in the EPRB experiment with photons (see Section~\ref{FAIL}), between every repetition,
a random number generator is used to choose between circuits with $(\alpha,\beta)$
in Figs.~\ref{CIRCfig2a} and~\ref{CIRCfig2b} corresponding to the choices
$(a,c)=(0,45)$, $(a,d)=(0,135)$, $(b,c)=(90,45)$ and $(b,d)=(90,135)$.
The experimental results are shown in Table~\ref{tab1}.

Alternatively, assuming rotational invariance, we can use the data in Fig.~\ref{fig4} to calculate
the value of the Bell-CHSH function  $|S|=|E(a-c,0)-E(a-d,0)+E(b-c,0)+E(b-d,0)|=|3E(0,45)-E(0,135)|$ and find
$|S|=2.3496$,
$|S|=2.5872$, and
$|S|=2.6529$
for the runs employing
qubits $(i,j)=(1,2)$ (Fig.~\ref{fig4}(a)), $(i,j)=(2,3)$ (Fig.~\ref{fig4}(b)), and $(i,j)=(3,4)$ (Fig.~\ref{fig4}(c)),
respectively.
Apparently, using the data yielding the ``best'' curve (Fig.~\ref{fig4}(b)) does not necessarily result in the largest violation of the Bell-CHSH inequality.

\begin{table*}[!htp]
\caption{
Raw data produced by the IBM-QE Manila QC device in September 2022 using qubits $(i,j)=(2,3)$,
obtained by randomly switching between circuits that use pairs of qubits which share one common qubit (see Fig.~\ref{CIRCfig3} in~\ref{APPQC}), yielding
$E(a,c)=E_1^{(12)}$, $E(a,d)=E_2^{(12)}$, $E(b,c)=E_3^{(12)}$, and $E(b,d)=E_4^{(12)}$.
In total, $N=10000$ repetitions were used to compute
the Bell-CHSH function $|S|=|E(a,c)-E(a,d)+E(b,c)+E(b,d)|=|E_1^{(12)}-E_2^{(12)}+E_3^{(12)}+E_4^{(12)}|$.
}
\begin{tabular*}{\textwidth}{@{\extracolsep{\fill} } crrrrrrrrrr}
\noalign{\medskip}
\hline\hline\noalign{\smallskip}
$s$& $a$  &$b$&   $N_s$ & $n_{00}$ &$n_{01}$& $n_{10}$ &$n_{11}$&   ${E}_{s}^{(12)}$ & ${E}_{s}^{(1)}$ &${E}_{s}^{(2)}$\\
\hline\noalign{\smallskip}
$ 1 $&$   0  $&$  45 $&$   2545 $&$   271 $&$  1018$&$   1038$&$    218 $&$ -0.61572 $&$  0.02868$&$  0.01297 $ \\
$ 2 $&$   0  $&$ 135 $&$   2447 $&$  1058 $&$   234$&$    209$&$    946 $&$  0.63792 $&$  0.03555$&$  0.05599 $ \\
$ 3 $&$  45 $&$   90 $&$   2536 $&$   264 $&$   994$&$   1072$&$    206 $&$ -0.62934 $&$  0.05363$&$ -0.00789 $ \\
$ 4 $&$  90  $&$ 135 $&$   2472 $&$   324 $&$   954$&$    978$&$    216 $&$ -0.56311 $&$  0.05340$&$  0.03398 $ \\
$   $&$      $&$    $&$N=10000 $&$       $&$      $&$       $&$        $&$|S|=2.4609$&$         $&$          $ \\
\hline\noalign{\smallskip}
\end{tabular*}
\label{tab1}
\end{table*}

As shown in Table~\ref{tab1}, randomly selecting between one of the four pairs of angles in-between each measurement reduces the value from $|S|=2.5872$ to $|S|=2.4609$.
A similar reduction is obtained if for each sequence we alternate between the circuit for fixed $(\alpha,\beta)$ and a circuit that performs $X$ operations on each qubit and discard the bitstrings obtained with the latter.
This reduction may be interpreted as some loss of coherence due to the switching between different circuits in-between each measurement but may equally well be within the statistical errors (which we could not test because of limitations on the use of the IBM-QE device).

We have also obtained estimates for
$\langle \bm\sigma_1^\alpha\rangle$, $\langle \bm\sigma_2^\alpha\rangle$, and $\langle \bm\sigma_1^\alpha \bm\sigma_2^\beta\rangle$
for $\alpha,\beta=x,y,z$ by running nine different experiments on the IBM-QE Manila QC device.
From the results presented in Table~\ref{tab2}, we may conclude that executing the circuit that, in theory,
yields a singlet state and performing measurements on the two qubits, produces data that
can be described by the density matrix $\bm\rho=(1-0.9\bm\sigma_1\cdot\bm\sigma_2)/4$,
which is close to the density matrix Eq.~(\ref{SOC10}) of the singlet state and
in concert with the results depicted in Fig.~\ref{fig4}(b).

\begin{table*}[!htp]
\caption{
The raw data for the expansion coefficients of the density matrix Eq.~(\ref{FAIL1})
describing two qubits / two spin-1/2 objects.
The data shows that the density matrix describing the data is, to a good approximation,
given by $\bm\rho=(1-0.9\bm\sigma_1\cdot\bm\sigma_2)/4$.
This data was produced by the IBM-QE Manila QC device in September 2022 using qubits $(i,j)=(2,3)$.
}
\begin{tabular*}{\textwidth}{@{\extracolsep{\fill} } cccc|ccc|ccc}
\noalign{\medskip}
\hline\hline\noalign{\smallskip}
       &\multicolumn{3}{c|}{ $\beta=x$} &\multicolumn{3}{c|}{$\beta=y$} &\multicolumn{3}{c}{$\beta=z$} \\
$\alpha$&   $u_{\alpha}$ &$v_{\beta}$ &$w_{\alpha,\beta}$&   $u_{\alpha}$ &$v_{\beta}$ &$w_{\alpha,\beta}$&   $u_{\alpha}$ &$v_{\beta}$ &$w_{\alpha,\beta}$\\
\hline\noalign{\smallskip}
$x$&$ -0.0384 $&$   0.0936 $&$ -0.9052            $&$\phantom{-}0.0392 $&$   0.0710 $&$-0.0790 $&$\phantom{-}0.0296  $&$  0.0658  $&$\phantom{-}0.0054 $\\
$y$&$ -0.0202 $&$   0.0050 $&$ \phantom{-}0.0192  $&$\phantom{-}0.0256 $&$   0.0252 $&$-0.8932 $&$          -0.0048  $&$  0.0420  $&$\phantom{-}0.0148 $\\
$z$&$ -0.0280 $&$   0.0314 $&$ \phantom{-}0.0142  $&$          -0.0022 $&$   0.0286 $&$-0.0384 $&$\phantom{-}0.0238  $&$  0.0242  $&$          -0.9072 $\\
\hline\noalign{\smallskip}
\end{tabular*}
\label{tab2}
\end{table*}

Finally, by removing the ``Controlled X'' from Figs.~\ref{CIRCfig2a} and~\ref{CIRCfig2b},
the resulting circuit computes the single- and two-qubit averages for a two-particle
system in a pure product state.
The data (not shown) are in good agreement with the results of the corresponding quantum-theoretical description.

In summary, the IBM-QE Manila device produces
data which is in good agreement with the quantum-theoretical description in terms of the singlet state and product state (after removing the controlled X from the circuit).
As mentioned earlier, these experiments have no bearing on the issue of separability of the two qubits/spin-1/2 objects, but they do show that
a many-body system can produce, in an event-by-event-like manner, data
that is in good agreement with the quantum-theoretical description of the EPRB experiment.

\section{Non-quantum models}\label{NQM}

We take as the operational definition of a non-quantum model (NQM) any MM for which:
\begin{enumerate}
\item
All variables of the model, including those which have not been or cannot be measured, always have definite,
not necessarily discrete, values.
\item
All variables change in time according to a process complying with Einstein's notion of local causality.
\end{enumerate}
Note that this operational definition applies to MMs only and does not relate to realism,
the belief that there exists an external reality independent of anyone's thought, knowledge or observation.

Part 1 of the operational definition rules out all probabilistic and quantum-theoretical models, which are M2C.
The reason for this has already been discussed in Section~\ref{section1}. We repeat it here in slightly different words.
By construction, neither probability theory nor quantum theory
contain a specification of a procedure that assigns values to the random variables that
connect these theories to the individual events that are at the core of, but external to, these MMs.
Thus, M2Cs do not comply with the first requirement listed above
but the combination of a M2C and pseudo-random number generators providing definite values
to these random variables does. Of course, the latter is a CM, no longer a M2C.
In this section, in formulating a MM, it is implicitly assumed that, when necessary,
the MM is turned into a model that generates discrete data (a CM)
by including, into the description, an appropriate algorithm (e.g., a pseudo-random number generator)
that generates these data.

Part 2 of the operational definition, the ``locality principle'',
asserts that all physical effects are propagated with finite, subluminal velocities,
so that no effects can be communicated between systems separated by space-like intervals~\cite{Howard1985}.
This principle is related to but not the same as the ``separability principle''.
The latter asserts that any two spatially separated systems possess their own separate real states~\cite{Howard1985},
or in Bell's words, that ``mutually distant systems are independent of one another''~\cite{Bell1966}.
The approach of separating conditions~\cite{RAED19b}, exemplified in Section~\ref{SOC}
extends the ``separability principle'' in that there is no reference to ``spatially separated''
or ``mutually distant''.

A key question in the foundations of quantum physics is whether there exist NQMs
that yield the statistical results of the quantum-theoretical description of the EPRB experiment.
The answer to this question is ``yes''~\cite{RAED06c,RAED07c,ZHAO08,RAED16c,RAED20a}.
In the subsections that follow, we review an LHVM that fails at producing
the (imagined) data of the EPRB experiment (Section~\ref{BELL}) and a M2C (Section~\ref{TPM}) and a CM (Section~\ref{EVENT}) that both succeed.

\subsection{Bell's models and theorem}\label{BELL}

The quantum-theoretical description of the EPRB experiment
in terms of two spin-1/2 objects described by the singlet state yields for the correlation
$\widehat{E}_{12}=-\ba\cdot\bc$ (see Section~\ref{SOC} or~\ref{QTDE} for more details).
Bell demonstrated that there is a conflict between the
quantum-theoretical model of the EPRB experiment and
a class of local realist models~\cite{BELL64,BELL71,BELL93}.

First, Bell considered a model for the correlation ${C}(\ba,\bc)$ of the form~\cite{BELL64}
\begin{eqnarray}
C(\ba,\bc)&=&-\int {A}(\ba,\lambda){A}(\bc,\lambda)\,\mu(\lambda)\,d\lambda
\;,\;
{A}(\ba,\lambda)=\pm1
\;,\;0\le\mu(\lambda)
\;,\;\int \mu(\lambda)\,d\lambda=1
\;,
\label{IN0a}
\end{eqnarray}
where the function ${A}(\ba,\lambda)$
is assumed to model the process that generates the discrete data in the EPRB experiment.

Bell gave a proof that ${C}(\ba,\bc)$
cannot arbitrarily closely approximate the correlation $-\ba\cdot\bc$ {\it for all}
unit vectors $\ba$ and $\bc$~\cite{BELL64}.
According to Bell himself (see Ref.~\cite{BELL93}(p.65)), {\bf this is the theorem.}
From Eq.~(\ref{IN0a}) and ${A}(\ba,\lambda)=\pm1$ it follows immediately
that ${C}(\ba,\bc)=-1$ for all $\ba=\bc$, a characteristic feature
of the correlation of two spin-1/2 objects described by a singlet state.

Subsequently, Bell showed that his theorem also holds for a
more general expression of the correlation Eq.~(\ref{IN0a}),
namely~\cite{BELL71,BELL93}
\begin{eqnarray}
{C}(\ba,\bc)&=&\int {A}(\ba,\lambda){B}(\bc,\lambda)\,\mu(\lambda)\,d\lambda
\;,\;
|{A}(\ba,\lambda)|\le1
\;,\;
|{B}(\bc,\lambda)|\le1
\;,\;0\le\mu(\lambda)
\;,\;\int \mu(\lambda)\,d\lambda=1
\;,
\label{IN0}
\end{eqnarray}
where the functions ${A}(\ba,\lambda)$ and ${B}(\bc,\lambda)$ are
assumed to model the process that generates the discrete data in the EPRB experiment.

Bell's theorem only applies to MMs of the type Eqs.~(\ref{IN0}) and, as
we show later by concrete examples, certainly {\bf not to all} MMs for the EPRB experiments.
In fact, Bell's theorem does not apply to
MMs for the raw, discrete data
generated by EPRB laboratory experiments, see Section~\ref{FAIL}.
For convenience of the reader, the standard proof of Bell's theorem is given in~\ref{BELLPROOF}.
Alternative proofs for specific choices of the functions ${A}(\ba,\lambda)$ and ${B}(\bc,\lambda)$
that are not based on Bell-type inequalities are given in appendices~\ref{NOINEQ} and~\ref{NOINEQII}.

MMs defined by Eq.~(\ref{IN0}) are commonly referred to as local hidden variable models (LHVMs).
The specification of the dependence of
${A}(\ba,\lambda)$ on $\ba$ and $\lambda$ and of ${B}(\bc,\lambda)$ on $\bc$ and $\lambda$
is an integral part of any LHVM.
The symbol $\lambda$ stands for all ``hidden variables'', taking values in a domain denoted by $\Lambda$.
The term ``local'' refers to the fact that for each value of $\lambda$, the value of
${A}(\ba,\lambda)$ (${B}(\bc,\lambda)$) does not depend on the value of $\bc$ ($\ba$)
(see Ref.~\cite{BELL93}, p.15,36).
This requirement of independence, made explicit by writing ${A}(\ba,\lambda)$ and ${B}(\bc,\lambda)$,
guarantees that the model satisfies Einstein's criterion of local causality.
Actually, this requirement ensures much more, namely that the choice of $\bc$ ($\ba$)
can {\sl never} have an effect on the value of
${A}(\ba,\lambda)$ (${B}(\bc,\lambda)$), not in the future nor in the past.
In this sense, the term ``local'' is somewhat unfortunate and misleading
but in the spirit of ``the child (Eq.~(\ref{IN0})) should have a name'',
the ``L'' in LHVM is fine as long as we
keep in mind that ``local'' does not refer to Einstein's notion of local causality.

In Eq.~(\ref{IN0}), the symbol $\lambda$ stands for one or more variables.
The integrals over $\lambda$ in Eq.~(\ref{IN0}) may be defined in terms of sums running
over a partition of the domain $\Lambda$ in $P$ elements $V_i$ and letting $P\to\infty$.
In particular we have
\begin{eqnarray}
{C}(\ba,\bc)&=& \sum_{i=1}^P {A}(\ba,\lambda_i){B}(\bc,\lambda_i)\kappa(V_i)
\;,
\label{IN2}
\end{eqnarray}
where $\lambda_i$ denotes an arbitrary point in the partition element $V_i$,
and $\kappa(V_i)\ge0$ is a measure of the ``volume'' of $V_i$ satisfying $\sum_{i=1}^P \kappa(V_i)=1$.

With an eye on the divide between discrete data and MMs describing these data,
consider the case where ${A}(\ba,\lambda)=\pm1$ and ${B}(\bc,\lambda)=\pm1$.
Once the dependence of ${A}(\ba,\lambda)$ and ${B}(\bc,\lambda)$
and the partition of the domain $\Lambda$ in $P$ parts have been fixed,
we can use a digital computer to generate the $\pm1$'s.
Therefore, in its discretized form Eqs.~(\ref{IN2}) and when implemented as a CM,
Bell's MM actually produces discrete data.
There is no barrier of the kind mentioned in Section~\ref{section1}.

\subsection{Extension of Bell's theorem to stochastic models}\label{EXTE}

Bell's theorem can be generalized to stochastic models.
With ${A}(\ba,\lambda)= \sum_{x=\pm1}xP(x|\ba,\lambda)$ where
$0\le P(x|\ba,\lambda)\le1$ denotes the probability for $x=\pm1$ conditional
on $\ba$ and $\lambda$ and similarly
${B}(\bc,\lambda)=- \sum_{y=\pm1}yP(y|\bc,\lambda)$, Eq.~(\ref{IN0}) becomes
\begin{eqnarray}
{C}(\ba,\bc)&=&-\sum_{x,y=\pm1}\int xyP(x|\ba,\lambda)P(y|\bc,\lambda)\,\mu(\lambda)\,d\lambda
\;,\;
0\le\mu(\lambda)
\;,\;
\int \mu(\lambda)\,d\lambda=1
\;,
\label{IN4}
\end{eqnarray}
and Bell's theorem now says that there do not exist probabilities $P(x|\ba,\lambda)$
and $P(y|\bc,\lambda)$ such that ${C}(\ba,\bc)=-\ba\cdot\bc$.

Once we move into the realm of probabilistic models, there are some new aspects that are
not present in a model that is formulated in terms of variables that take values $\pm1$ only.
One can argue that the values of the variables $x$ and $y$ appearing
in Eq.~(\ref{IN4}) are random and therefore not known at all times.
Then, according to our definition, Eq.~(\ref{IN4}) does not qualify as an NQM
but Eq.~(\ref{IN4}) still qualifies as a local, factorable stochastic model~\cite{FINE82}.
The latter is ``local'' in the sense that mathematical entities, $P(x|\ba,\lambda)$ and $P(y|\bc,\lambda)$
do not depend on $\bc$ and $\ba$, respectively.
However, as probabilities express logical relations, not always physical or causal relations~\cite{JAYN89},
the fact that $\ba$ and $\bc$ appear separated in Eq.~(\ref{IN4}) has no
bearing on Einstein's criterion of local causality being satisfied or not.

\subsection{Proper probabilistic models}\label{PROP}

Any correct probabilistic description of the data collected in an EPRB experiment
has to start from the probability $P(x,y|\ba,\bc)$ for the joined event $(x,y)$, conditional on $\ba$ and $\bc$.
Let us try to express $P(x,y|\ba,\bc)$ in terms of $P(x|\ba,\lambda)$ and $P(y|\bc,\lambda)$.
According to the rules of probability theory~\cite{GRIM01,JAYN03}, we may write
\begin{equation}
P(x,y|\ba,\bc)=\int P(x,y|\ba,\bc,\lambda)\,\mu(\lambda)\ \,d\lambda
=\int P(x|y,\ba,\bc,\lambda)P(y|\ba,\bc,\lambda)\,\mu(\lambda)\ \,d\lambda
\not=
\int P(x|\ba,\lambda)P(y|\bc,\lambda)\,\mu(\lambda)\ \,d\lambda
\;,
\label{IN5}
\end{equation}
where $0\le\mu(\lambda)$ and $\int \mu(\lambda)\,d\lambda =1$.

To proceed, we have to make some assumptions, namely that it is allowed (or a good approximation) to
\begin{itemize}
\item
replace $P(y|\ba,\bc,\lambda)$ by $P(y|\bc,\lambda)$,
\item
replace $P(x|y,\ba,\bc,\lambda)$ by $P(x|\ba,\lambda)$.
\end{itemize}
Probabilities express logical relations, not always physical or causal relations,
a fact that is easily proven by considering an experiment involving a red and a blue ball (see e.g.~Ref.~\cite{JAYN89}): It is perfectly fine to compute $P(\text{``first ball red''}\mid\text{``second ball blue''})$ and $P(\text{``first ball red''}$), which are in general not equal, even though, obviously, the second ball drawn cannot physically influence the first ball drawn.
Therefore, we cannot call upon the requirement of Einstein's notion of local causality to justify
replacing $P(y|\ba,\bc,\lambda)$ by $P(y|\bc,\lambda)$, see for instance~\cite{JAYN89}.
As a matter of fact, within a probabilistic setting, given that one has to start from Eq.~(\ref{IN5}), it is simply impossible to
justify an expression such as Eq.~(\ref{IN4}) on mathematical grounds.
Of course, Eq.~(\ref{IN4}) may be useful as an ``out-of-the-blue'', uncontrolled approximation.
The arguments against the justification of Eq.~(\ref{IN4})
pertain to the probabilistic setting only. They cannot be used against the justification
of Eq.~(\ref{IN0}) as an NQM of the correlation between the functions ${A}(\ba,\lambda)$ and ${B}(\bc,\lambda)$.

It is easy to write $-\ba\cdot\bc$ in the factorized form Eq.~(\ref{IN4}).
For instance, in the case of polarized photons we may write
\begin{eqnarray}
-\cos2(a-c)=\frac{1}{2\pi}\sum_{x,y=\pm1}\int_0^{2\pi}
xy\frac{1 - x\sqrt{2}\cos2(a-\phi)}{2}
\frac{1 - y\sqrt{2}\cos2(c-\phi+\pi/2)}{2}\,d\phi
\;.
\label{IN6}
\end{eqnarray}
The fractions that appear in Eq.~(\ref{IN6}) can take negative values
and therefore do not qualify as probabilities.
In other words, model Eq.~(\ref{IN6}) has to be rejected on elementary grounds.

\subsection{Probabilities and Bell-type inequalities}\label{PROB}

For completeness and because probabilistic models are often used to
argue that Bell-type inequalities only say something about the existence
of joint probabilities and probability spaces~\cite{SUPP81,FINE82a,FINE82b,MUYN86,KUPC86,BROD89,BROD93,KHRE09z,HESS01a,HESS01b,
KUPC05,KHRE07,Khrennikov2008,NIEU09,MATZ09,KHRE09,KHRE11,NIEU11,KUPC16z,KUPC17,HESS17a,NIEU17,Khrennikov2018,Khrennikov2022},
we collect some known results about the Bell-type inequalities in a
probabilistic setting~\cite{SUPP81,FINE82a,FINE82b,BROD89}.
Our presentation differs from earlier ones~\cite{SUPP81,FINE82a,FINE82b,BROD89}
in that we extensively use the representation of the frequencies in terms of their moments.

The first step in formulating a probabilistic model
that describes the data generated by an EPRB experiment with settings $\ba$ and $\bc$ is
to introduce the probability $P(x_1,x_2|\ba,\bc)$
of the event $(x_1,x_2)$, where $x_1,x_2=\pm1$.
As before, we use the notation $|\ba,\bc)$ to keep track of the context (condition) $(\ba,\bc)$
in (under) which the experiment was performed.

With the aim of testing for violations of e.g., the Bell-CHSH inequality,
repeating the experiment for different pairs of settings $(\ba,\bd)$, $(\bb,\bc)$ and  $(\bb,\bd)$
yields data that, within the probabilistic model, are described by
$P(x_1,x_2|\ba,\bd)$, $P(x_1,x_2|\bb,\bc)$, and $P(x_1,x_2|\bb,\bd)$, respectively.

In applying Kolmogorov's probability theory, it is often silently assumed
that the context for which the Kolmogorov probability space (KPS)
$(\Omega,{\cal F},P)$ has been constructed
is fixed for the remainder of the discourse~\cite{KOLM56,GRIM01}.
However, a probabilistic model of the data obtained
by four EPRB experiments with different settings
$(\ba,\bc)$, $(\ba,\bd)$, $(\bb,\bc)$ and $(\bb,\bd)$
has to explicitly account for these four different contexts.
In general, each of the four bivariates
$P(x_1,x_2|\ba,\bc)$, $P(x_1,x_2|\ba,\bd)$, $P(x_1,x_2|\bb,\bc)$, and $P(x_1,x_2|\bb,\bd)$
has its own KPS.

In the case of the EEPRB experiment (see~\ref{EEPRB}), there is only {\bf one single} experiment being performed
in the context $(\ba,\bb,\bc,\bd)$. Therefore, the probability $P(x_1,x_2,x_3,x_4|\ba,\bb,\bc,\bd)$
is well-defined and so is the associated KPS.
In contrast, in principle only bivariates can be used
to model data originating from an EPRB experiment
simply because conceptually and physically, it is impossible to perform the four EPRB experiments as a single experiment.
Consequently, in this case there does not exist a probability $P(x_1,x_2,x_3,x_4|\ba,\bb,\bc,\bd)$
and a single KPS that describes the data of the collective of these four EPRB experiments.
However, we may relax the requirement of a joint KPS a little and
ask the mathematically interesting question if there are conditions that guarantee the existence
of a joint distribution ${\widetilde f}(x_1,x_2,x_3,x_4)$
such that only some of its marginals, namely
$\sum_{x_3,x_4=\pm1}{\widetilde f}(x_1,x_2,x_3,x_4)=P(x_1,x_2|\ba,\bc)$ etc.,
describe the data of the four EPRB experiments.
Such conditions were first established by A. Fine~\cite{FINE82a,FINE82b}.
Conditions for the existence of a joint distribution of three variables and a generalization to the many-variable case
are also given in Refs.~\cite{SUPP81} and~\cite{Kujala2015}, respectively.

The purpose of this subsection is to give an alternative derivation
of these conditions in terms of moments such as
$K_1=\sum_{x_2=\pm1}x_1P(x_1,x_2|\ba,\bc)$ and
$K_{12}=\sum_{x_1,x_2=\pm1}x_1x_2P(x_1,x_2|\ba,\bc)$
of the probabilities.
In particular, we focus on the statement
that the Bell inequalities hold if and only if
there exist a joint distribution ${\widetilde f}(x_1,x_2,x_3,x_4)$
for all observables of the experiment, returning the
marginals $P(x_1,x_2|\ba,\bc)$, $P(x_1,x_2|\ba,\bd)$, $P(x_1,x_2|\bb,\bc)$,
and $P(x_1,x_2|\bb,\bd)$ which describe the data of the four EPRB experiments
with the corresponding settings.

{\bf From the viewpoint of modeling experimental data,
the existence of ${\widetilde f}(x_1,x_2,x_3,x_4)$ does not bring additional insight in the physical processes that generated the data. 
Whenever the conditions change and new data is collected, we have to recompute ${\widetilde f}(x_1,x_2,x_3,x_4)$ from the new data.
Importantly, even if such a joint distribution ${\widetilde f}(x_1,x_2,x_3,x_4)$ is found to exist, this description of EPRB data accomplishes exactly the opposite of
the separation in parts, facilitated by the quantum-theoretical description.
Indeed, the joint distribution provides a description of the four particular experiments as whole.}

In the following, we only consider real-valued functions
$f(x_1,x_2,\ldots)$ of two-valued variables $\{x_1=\pm1,x_2=\pm1,\ldots\}$
for which ${\cal N}=\sum_{\{x_1=\pm1,x_2=\pm1,\ldots\}}f(x_1,x_2,\ldots)\not=0$.
Then, we can, without loss of generality,
replace $f(x_1,x_2,\ldots)$ by $f(x_1,x_2,\ldots)/{\cal N}$
such that the new $f(x_1,x_2,\ldots)$ is normalized to one.
The existence of a normalized, nonnegative real-valued function $f(x_1,x_2,\ldots)$
is a prerequisite for constructing a KPS~\cite{KOLM56,GRIM01}.
Our aim is to establish the equivalence between
the existence of a nonnegative, normalized function $f(x_1,x_2,\ldots)$
and inequalities involving the moments of $f(x_1,x_2,\ldots)$.

Recall that (relative) frequencies are restricted to be ratios of finite
integers, and are therefore discrete data belonging to the domain of ``reality'',
as defined in the introduction. As also explained in the introduction,
probability theory belongs to the class M2C which requires the
introduction of concepts (e.g., ``real'' real numbers such as the square root of
2) that are outside the domain of discrete data. Notwithstanding these
conceptual differences, the equivalences established in this section hold for
both frequencies (discrete data) and probabilities.

\subsubsection{Bivariates of two-valued variables}

Without loss of generality, any real-valued, normalized function $f(x_1,x_2)$ of the two-valued variables
$x_1=\pm1$ and $x_2=\pm1$ can be written as
\begin{eqnarray}
f(x_1,x_2)&=&\frac{1 + K_{1}\,x_1 + K_{2}\,x_2 + K_{12}\,x_1x_2 }{4}
=\frac{1 + x_1(K_{1} + K_{2}\,x_1x_2) + K_{12}\,x_1x_2 }{4}
\;.
\label{TWO0}
\end{eqnarray}
From Eq.~(\ref{TWO0}) it follows that
\begin{subequations}
\label{TWO1}
\begin{eqnarray}
1&=&\sum_{x_1=\pm1}\sum_{x_2=\pm1} f(x_1,x_2)
\;,
\label{TWO1a}
\\
K_{i}&=&\sum_{x_1=\pm1}\sum_{x_2=\pm1} x_i f(x_1,x_2)\;,\; i\in\{1,2\}
\;,
\label{TWO1b}
\\
K_{ij}&=&\sum_{x_1=\pm1}\sum_{x_2=\pm1} x_1x_2 f(x_1,x_2)
\;,
\label{TWO1c}
\end{eqnarray}
\end{subequations}
where the normalization of $f(x_1,x_2)$ implies Eq.~(\ref{TWO1a})
and the $K$'s are the moments of $f(x_1,x_2)$.

If $0\le f(x_1,x_2)\le 1$, application of the triangle inequality to Eqs.~(\ref{TWO1b}) and~(\ref{TWO1c})
yields $|K_{1}|\le1$, $|K_{2}|\le1$, $|K_{12}|\le1$, and
$\vert K_{1}\pm K_{2}\vert \le 1 \pm K_{12}$.
Conversely,
from $|K_1|\le1$, $|K_2|\le1$, and $|K_{12}|\le1$
it immediately follows from Eq.~(\ref{TWO0}) that $f(x_1,x_2)\le1$.
As $\vert K_{1}\pm K_{2}\vert \le 1 \pm K_{12}$ implies
$-x_1(K_{1}x_1 + K_{2}x_1x_2) \le 1 +  K_{12}x_1x_2$,
it immediately follows from Eq.~(\ref{TWO0}) that $f(x_1,x_2)\ge0$.
Therefore, we have
\newsavebox\THEOREM
\savebox\THEOREM{{\bf Theorem I:\ }}
\begin{center}
\framebox{
\parbox[t]{0.9\hsize}{%
\hangindent=\wd\THEOREM
\hangafter=1
\usebox\THEOREM
There exists a real-valued, normalized, nonnegative function $f(x_1,x_2)$ of two-valued variables
with moments $K_1$, $K_2$ and $K_{12}$
if and only if all the inequalities
\begin{eqnarray}
|K_1|\le1\;,\; |K_2|&\le&1\;,\; |K_{12}|\le1\;,
\label{TWO2c}
\\
\vert K_{1}\pm K_{2}\vert &\le& 1 \pm K_{12}\;,
\label{TWO2d}
\end{eqnarray}
are satisfied.
The explicit form of $f(x_1,x_2)$ in terms of its moments is given by Eq.~(\ref{TWO0}).
}}%
\end{center}

\subsubsection{Trivariate of two-valued variables}
Without loss of generality, any real-valued, normalized function $f(x_1,x_2,x_3)$ of the two-valued variables
$x_1=\pm1$, $x_2=\pm1$, and $x_3=\pm1$ can be written as
\begin{eqnarray}
f(x_1,x_2,x_3)&=&\frac{1 + K_{1}\,x_1 + K_{2}\,x_2 + K_{3}\,x_3+ K_{12}\,x_1x_2
+ K_{13}\,x_1x_3 + K_{23}\,x_2x_3 + K_{123}\,x_1x_2x_3 }{8}
\;.
\label{LGI0}
\end{eqnarray}
From Eq.~(\ref{LGI0}) it follows that
\begin{subequations}
\label{LGI1}
\begin{eqnarray}
1&=&\sum_{x_1=\pm1}\sum_{x_2=\pm1}\sum_{x_3=\pm1} f(x_1,x_2,x_3)
\;,
\label{LGI1a}
\\
K_{i}&=&\sum_{x_1=\pm1}\sum_{x_2=\pm1}\sum_{x_3=\pm1} x_i f(x_1,x_2,x_3)\;,\; i\in\{1,2,3\}
\;,
\label{LGI1b}
\\
K_{ij}&=&\sum_{x_1=\pm1}\sum_{x_2=\pm1}\sum_{x_3=\pm1} x_ix_j f(x_1,x_2,x_3)
\;,\; (i,j)\in\{(1,2),(1,3),(2,3)\}
\;,
\label{LGI1c}
\\
K_{123}&=&\sum_{x_1=\pm1}\sum_{x_2=\pm1}\sum_{x_3=\pm1} x_1x_2x_3 f(x_1,x_2,x_3)
\;,
\label{LGI1d}
\end{eqnarray}
\end{subequations}
where the $K$'s are the moments of $f(x_1,x_2,x_3)$ and Eq.~(\ref{LGI1a})
is a restatement of the normalization of $f(x_1,x_2,x_3)$.

If $f(x_1,x_2,x_3)$ is going to be used as a model for empirical frequencies,
it must satisfy $0\le f(x_1,x_2,x_3) \le 1$ and $\sum_{x_1,x_2,x_3=\pm1} f(x_1,x_2,x_3)=1$.
From $0\le f(x_1,x_2,x_3)\le 1$, it follows immediately that all the $K$'s in
Eq.~(\ref{LGI1}) are smaller than one in absolute value.
Furthermore the marginals
$f_3(x_1,x_2)=\sum_{x_3=\pm1}f(x_1,x_2,x_3)$,
$f_2(x_1,x_3)=\sum_{x_2=\pm1}f(x_1,x_2,x_3)$, and
$f_1(x_2,x_3)=\sum_{x_1=\pm1}f(x_1,x_2,x_3)$
are real-valued, normalized and nonnegative bivariates
that is $0\le f_3(x_1,x_2) \le 1$, $\sum_{x_1,x_2=\pm1} f_3(x_1,x_2)=1$, etc.
From the nonnegativity of these marginals, it follows that
$|K_i\pm K_j|\le 1 \pm K_{ij}$ for $(i,j)\in\{(1,2),(1,3),(2,3)\}$~\cite{RAED23}.
Other inequalities involving moments follow by making linear combinations of the inequalities $f(x_1,x_2,x_3)\ge0$
for different values of $(x_1,x_2,x_3)$.
For instance, from
$4[f(+1,+1,+1) + f(-1,-1,-1)] = 1+K_{12}+K_{13}+K_{23} \ge 0$
and
$4[f(-1,+1,+1) + f(+1,-1,-1)] = 1-K_{12}-K_{13}+K_{23} \ge 0$
it follows that $|K_{12}+K_{13}|\le 1 + K_{23}$, one instance the Boole-Bell inequality.
Recall that the latter implies that the inequalities
$\vert K_{12}\pm K_{23}\vert \le 1 \pm K_{13}$ and
$\vert K_{13}\pm K_{23}\vert \le 1 \pm K_{12}$ are also satisfied, see Eq.~(\ref{TRIPLE}).

Summarizing: if the data can be modeled by
a nonnegative, normalized trivariate $f(x_1,x_2,x_3)$, all inequalities
\begin{subequations}
\label{THREE3}
\begin{align}
|K_{1}|&\le1\;,\; |K_{2}|\le1 \;,\;|K_{3}|\le1\;,
|K_{12}|\le1\;,|K_{13}|\le1\;,\;|K_{23}|\le1
\;,
\label{THREE3a}
\\
\vert K_{1}\pm K_{2}\vert &\le 1 \pm K_{12}\;,\;
\vert K_{1}\pm K_{3}\vert \le 1 \pm K_{13}\;,\;
\vert K_{2}\pm K_{3}\vert \le 1 \pm K_{23}\;,\;
\label{THREE3b}
\\
\vert K_{12}\pm K_{13}\vert &\le 1 \pm K_{23}\;,\;
\label{THREE3c}
\end{align}
\end{subequations}
which include the Boole-Bell inequalities are satisfied.
Thus, the inequalities Eq.~(\ref{THREE3}) do not only derive from Bell's model Eq.~(\ref{IN0}) but
also from a much more general model defined by the trivariate Eq.~(\ref{LGI0}).

A remarkable fact, first shown by Fine~\cite{FINE82a,FINE82b} by a different approach than the one taken here,
is that if all inequalities Eq.~(\ref{THREE3})
are satisfied, it is possible to construct a real-valued,
normalized trivariate $0\le f(x_1,x_2,x_3) \le 1$
of the two-valued variables $x_1=\pm1$, $x_2=\pm1$, and $x_3=\pm1$,
yielding all the moments that appear in Eq.~(\ref{THREE3}).
The text that follows replaces the corresponding part and theorem in Ref.~\cite{RAED23}, which are not correct.

First note that if Eqs.~(\ref{THREE3a}) and~(\ref{THREE3b}) hold,
the existence of the three normalized bivariates $0\le f_3(x_1,x_2) \le 1$,
$0\le f_2(x_1,x_3) \le 1$, and $0\le f_1(x_2,x_3) \le 1$ is guaranteed~\cite{RAED23}.
Indeed, for instance, if
$|K_{1}|\le1$, $|K_{2}|\le1$, $|K_{12}|\le1$, and $\vert K_{1}\pm K_{2}\vert \le 1 \pm K_{12}$
it follows immediately that
$0\le f_3(x_1,x_2)=(1 + K_{1}\,x_1 + K_{2}\,x_2 + K_{12}\,x_1x_2)/4\le1$
is the desired normalized bivariate.
Therefore, what remains to be proven is the existence of a real-valued trivariate $g(x_1,x_2,x_3)$ that
(i) takes values in the interval $[0,1]$ and (ii)
yields the three named bivariates with their respective moments $K_1,\ldots,K_{23}$ that appear
in Eqs.~(\ref{THREE3a}) and~(\ref{THREE3b}) as marginals.

Second, note that without loss of generality, any real-valued trivariate $g(x_1,x_2,x_3)$ of the two-valued variables
$x_1=\pm1$, $x_2=\pm1$, and $x_3=\pm1$ can be written as

\begin{eqnarray}
g(x_1,x_2,x_3)&=&\frac{K'_0 + K'_{1}\,x_1 + K'_{2}\,x_2 + K'_{3}\,x_3+ K'_{12}\,x_1x_2
+ K'_{13}\,x_1x_3 + K'_{23}\,x_2x_3 + K'_{123}\,x_1x_2x_3 }{8}
\;.
\label{g0}
\end{eqnarray}
Imposing requirement (ii) immediately yields $K'_0=1$,
$K'_{1}=K_{1}$,
$K'_{2}=K_{2}$,
$K'_{3}=K_{3}$,
$K'_{12}=K_{12}$,
$K'_{13}=K_{13}$, and
$K'_{23}=K_{23}$,
leaving only $K'_{123}$ as unknown.

Third, the requirement that $0\le g(x_1,x_2,x_3)$ is used to
derive conditions for the unknown $K'_{123}$ in terms of all the moments that appear in Eq.~(\ref{THREE3}).
This can be accomplished as follows.
The eight inequalities $g(x_1,x_2,x_3)\ge0$ are rewritten as bounds on $K'_{123}$.
For instance
$g(+1,+1,+1)\ge0 \iff -1 - K_{1} - K_{2} - K_{3}-  K_{12}- K_{13} - K_{23}\le K'_{123}$
and $g(-1,-1,-1)\ge0 \iff K'_{123}\le 1 - K_{1} - K_{2} - K_{3} + K_{12}+ K_{13} + K_{23}$,
and so on.
The resulting eight inequalities can be summarized as
\begin{align}
\max\big[&-1+
K_1-K_2-K_3-K_{12}-K_{13}+K_{23},
-1-K_1+K_2-K_3-K_{12}+K_{13}-K_{23},\nonumber \\
&-1-K_1-K_2+K_3+K_{12}-K_{13}-K_{23},
-1+K_1+K_2+K_3+K_{12}+K_{13}+K_{23}
\big]\nonumber \\
&
\hbox to -.5cm{}\le K_{123}\le
\min\big[
1-K_1-K_2-K_3+K_{12}+K_{13}+K_{23},
1+K_1+K_2-K_3+K_{12}-K_{13}-K_{23},\nonumber \\
&\hbox to 1.72cm{}
1+K_1-K_2+K_3-K_{12}+K_{13}-K_{23},
1-K_1+K_2+K_3-K_{12}-K_{13}+K_{23}
\big]
\;.
\label{K123}
\end{align}
Using the inequalities $\max(a,b,c,d)\ge (a+b+c+d)/4$ and $\min(a,b,c,d)\le (a+b+c+d)/4$
it immediately follows from Eq.~(\ref{K123}) that $-1\le K'_{123}\le1$,
which together with Eq.~(\ref{THREE3a}), guarantees that $g(x_1,x_2,x_3)\le1$.

To prove that there exists at least one value of $K'_{123}$ satisfying Eq.~(\ref{K123})
if all inequalities in Eq.~(\ref{THREE3}) are satisfied, it is sufficient to show that the difference
$\min(\ldots) - \max(\ldots)$ cannot be negative.
One straighforward way to do this is to
use the sixteen inequalities Eqs.~(\ref{THREE3b}) and Eq.~(\ref{THREE3c}) to show
that all sixteen possible differences deriving from Eq.~(\ref{K123}) are nonnegative.
Instead, it is more elegant to rewrite the arguments of $\max(\ldots)$ and $\min(\ldots)$ in Eq.~(\ref{K123}) as
\begin{subequations}
\label{g2}
\begin{align}
\label{g2a}
  K'_{123} &\ge \max( -1-K_3-K_{12}+| K_1+K_2+K_{13}+K_{23} |, -1+K_3+K_{12}+| K_1-K_2-K_{13}+K_{23} | ) & =:&\ \LHS\;,\\
\label{g2b}
  K'_{123} &\le \min( 1-K_3+K_{12}-| K_1+K_2-K_{13}-K_{23} |,  1+K_3-K_{12}-| K_1-K_2+K_{13}-K_{23} | ) &  =:&\ \RHS\;.
\end{align}
\end{subequations}
The final step is then to prove that $\RHS-\LHS\ge0$.
Combining Eqs.~(\ref{g2a}) and~(\ref{g2b}) and using $\min(a,b)+\min(c,d)=\min(a+c,a+d,b+c,b+d)$ yields
\begin{align}
  \RHS - \LHS =
\min (&2+2K_{12}-|K_1+K_2-K_{13}-K_{23}|-|K_1+K_2+K_{13}+K_{23}|\;,\nonumber\\
 &2-2K_{12}-|K_1-K_2+K_{13}-K_{23}|-|K_1-K_2-K_{13}+K_{23}|\;, \nonumber\\
 &2-2K_3-|K_1+K_2-K_{13}-K_{23}|-|K_1-K_2-K_{13}+K_{23}|\;, \nonumber\\
 &2+2K_3-|K_1-K_2+K_{13}-K_{23}|-|K_1+K_2+K_{13}+K_{23}|)
 \;.
 \label{g4}
\end{align}
Using the inequalities Eq.~(\ref{THREE3}) and the identity  $-|a-b|-|a+b| = \min(-a+b,-b+a)+\min(-a-b,a+b)= 2\min(-|a|,-|b|)$,
it can be shown that each of the four arguments of $\min(.)$ in Eq.~(\ref{g4}) is nonnegative.
In detail
\begin{subequations}
\begin{align}
 2+2K_{12}-|K_1+K_2-(K_{13}+K_{23})|-|K_1+K_2+K_{13}+K_{23}| &= 2+2K_{12}+2\min(-|K_1+K_2|,-|K_{13}+K_{23}|)\nonumber\\
 &\ge 2+2K_{12} +2(-1-K_{12}) = 0\; ,\\
 2-2K_{12}-|K_1-K_2+K_{13}-K_{23}|-|K_1-K_2-(K_{13}-K_{23})| &= 2-2K_{12}+2\min(-|K_1-K_2|, -|K_{13}-K_{23}|)\nonumber\\
 &\ge 2-K_{12}+2(-1+K_{12}) = 0\; ,\\
 2-2K_3-|K_1-K_{13}+K_2-K_{23}|-|K_1-K_{13}-(K_2-K_{23})| &= 2-2K_3+2\min(-|K_1-K_{13}|,-|K_2-K_{23}|) \nonumber\\
 &\ge 2-2K_3+2(-1+K_3) = 0\; ,\\
 2+2K_3-|K_1+K_{13}-(K_2+K_{23})|-|K_1+K_{13}+K_2+K_{23}| &= 2+2K_3+2\min(-|K_1+K_{13}|,-|K_2+K_{23}|)\nonumber\\
 &\ge 2+2K_3+2(-1-K_3) = 0\; .
\end{align}
\end{subequations}
This completes the proof that if all inequalities Eq.~(\ref{THREE3}) are satisfied,
there exists a normalized, nonnegative trivariate $0\le g(x_1,x_2,x_3)\le1$ given by
Eq.~(\ref{g0}) with $K'_{123}\in[\LHS,\RHS]$.
Summarizing we have proven

\savebox\THEOREM{{\bf Theorem II:\ }} 
\begin{center}
\framebox{
\parbox[t]{0.85\hsize}{%
\hangindent=\wd\THEOREM
\hangafter=1
\usebox\THEOREM
Given a real-valued, normalized function
$0\le f(x_1,x_2,x_3)\le 1$ of two-valued variables, its moments Eq.~(\ref{LGI1a})--(\ref{LGI1c})
satisfy all the inequalities Eq.~(\ref{THREE3}).
Conversely,
given the values of the moments Eq.~(\ref{LGI1a})--(\ref{LGI1c})
satisfying all the inequalities Eq.~(\ref{THREE3}), it is always possible to choose $K_{123}$ in the range $[\LHS,\RHS]$
(defined by the right hand sides of Eqs.~(\ref{g2a}) and~(\ref{g2b}), respectively),
and construct a real-valued, normalized function $0\le f(x_1,x_2,x_3)\le 1$ of two-valued variables
which yields the specified values of the moments Eq.~(\ref{LGI1a})--(\ref{LGI1c}).
}}
\end{center}
A different strategy was implemented in Mathematica\textsuperscript{\textregistered},
providing an independent proof of the theorem.
The explicit form of $f(x_1,x_2,x_3)$ in terms of its moments is given by Eq.~(\ref{LGI0}).

\subsubsection{Quadrivariate of two-valued variables}

EPRB experiments with photons aiming at demonstrating a violation of
the CHSH inequality require collecting data for four different pairs of contexts/conditions.
As it is physically impossible to perform the four EPRB experiments in one run of an experiment,
there does not exist a probability $P(x_1,x_2,x_3,x_4|\ba,\bb,\bc,\bd)$
and a single KPS that describes the data of these four EPRB experiments.
As a matter of principle, four bivariates with different KPS's are required
to model the data originating from the four runs of the EPRB experiments.

However, as mentioned earlier, it is of interest to ask for the conditions that guarantee the existence
of a joint probability ${\widetilde f}(x_1,x_2,x_3,x_4)$
such that some of its bivariate marginals yield moments
that describe the data of the four EPRB experiments.
This question was first considered and answered by A. Fine~\cite{FINE82a,FINE82b}.

Without loss of generality, any real-valued, normalized function $f(x_1,x_2,x_3,x_4)$ of the two-valued variables
$x_1=\pm1$, $x_2=\pm1$, $x_3=\pm1$, and $x_4=\pm1$ can be written as
\begin{eqnarray}
f(x_1,x_2,x_3,x_4)&=&
\frac{1 + K_{1}x_1 + K_{2}x_2 + K_{3}x_3 + K_{4}x_4}{16}
\nonumber \\
&&+\frac{K_{12}x_1x_2 + K_{13}x_1x_3 + K_{14}x_1x_4 + K_{23}x_2x_3 + K_{24}x_2x_4 + K_{34}x_3x_4}{16}
\nonumber \\
&&+\frac{K_{123}x_1x_2x_3+K_{124}x_1x_2x_4+K_{134}x_1x_3x_4+K_{234}x_2x_3x_4 +K_{1234}x_1x_2x_3x_4}{16}
\;,
\label{CHSH0}
\end{eqnarray}
where the moments are given by
\begin{subequations}
\begin{eqnarray}
1&=&\sum_{x_1=\pm1}\sum_{x_2=\pm1}\sum_{x_3=\pm1} \sum_{x_4=\pm1}f(x_1,x_2,x_3,x_4)
\;,\label{CHSH1a}\\
K_{i}&=&\sum_{x_1=\pm1}\sum_{x_2=\pm1}\sum_{x_3=\pm1}\sum_{x_4=\pm1} x_i\, f(x_1,x_2,x_3,x_4)\;,\; i\in\{1,2,3,4\}
\;,\label{CHSH1b}\\
K_{ij}&=&\sum_{x_1=\pm1}\sum_{x_2=\pm1}\sum_{x_3=\pm1}\sum_{x_4=\pm1}
 x_ix_j\,f(x_1,x_2,x_3,x_4)\;,\; (i,j)\in\{(1,2),(1,3),(1,4),(2,3),(2,4),(3,4)\}
\;,\label{CHSH1c}\\
K_{ijk}&=&\sum_{x_1=\pm1}\sum_{x_2=\pm1}\sum_{x_3=\pm1} \sum_{x_4=\pm1} x_ix_jx_k\, f(x_1,x_2,x_3,x_4)
\;,\; (i,j,k)\in\{(1,2,3),(1,2,4),(1,3,4),(2,3,4)\}
\;,\label{CHSH1d}\\
K_{1234}&=&\sum_{x_1=\pm1}\sum_{x_2=\pm1}\sum_{x_3=\pm1} \sum_{x_4=\pm1} x_1x_2x_3x_4\, f(x_1,x_2,x_3,x_4)
\;.\label{CHSH1e}
\end{eqnarray}
\end{subequations}
Using the triangle inequality, it is easy to show that none of the $K$'s exceeds one in absolute value.

Summing Eq.~(\ref{CHSH0}) over one of the four variables yields the trivariate marginals
\begin{subequations}
\label{CHSH2}
\begin{eqnarray}
f_4(x_1,x_2,x_3)&=&\frac{1 + K_{1}\,x_1 + K_{2}\,x_2 + K_{3}\,x_3+ K_{12}\,x_1x_2 + K_{13}\,x_1x_3 + K_{23}\,x_2x_3 + K_{123}\,x_1x_2x_3 }{8}
\;,\label{CHSH2a}
\\
f_3(x_1,x_2,x_4)&=&\frac{1 + K_{1}\,x_1 + K_{2}\,x_2 + K_{4}\,x_4+ K_{12}\,x_1x_2 + K_{14}\,x_1x_4 + K_{24}\,x_2x_4 + K_{124}\,x_1x_2x_4 }{8}
\;,\label{CHSH2b}
\\
f_2(x_1,x_3,x_4)&=&\frac{1 + K_{1}\,x_1 + K_{3}\,x_3 + K_{4}\,x_4+ K_{13}\,x_1x_3 + K_{14}\,x_1x_4 + K_{34}\,x_3x_4 + K_{134}\,x_1x_3x_4 }{8}
\;,\label{CHSH2c}
\\
f_1(x_2,x_3,x_4)&=&\frac{1 + K_{2}\,x_2 + K_{3}\,x_3 + K_{4}\,x_4+ K_{23}\,x_2x_3 + K_{24}\,x_2x_4 + K_{34}\,x_3x_4 + K_{234}\,x_2x_3x_4 }{8}
\;,\label{CHSH2d}
\end{eqnarray}
\end{subequations}
Trivariates are called ``compatible'' if their common first and second moments are the same.
The trivariates in Eq.~(\ref{CHSH2}) are pairwise compatible. For instance, we have $\sum_{x_2=\pm1}f_1(x_2,x_3,x_4) = \sum_{x_1=\pm1}f_2(x_1,x_3,x_4)
=(1 + K_{3}\,x_3 + K_{4}\,x_4 + K_{34}\,x_3x_4)/4$,
showing that $f_1(x_2,x_3,x_4)$ and $f_2(x_2,x_3,x_4)$ have the same moments
$K_3$, $K_4$, and $K_{34}$.

As in the case of the trivariate, linear combinations of $f(x_1,x_2,x_3,x_4)\ge0$
with different $(x_1,x_2,x_3,x_4)$ yield inequalities in terms of the $K$'s.
However, it saves work to use the inequalities Eq.~(\ref{THREE3}) and simply
change the subscripts properly.
In addition, from $|K_{13}-K_{14}|\le 1-K_{34}$ and $|K_{23}+K_{24}|\le1+K_{34}$ it follows that
\begin{eqnarray}
|K_{13}-K_{14}+K_{23}+K_{24}|\le|K_{13}-K_{14}|+|K_{23}+K_{24}|\le2
\;,
\label{CHSH4z}
\end{eqnarray}
which is one of the Bell-CHSH inequalities~\cite{BELL93,CLAU74}.
The other forms of Bell-CHSH inequalities follow by interchanging subscripts.
Therefore, the existence of the quadrivariate $0\le f(x_1,x_2,x_3,x_4)\le 1$
implies that all Bell-type inequalities hold.

Let us assume that four EPRB experiments yield
discrete data which, to good approximation, can be described by the frequencies
$f(x_1,x_3)=(1 + K_{1}\,x_1 + K_{3}\,x_3 + K_{13}\,x_1x_3)/4$,
$f(x_1,x_4)=(1 + K_{1}\,x_1 + K_{4}\,x_4 + K_{14}\,x_1x_4)/4$,
$f(x_2,x_3)=(1 + K_{2}\,x_2 + K_{3}\,x_3 + K_{23}\,x_2x_4)/4$, and
$f(x_2,x_4)=(1 + K_{2}\,x_2 + K_{4}\,x_4 + K_{24}\,x_2x_4)/4$,
and that all Bell-CHSH inequalities such Eq.~(\ref{CHSH4z}) hold.
Note that these four frequencies are nonnegative, normalized,
pairwise compatible bivariates of their respective arguments.
Therefore, Theorem I applies to each of them, implying that their moments
satisfy the inequalities Eqs.~(\ref{TWO2c}) and~(\ref{TWO2d}) (with appropriate changes of subscripts).

The mathematical problem we now pose ourselves is under which conditions there exists
a nonnegative, normalized function ${\widetilde f}(x_1,x_2,x_3,x_4)$ with
moments $K_{1}$, $K_{2}$, $K_{3}$, $K_{4}$,
$K_{13}$, $K_{14}$, $K_{23}$, and $K_{24}$.
If we can find/construct ${\widetilde f}(x_1,x_2,x_3,x_4)$, we have succeeded
to describe the outcomes of the four EPRB experiments with different contexts/conditions by one joint distribution,
which can then be used to construct a probabilistic model with a common KPS.

As an intermediate step, we prove that given the four, pair-wise compatible bivariates
$f(x_1,x_3)$, $f(x_1,x_4)$, $f(x_2,x_3)$, and $f(x_2,x_4)$
and the assumption that Eq.~(\ref{CHSH4z}) hold,
there exist two compatible trivariates $f_2(x_1,x_3,x_4)$ and $f_1(x_2,x_3,x_4)$
with moments $K_{1}$, $K_{3}$, $K_{4}$, $K_{13}$, $K_{14}$, $K_{34}$,
and $K_{2}$, $K_{3}$, $K_{4}$, $K_{23}$, $K_{24}$, $K_{34}$, respectively.
Except for $K_{34}$, all moments are already known, derivable from the four, pair-wise compatible bivariates.
From the data of the four EPRB experiments, we cannot infer the value of $K_{34}$.

Therefore, the existence of $f_2(x_1,x_3,x_4)$ and $f_1(x_2,x_3,x_4)$ depends on whether
is possible to assign a value to $K_{34}$ in the interval $[-1,1]$ such that
$f_2(x_1,x_3,x_4)$ and $f_1(x_2,x_3,x_4)$ are nonnegative, normalized trivariates.
Lemma I (see~\ref{LEMMA})
shows that there exists such a value of the moment $-1\le K_{34}\le 1$ if there exist four
nonnegative, normalized, pair-wise compatible bivariates
$f(x_1,x_3)$, $f(x_1,x_4)$, $f(x_2,x_3)$, and $f(x_2,x_4)$
and if Eq.~(\ref{CHSH4z}) holds.
Furthermore, the moments
$K_{1}$, $K_{3}$, $K_{4}$, $K_{13}$, $K_{14}$, and $K_{34}$
satisfy the inequalities Eqs.~(\ref{THREE3a})--(\ref{THREE3c}) (with appropriate change of subscripts).
From Theorem II it then follows that we can always find
at least one value of $K_{134}$ such that there exists a nonnegative, normalized trivariate
$f_2(x_1,x_3,x_4)$ with these moments.
Similarly, $K_{2}$, $K_{3}$, $K_{4}$, $K_{23}$, $K_{24}$, and $K_{34}$
satisfy the inequalities Eqs.~(\ref{THREE3a})--(\ref{THREE3c}) (with appropriate change of subscripts),
implying that there exists a nonnegative, normalized trivariate $f_1(x_2,x_3,x_4)$ with these moments.

Following Fine~\cite{FINE82b}, we use these compatible trivariates
$f_2(x_1,x_3,x_4)$ and $f_1(x_2,x_3,x_4)$ to define the quadrivariate
\begin{eqnarray}
{\widetilde f}(x_1,x_2,x_3,x_4)&=&
\left\{
\begin{array}{lcl}
\frac{f_2(x_1,x_3,x_4)f_1(x_2,x_3,x_4)}{\sum_{x_1=\pm1}f_2(x_1,x_3,x_4)}
&,&\sum_{x_1=\pm1}f_2(x_1,x_3,x_4)>0\\
\\
0&,&\sum_{x_1=\pm1}f_2(x_1,x_3,x_4)=0
\end{array}
\right.
\;.
\label{CHSH4}
\end{eqnarray}
As
$0\le f_2(x_1,x_3,x_4)\le\sum_{x_1=\pm1}f_2(x_1,x_3,x_4)$ and
$0\le f_1(x_2,x_3,x_4)\le\sum_{x_1=\pm1}f_2(x_1,x_3,x_4)$
imply that
$f_2(x_1,x_3,x_4)\times$ $f_1(x_2,x_3,x_4)\le\left(\sum_{x_1=\pm1}f_2(x_1,x_3,x_4)\right)^2
\le \sum_{x_1=\pm1}f_2(x_1,x_3,x_4)\le1$.
Therefore, $0\le{\widetilde f}(x_1,x_2,x_3,x_4)\le1$
showing that Eq.~(\ref{CHSH4}) defines a normalized, nonnegative quadrivariate.
From Eq.~(\ref{CHSH4}) it follows immediately that
\begin{subequations}
\label{CHSH5}
\begin{eqnarray}
{\widetilde K}_{i}&=&\sum_{x_1=\pm1}\sum_{x_2=\pm1}\sum_{x_3=\pm1}\sum_{x_4=\pm1} x_i\, {\widetilde f}(x_1,x_2,x_3,x_4)
=K_{i}\;,\; i\in\{1,2,3,4\}
\;,\label{CHSH5a}\\
{\widetilde K}_{ij}&=&\sum_{x_1=\pm1}\sum_{x_2=\pm1}\sum_{x_3=\pm1}\sum_{x_4=\pm1}
 x_ix_j\,{\widetilde f}(x_1,x_2,x_3,x_4)=K_{ij}\;,\; (i,j)\in\{(1,3),(1,4),(2,3),(2,4),(3,4)\}
\;,\label{CHSH5b}
\end{eqnarray}
\end{subequations}
as required.
The expression of the second moment $-1\le{\widetilde K}_{12}\le1$
in terms of the other first and second moments is rather lengthy and therefore not given here.
By construction, the ${\widetilde K}$'s (with proper combination of subscripts) satisfy
the inequalities Eqs.~(\ref{THREE3a})--(\ref{THREE3c}), which include the Bell inequalities.
Conversely,
if ($K_{1}$, $K_{3}$, $K_{4}$, $K_{13}$, $K_{14}$, $K_{34}$)  and
($K_{2}$, $K_{3}$, $K_{4}$, $K_{23}$, $K_{24}$, $K_{34}$)
satisfy the inequalities Eqs.~(\ref{THREE3a})--(\ref{THREE3c}),
we can construct a normalized, nonnegative quadrivariate with these moments.


\subsubsection{Discussion}
From the foregoing, it is abundantly clear that in contrast to Bell's original derivation, the derivation of Bell-type inequalities in the probabilistic setting does not rely on assumptions about ``locality'', ``macroscopic realism'', ``non-invasive measurements'' and the like.
Violations of Bell-type inequalities derived within the framework of a probabilistic model
are a signature of the non-existence of a joint distribution rather
than some signature of  ``quantum physics''.

Most importantly, describing two-valued data of EPRB experiments performed under four different conditions
in terms of a joint distribution (if it exists) accomplishes exactly the opposite
of the description in separated parts provided by quantum theory,
see the text in boldface in Section~\ref{PROB}.

\subsection{Stochastic hidden-variables model for
the data collected in EPRB laboratory experiments}\label{TPM}

The fundamental problem with applying Bell's model Eq.~(\ref{IN0}) to the description of
experimental data is the following. Evidently, in any laboratory EPRB experiment, before
one can even think about computing correlations of particle properties, it is
necessary to first classify a detection event as corresponding to the arrival of a particle or as something else.
Such a procedure is missing in both the EPRB thought experiment
and Bell's model Eq.~(\ref{IN0}).
If the aim is to describe the data of an EPRB laboratory experiment,
it is necessary to generalize Bell's model, for instance by incorporating such an identification procedure.
To the best of our knowledge, such a generalization was first studied in Refs.~\cite{PASC86,PASC87}.

As illustrated in Fig.~\ref{weihsexp} and further explained in Section~\ref{FAIL},
for any particular choice of settings, the raw data produced by EPRB experiments comes
in pairs $\{(x_m,t_m)|\,m=1,\ldots,M\}$ and $\{(y_n,t'_n)|\,n=1,\dots,N\}$
where, in practice, the numbers of detected events $M$ and $N$ are unlikely to be same.
In most EPRB experiments, the $t$'s represent time tags, times at which the corresponding detectors fired.
More recent experiments~\cite{SHAL15,GIUS15} employ so-called event-ready detection techniques, in which case
the $t$'s represent voltage pulses. In this case, the detection of a single photon is defined as the voltage exceeding
a voltage threshold, tuned to maximize the violation~\cite{SHAL15,GIUS15} of the Clauser-Horn (CH) inequality~\cite{CLAU74}.
These experiments have demonstrated a violation of the latter but,
unlike for the experiments discussed in Section~(\ref{FAIL}), did not show results as a function of the rotation angles.
Conceptually and mathematically, the voltage threshold is just
another mechanism to identify pairs by rejecting events~\cite{RAED17a}.
As the model presented in this section can be tailored to these
experiments by a minor modification~\cite{RAED17a}, we focus on building
a probabilistic model for the EPRB experiment with polarized photons depicted in Fig.~\ref{weihsexp}
and discussed in Section~\ref{FAIL}.
A more general treatment can be found elsewhere~\cite{RAED07c}.

We idealize matters a little by assuming that $N=M$.
Then, the data set looks like ${\cal E}=\{ (x_n,t_n,y_n,t'_n)|n=1,\ldots, N\}$.
As before, for simplicity, for polarized photons
we use the angles $a$ and $c$ instead of $\ba$ and $\bc$, respectively.
Our goal is to construct the simplest probabilistic model which
\begin{enumerate}
\item
describes the compound event $(x_n,t_n,y_n,t'_n)$
in terms of processes that are local to station 1 and 2,
\item
yields Malus' law (by construction),
\item
yields the averages and the correlation of two spin-1/2 objects in the singlet state,
\item
describes the dependence of the correlation on the time-coincidence window $W$,
observed experimentally (see Section~\ref{FAIL}),
\item
yields the averages and the correlation that are obtained from the quantum-theoretical description
of two spin-1/2 objects in the product state.
\end{enumerate}

For simplicity and in concert with features of the raw data listed in Section~\ref{LABE},
we assume that $(x_m,y_m,t_m,t'_m)$ and $(x_n,y_n,t_n,t'_n)$ are uncorrelated  for all $m\not=n$.
Then the probability for the $(x_n,y_n,t_n,t'_n)$ does not depend on the subscript $n$,
which we omit in what follows.

Without infringing on the axioms of probability theory and
without loss of generality, the probability $P(x,y,t,t'|{a},{b})$
can always be expressed as an integral over
$\xi $, $\zeta $, representing the polarization of the photons arriving in stations 1 and 2, respectively.
We have
\begin{equation}
P(x,y,t,t'|{a},{c})=\frac{1}{4\pi^2}\int_0^{2\pi}\int_0^{2\pi}
P(x,y,t,t'|{a},{c},\xi ,\zeta )p(\xi ,\zeta |{a},{c})d\xi d\zeta
\;.
\label{TPM1}
\end{equation}
where $p(\xi ,\zeta |a,c)$ denotes the probability density
with which the source emits photons with polarizations $(\xi ,\zeta )$ under
the conditions $({a},{c})$.

With our eye on constructing a model having the same features as those
used by Bell to prove his theorem, that is independence/separability, we assume that
(i) the values of
$x$, $y$, $t$, and $t'$ are mutually independent,
(ii) the values of $x$ and $t$ ($y$ and $t'$) are independent
of ${c}$ and $\zeta $ (${a}$ and $\xi $),
(iii) $\xi $ and $\zeta $ are independent of ${a}$ or ${c}$.
With these assumptions Eq.~(\ref{TPM1}) becomes~\cite{RAED07c}
\begin{eqnarray}
P(x,y,t,t'|{a},{c})
&=&\frac{1}{4\pi^2}\int_0^{2\pi}\int_0^{2\pi}P(x|{a},\xi )P(t|{a},\xi )
P(y|{c},\zeta )  P(t'|{c},\zeta )p(\xi ,\zeta )d\xi d\zeta
\;.
\label{TPM2}
\end{eqnarray}
Note that $xy$-correlation calculated using  Eq.~(\ref{TPM2}) has the same mathematical structure as Eq.~(\ref{IN4})
and therefore, according to Bell's theorem, this correlation can never be arbitrarily close to $-\cos2(a-c)$ for all $a$ and $c$.

However, EPRB experiments~\cite{WEIH98,WEIH00,ADEN12,AGUE09}
employing a time-coincidence window $W$ to identify pairs of photons
are not described by Eq.~(\ref{TPM2}).
Accounting for the time-coincidences requires that we multiply Eq.~(\ref{TPM2})
by the unit-step function $\Theta(W-|t-t'|)$ and integrate over all possible $t$ and $t'$.
Therefore, the data used to compute the correlations are described by the probability
\begin{equation}
P(x,y|{a},{c})
=
\int_0^{2\pi}\int_0^{2\pi}P(x|{a},\xi )P(y|{c},\zeta ) {\widehat \mu}(\xi ,\zeta |{a},{c})d\xi d\zeta
\;,
\label{TPM3}
\end{equation}
where
\begin{equation}
{\widehat \mu}(\xi ,\zeta |{a},{c})=\frac{w(\xi ,\zeta |{a},{c})p(\xi ,\zeta )}{
\int_0^{2\pi}\int_{0}^{2\pi}w(\xi' ,\zeta' |{a},{c})p(\xi' ,\zeta' )d\xi' d\zeta'}
\;,
\label{TPM3a}
\end{equation}
and the weight $w(\xi ,\zeta |{a},{c})$ is given by
\begin{eqnarray}
w(\xi ,\zeta |{a},{c})&=&
\int_{-\infty}^{+\infty}\int_{-\infty}^{+\infty}
\Theta (W-|t-t'|)P(t|{a},\xi )P(t'|{c},\zeta )\,dt\,dt'
\;.
\label{TPM4}
\end{eqnarray}
Equation~(\ref{TPM3}) no longer has the structure of Bell's model
because the probability density Eq.~(\ref{TPM3a}) may depend on $({a},{c})$ (violating the notion of separability).
In other words, Bell's theorem does not apply to the probabilistic model defined by Eq.~(\ref{TPM3}).

In order to proceed, we have to make a choice for the distribution of time delays.
As a very simply choice we take $P(t|{a},\xi )=\Theta(t)\Theta(T(\xi -a)-t)/T(\xi -a)$, where $T$ is a function specified below, and obtain
\begin{eqnarray}
w(\xi ,\zeta |{a},{c})&=&\frac{1}{T(\xi -a)T(\zeta -c)}\int_{0}^{T(\xi -a)}dt \int_{0}^{T(\zeta -c)}
\Theta(W-|t -t'|)\,dt\,dt'
\;.
\label{TPM6}
\end{eqnarray}
With this definition  of $P(t|{a},\xi )$, the integrals in Eq.~(\ref{TPM6}) can be worked out analytically~\cite{RAED07c}.
Inspired by the results of event-by-event simulations~\cite{RAED06c,RAED07c,ZHAO08},
we set $T(\xi -a)=T_0 |\sin2(\xi -a)|^{d/2}$ where $T_0$ and $d$ are parameters of the model.
As the full expression is not of immediate interest,
we only give the expressions for a few interesting limiting cases~\cite{RAED07c}:
\begin{eqnarray}
w(\xi ,\zeta |{a},{c})&=&
\left\{\begin{array}{ccl}
1 &, & W\ge T_0\\ \\
\frac{W(2T_0-W)}{T_0^2}& , &  d=0 \;\; \hbox{and}\;\;W < T_0\\ \\
\frac{2W}{\max(T(\xi -a),T(\zeta -c))}+{\cal O}(W^2)&, & W \rightarrow 0
\end{array}
\right.
\;.
\label{TPM7}
\end{eqnarray}

Let us first demonstrate how Eqs.~(\ref{TPM3}) and~(\ref{TPM6}) yield the correct
quantum-theoretical expression for two spin-1/2 objects in a product state.
Assume that  $p(\xi ,\zeta )=\delta(\xi -\alpha)\delta(\zeta -\beta)$.
Then, independent of the distributions of $t$ and $t'$, we find
\begin{equation}
P(x,y|{a},{c})=P(x|{a},\alpha)P(y|{c},\beta)
\;,
\label{TPM5}
\end{equation}
which has the structure of the probability for a quantum system in product state, see~\ref{PRODUCTSTATE}.
Furthermore, independent of the choice of $p(\xi ,\zeta )$, we also recover Eq.~(\ref{TPM5})
if Eq.~(\ref{TPM7}) does not dependent on $\xi$ or $\zeta$.
Thus, as the first two rows of Eq.~(\ref{TPM7}) show,
we also recover Eq.~(\ref{TPM5}) if $W$ is larger than the maximum time delay $T_0$ or if the parameter $d=0$.

Next, we demonstrate that Eqs.~(\ref{TPM3}) and~(\ref{TPM6}) yield the quantum-theoretical expressions
for the averages and the correlation of two photon-polarizations in the singlet state.
We start by assuming that the source emits pairs of photons with orthogonal polarizations,
in symbols $p(\xi ,\zeta )=\delta(\xi +\pi/2-\zeta )$.

For small $W/T_0$, we use the expression in the third row of Eq.~(\ref{TPM7})
and insert the expressions $P(x|{a},\xi )=(1+x\cos2(\xi -a))/2$
and $P(y|{c},\zeta )=(1+y\cos2(\zeta -c))/2$ known from Malus' law.
By symmetry, it follows immediately that $E_1(a,c,W\rightarrow0)=E_2(a,c,W\rightarrow0)=0$.
For even integer values of $d$,
the expression of the correlation $E_{12}(a,c,W\rightarrow0)$ can be obtained analytically.
We find~\cite{RAED07c}
\begin{eqnarray}
E_{12}(a,c,W\rightarrow0)
&=&
\left\{
\begin{array}{lcl}
-\frac{1}{2}\cos 2\theta&,& d=0\\
\frac{\pi}{4}\sin2\theta\cos2\theta - \cos2\theta +\ln[|\tan \theta|^{\sin^22\theta/2}]&,& d=2\\
-\cos 2\theta&,& d=4\\
-\frac{1}{2}\cos 2\theta\left[1+24(19+5\cos4\theta)^{-1}\right]&,& d=6\\
-(53\cos2\theta+7\cos6\theta)(39+21\cos4\theta)^{-1}&,& d=8
\end{array}
\right.
,
\label{TPM8}
\end{eqnarray}
where $\theta=a-c$.
Clearly, for $d=4$ and $W\rightarrow 0$, the probabilistic model yields
the desired correlation $\widehat{E}_{12}(\ba,\bc)=-\cos2(a-c)$
for two photon polarizations described by the singlet state.
For $d>4$, the resulting correlation is outside the scope
of what a quantum-theoretical model of two spin-1/2 objects can describe
because it violates the Cirel'son bound~\cite{CIRE80}, see Eq.~(\ref{EXQT5}).

Repeating the calculation with parallel instead of antiparallel polarizations,
that is with $p(\xi ,\zeta )=\delta(\xi -\zeta )$, simply changes the sign
of $E(a,b,W\rightarrow0)$ in Eq.~(\ref{TPM8}).
The resulting correlation is outside the scope of what a quantum-theoretical model of two spin-1/2 objects can describe, (because $q<-1/3$, see Section~\ref{NOGO}).

In conclusion, the probabilistic model for the raw data produced by EPRB experiments, in combination
with time coincidence counting employed in most of these experiments,
does indeed (for $W\to0$ and $d=4$) yield
the averages and the correlation of two spins-1/2 objects in the singlet state,
complies with Malus' law and can (if $d=0$ or $W\ge T_0$) also describe two spins-1/2 objects in the product state.

EPRB experiments that do not rely on coincidence
counting but identify photons by a process that is local to the observation station~\cite{GIUS15,SHAL15}
can be treated similarly~\cite{RAED17a}.
We only have to replace Eq.~(\ref{TPM6}) by
\begin{eqnarray}
w(\xi ,\zeta |{a},{c})=w(\xi |{a})w(\zeta |{c})
\;,
\label{TPM9}
\end{eqnarray}
where
\begin{eqnarray}
w(\xi |{a})=\left[\frac{1}{T(\xi -a)}\int_{0}^{T(\xi-a)}\Theta(W-t)\,dt\right]
=\Theta(W-T(\xi-a))+\frac{W\Theta(T(\xi-a)-W)}{T(\xi-a)}
\;
\label{TPM10}
\end{eqnarray}
and work out the details.

\subsection{Subquantum model: event-by-event simulation}\label{EVENT}

In this section, we briefly discuss results obtained by event-by-event simulations
that comply with the operational definition of an NQM, as given in the beginning of this section.
In the appropriate limits, event-by-event simulation can reproduce,
to very good approximation, all results of the quantum theory
for two spins-1/2 objects~\cite{RAED06c,RAED07c,ZHAO08,RAED16c,RAED20a}.
As there are abundant publications about this simulation work, we refrain from describing
the algorithm and refer the interested reader to these publications.

Unlike real experiments, an event-by-event CM of the EPRB experiment can operate in a mode
in which all the data in
${\cal D}_1$,
${\cal D}_2$,
${\cal D}_3$, and
${\cal D}_4$
can be extracted from quadruples~\cite{RAED16c}.
Computationally, this feat is realized by using the same pseudo-random numbers
for each of the four different experiments~\cite{RAED16c}.
In this case, the process that generates the data is said to comply
with the notion of counterfactual definiteness~\cite{PEAR09}.

In the counterfactual definite mode of simulation,
the four sets of data originate from one set of quadruples, and therefore we have $\QUAD=1$.
It then follows from Eq.~(\ref{CHSH}) that $S_{\mathrm{CHSH}}\le 2$.
Clearly, the counterfactual definite mode of simulation can {\bf never}
generate data that leads to a violation of the Bell-CHSH inequality.
Consequently,  Bell's theorem guarantees that the
counterfactual definite mode of simulation can {\bf never} yield
a correlation that closely resembles $E_{12}=-\ba\cdot\bc$.

Generating raw data ${\cal E}=\{ (x_1,t_1,y_1,t'_1),\ldots,(x_{N},t_N,x_{N},t'_N)\}$
with the perfect, counterfactual definite mode of simulation
yielding a correlation that closely resembles $E_{12}=-\ba\cdot\bc$
can {\bf only} be accomplished by discarding raw data depending on the conditions $\ba$ and $\bc$.
Then the new data sets
${\cal D}'_1$,
${\cal D}'_2$,
${\cal D}'_3$, and
${\cal D}'_4$
no longer have the property that all contribution to the correlations can be reshuffled to form quadruples.

\begin{figure}[!htp]
\centering
\includegraphics[width=0.7\hsize]{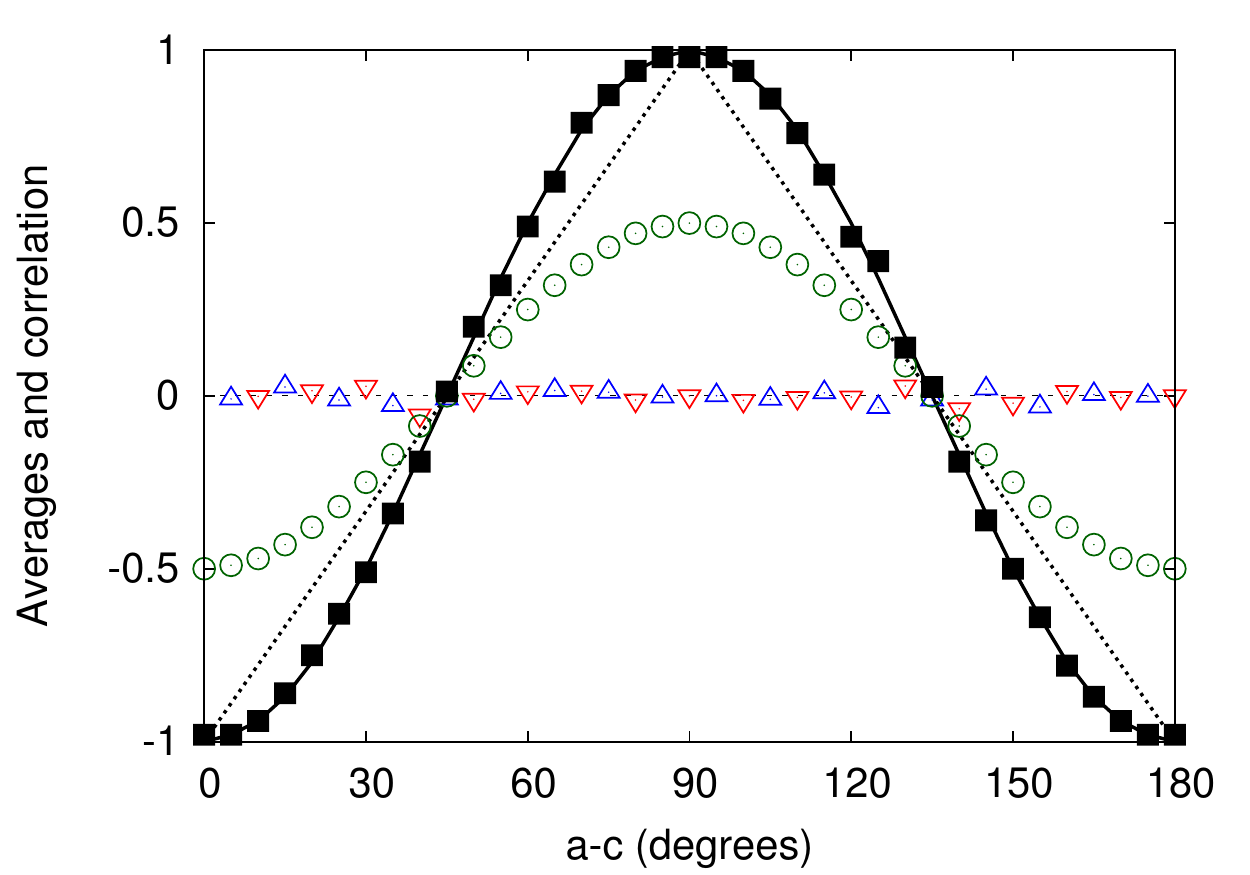}    
\caption{(color online)
The averages and correlations
as obtained by an event-by-event simulation of an EPRB laboratory experiment
in which the detection of a photon in a station 1 or 2 depends
on a threshold that is local to the respective stations~\cite{GIUS15,SHAL15}.
There is no coincidence counting in this set up.
Solid line: the correlation $E_{12}(a,c)=-\cos2(a-c)$, as obtained from LI applied to
a reproducible and robust experiment (see Section~\ref{LI}) and from the
quantum theory of two spins-1/2 objects in the singlet state;
dotted line: the correlation $E_{12}(a,c)=-1+(2/\pi)\arccos(\cos2(a-c))$
as obtained from Bell's toy model (see~\ref{CBEL});
solid squares: the correlation $E^{(12)}_1$ (see Eq.~(\ref{CORR}))
computed from the data generated by the event-by-event simulation;
open triangles: the averages (see Eq.~(\ref{CORR})) $E^{(1)}_1$ ($\triangle$) and $E^{(2)}_1$ ($\bigtriangledown$)
computed from the data generated by the event-by-event simulation;
open circles: the correlation $E^{(12)}_1$
computed from the data generated by the event-by-event simulation without accounting for the
local thresholds. In the latter case, the simulation data is in excellent agreement
with the correlation obtained from Bell's modified toy model (see Section~\ref{CBELM}).
Using a threshold mechanism similar to the one employed in the so-called loophole free EPRB
experiments~\cite{GIUS15,SHAL15}, there is excellent agreement between the simulation data (solid squares and
triangles) obtained with the quantum-theoretical description of the EPRB experiment.
}
\label{subq}
\end{figure}

Figure~\ref{subq} shows the results of an event-by-event simulation of EPRB laboratory
experiments with photons in which data is discarded, not by coincidence counting, but
by a local procedure mimicking the one used in the analysis of EPRB laboratory experiments~\cite{GIUS15,SHAL15},
a procedure similar to the one described by Eq.~(\ref{TPM9}).
For the complete specification of the simulation algorithm the reader should consult Ref.~\cite{RAED20a}.

Figure~\ref{subq} demonstrates that the correlation (solid squares) and averages (triangles)
obtained by the event-by-event simulation are in excellent agreement
with the corresponding quantum-theoretical results.
Also shown by the open circles is the correlation obtained without discarding raw data.
This correlation matches the one obtained from Bell's modified toy model, see~\ref{CBELM}.
For reference, the dotted line shows the correlation obtained from Bell's toy model, see~\ref{CBEL}.

In summary, the counterfactual definite mode of simulation can only generate raw data
for the $A$'s, and $B$'s which can be extracted from quadruples.
Therefore, the correlation computed with this data can never violate the Bell-CHSH inequality
and, by appeal to Bell's theorem, can never closely resemble $E_{12}=-\cos2(a-c)$.
In complete agreement with model-free inequality Eq.~(\ref{DATA4}), there is
only one way out of the conundrum, namely one has to discard, with whatever procedure, data.
Bell's model Eq.~(\ref{IN0}) is too primitive. It does not contain such a procedure.
On the other hand, all EPRB laboratory experiments
(including the so-called loophole free experiments~\cite{HENS15,GIUS15,SHAL15}), one way or another, never use
all the raw data to check for violations of Bell-type inequalities but have their ways to identify a fraction of them as photons.
It is this selection process that can give rise to the violation of a Bell-type inequality (beyond the usual statistical fluctuations)
and, in some cases such as the one discussed in the present and the previous section, yields
a correlation that resembles $E_{12}=-\cos2(a-c)$.

\section{Conclusion}\label{CONC}

The view taken in this paper is that discrete data recorded by experiments
and mathematical models envisaged to describe
relevant features (e.g., averages, correlations, etc.) of these data
belong in their own, separate universes and should be treated accordingly.
Using EPRB experiments as an example, we have scrutinized various aspects of treating
discrete data and mathematical models thereof separately.

In the universe of discrete data, the number of operations one can carry out on the data
without changing the numerical values of the relevant features is very limited.
In fact, we can only resort to the commutativity of the addition operation
to reshuffle the order in which the contributions to a sum appear.
In this paper, we demonstrate that this property suffices to prove new nontrivial bounds on
the value of certain linear combinations (e.g., as they appear in the Bell-CHSH inequality) of correlations.
Being the result of elementary arithmetic,
that is being the result of multiplying and adding numbers which take values plus and minus one only,
these nontrivial bounds cannot be violated by data obtained from EPRB experiments.

Most importantly, the proof of these nontrivial bounds does not require making assumptions about the process
that produced the data. These bounds are ``model-free'', linear functions
of the number of quadruples one can create by reordering the contributions to each of the correlations.
If the number of quadruples is equal to the number of contributions to the correlations,
the {\sl values} of these
nontrivial bounds coincide with the {\sl values} known from the Bell, CHSH, or CH inequalities.
However, the latter have been derived in the context of a mathematical (Bell-type) model, not in the context of experimental data.
The model-free Bell-type bounds for discrete data, a special case of the general model-free inequality,
may be violated. However, this violation has no bearing on the validity of the assumptions that
may have provided the motivation to construct the mathematical (Bell-type) model.

The existence of the divide between the universes of discrete data and mathematical models thereof
is further supported by Fine's theorem for a probabilistic model of the EPRB experiment.
Of particular relevance to the present discussion is the part of the theorem
that establishes the Bell-CHSH inequalities (plus compatibility) as being the necessary and sufficient conditions
for the existence of a joint distribution of the four observables involved in these inequalities.
This four-variable joint distribution returns
the pair distributions describing the four EPRB experiments required to test for a violation of these inequalities.
Fine's theorem holds in the mathematical-model universe only.
Only in the unattainable limit of an infinite number of measurements (that is by leaving the universe of discrete data),
and in the special case that the Bell-CHSH inequalities hold,
it may be possible to prove the mathematical equivalence between the model-free and Bell-type bounds.

Having established that the discrete-data and mathematical-model universes are separate entities,
a major question is how to establish the applicability and validity of a particular mathematical model.
In the case that the results of individual measurements are assumed to be unpredictable, as in EPRB experiments,
the comparison is through the averages, correlations, etc.,
predicted by the mathematical model with those of the experimental data.

Starting from discrete data, we have demonstrated that the construction of a mathematical
model giving a concise description of the averages and correlations computed from these data yields
the quantum-theoretical model of the EPRB thought experiment.
Essential ingredients of this construction are separation of conditions and the assumptions that the relevant features
are reproducible and change smoothly with smooth variations (robustness) of the conditions under which the experiments
are carried out.

In contrast to the conventional approach which postulates that quantum
theory models the process by which experiments produce data,
the constructive approach adopted in this paper is free of all issues
created by attempts to attach an interpretation to the symbols appearing in the mathematical model
and to ``explain'' that each laboratory measurement yields a definite result.

Our construction shows that the quantum model of the EPRB experiments
is just one particular description of the relevant features of the discrete data.
We have argued that compared to other representations of data obtained from EPRB experiments,
quantum theory is exceptionally powerful in that it can represent the relevant features of the data of
EPRB thought experiments with many different settings in terms of only fifteen numbers.

In summary, we have shown that by starting from the discrete data produced by EPRB experiments,
all controversies about the meaning of the violation of Bell-type inequalities evaporate
and that the quantum-theoretical model of the EPRB experiments emerges (not by postulate)
as a very powerful description of the data.

In conclusion, there seem to be two non-compatible alternatives to model
physical phenomena:

\begin{itemize}
\item
In the data-driven approach, the experimental data are considered to be immutable facts.
The first step is to analyze the data without committing to a particular mathematical model for
the process that is imagined to have created the data.
The second step is to construct, not postulate, mathematical models which provide a concise, accurate
picture of how the relevant features of the data change as the conditions
under which the experiments have been performed change.
Being data-driven, the applicability of the mathematical model is measured by
the degree by which the model prediction fits (not produces) the data.
The quantum-theoretical description emerges as a powerful, compact,
separated, interpretation-free model of the data generated by EPRB thought experiments.

Alternatively, one can develop event-based computer models
that are capable of mimicking the processes that give rise to the experimental data.
Essential for the success of this step is that the model accounts for all
processes, including the data processing part, that are critical for the experiment to succeed.
There is no obstacle for constructing probabilistic models and computer models for EPRB laboratory experiments
which produce data to which the quantum-theoretical description of
EPRB thought experiments fits very well.

\item
In the theory-driven approach, it is (often implicitly) assumed that
the production of experimental data is governed by a (possibly yet to be discovered)
pre-existing mathematical model.
Starting from a set of axioms, this approach proves theorems about these mathematical models and
by assumption, also about the phenomena that we experience with our senses.
In the case of the EPRB experiment, the failure of Bell's model
to describe the experimental data is regarded as a proof that one or more of the
assumptions (e.g., locality, etc.) underlying the model do not hold in the universe that we experience with our senses.
This failure has given birth to notions such as ``spooky action on a distance'', ``nonlocality'', etc.,
injecting new elements to the already vast universe of interpretations of quantum theory.
The theory-driven approach leads to endless discussions about
the interpretation of the mathematical symbols used.

\end{itemize}

If one is mainly interested in modeling natural phenomena, the data-driven approach has some outstanding merits.
If one is mainly interested in discussing ``interpretations'', the theory-driven approach is the method of choice.
The choice between these two alternatives depends on one's world view and interests.
Summarizing the above discussion of the two non-compatible alternatives to model
physical phenomena, one should exert extreme caution when switching between these two alternatives
and, in the authors' opinion, failing to do so is what plagues the interpretation of the results of EPRB experiments.

\noindent
\medskip
\section*{Acknowledgments}
We are grateful to Bart De Raedt for critical reading of parts of the manuscript,
suggesting that finding the maximum number of quadruples can
be cast into a linear optimization problem, and for making pertinent comments.
We thank Koen De Raedt for many discussions and continuous support.

The work of M.I.K. was supported by the European Research Council (ERC) under the European Union’s
Horizon 2020  research and innovation programme, grant agreement 854843 FASTCORR.
M.S.J. acknowledges support from the project OpenSuperQ (820363) of the EU Quantum Flagship.
V.M., D.W. and M.W. acknowledge support from the project J\"ulich UNified Infrastructure for Quantum computing (JUNIQ)
that has received funding from the German Federal Ministry of Education and Research (BMBF)
and the Ministry of Culture and Science of the State of North Rhine-Westphalia.

We acknowledge the use of IBM Quantum services for the work presented in Section~\ref{QCE}.
The views expressed in that section are ours and do not reflect the official policy or position of IBM or the IBM Quantum team.

\appendix
\section{Pairs, triples, quadruples, and octuples}\label{PAIRS}

In mathematics, an $n$-tuple is an ordered list of $n$ elements, denoted by $(A,B,\ldots)$.
In this paper we call $(A,B)$ a pair, implicitly assuming it is ordered, a 2-tuple.
Similarly, the lists $(A,B,C)$ and $(A,B,C,D)$ are
referred to as a triple (3-tuple) or quadruple (4-tuple), and an ordered list of 8 elements is an octuple.

\section{Proof of the model-free inequality for correlations of discrete data}\label{DISD}

The inequality derived in this section applies to any experiment that
produces {\bf discrete} data, which without loss of generality, can always
be thought of as being rescaled to lie in the interval $[-1,+1]$.
If the data were real-valued, it is no longer possible
to uniquely identify the quadruples which are essential for the proof of the inequality.
We derive a bound on a certain combination of correlations,
each one computed from data gathered under different conditions.
The $n$th data item obtained under condition $\bx$ is denoted
by $A_{\bx,n}$ for $n=1,\ldots,N$ where $|A_{\bx,n}|\le 1$.
Recall that the subscript $\bx$ labels the condition only and does not, in any way,
implicitly imply a dependence of $A_{\bx,n}$ on $\bx$ in terms of a MM.
As before, the symbols $A$ and $B$ represent discrete data.

Inspiration to derive the model-free inequality stems
from the standard procedure of demonstrating a violation of a Bell-CHSH inequality.
The latter consists of performing EPRB laboratory experiments
with four different pairs of settings $(\ba,\bc)$, $(\ba,\bd)$, $(\bb,\bc)$ and $(\bb,\bd)$.
As mentioned earlier, all EPRB laboratory experiments
use, e.g. time-coincidence, local time windows, voltage thresholds, etc. to identify pairs.
This results in a further, often substantial reduction of the number of pairs.
Furthermore, in practice, the number of observed pairs may depend on the setting.
However, data sets can always be truncated such that the number of pairs of the four sets is the same.
Thus, an EPRB laboratory experiment, with or without some
post-processing of the data,
produces {\bf discrete} data $A_{\Cac,n}$, etc., for $n=1,\ldots,N$ in four independent runs of length $N$.
The procedure of how this data was obtained is irrelevant for the derivation of the inequality presented in this section.

The data sets obtained for
four different conditions denoted by $\Cac$, $\Cad$, $\Cbc$ and $\Cbd$ read
\begin{subequations}
\label{DISD0}
\begin{eqnarray}
{\cal D}_{\Cac}&=&\{(A_{\Cac,n},B_{\Cac,n})\,|\,
|A_{\Cac,n}|\le1,|B_{\Cac,n}|\le1\,;\,n=1,\ldots,N\}
\;,
\label{DISD0a}
\\
{\cal D}_{\Cad}&=&\{(A_{\Cad,n},B_{\Cad,n})\,|\,
|A_{\Cad,n}|\le1,|B_{\Cad,n}|\le1\,;\,n=1,\ldots,N\}
\;,
\label{DISD0b}
\\
{\cal D}_{\Cbc}&=&\{(A_{\Cbc,n},B_{\Cbc,n})\,|\,
|A_{\Cbc,n}|\le1,|B_{\Cbc,n}|\le1\,;\,n=1,\ldots,N\}
\;,
\label{DISD0c}
\\
{\cal D}_{\Cbd}&=&\{(A_{\Cbd,n},B_{\Cbd,n})\,|\,|A_{\Cbd,n}|\le1,|B_{\Cbd,n}|\le1\,;\,n=1,\ldots,N\}
\;,
\label{DISD0d}
\end{eqnarray}
\end{subequations}
where $N$ is the number of pairs.
From the discrete data Eq.~(\ref{DISD0}), we compute the correlations

\begin{eqnarray}
C_{\Cac}=\frac{1}{N}\sum_{n=1}^N A_{\Cac,n}B_{\Cac,n}\;,\;
C_{\Cad}=\frac{1}{N}\sum_{n=1}^N A_{\Cad,n}B_{\Cad,n}\;,\;
C_{\Cbc}=\frac{1}{N}\sum_{n=1}^N A_{\Cbc,n}B_{\Cbc,n}\;,\;
C_{\Cbd}=\frac{1}{N}\sum_{n=1}^N A_{\Cbd,n}B_{\Cbd,n}
\;.
\label{DISD1}
\end{eqnarray}
To simplify the notation somewhat, in this and all other appendices, we use the symbols $C_s$ instead of $E_s^{(12)}$ for $s=1,2,3,4$ to denote correlations.

In general, each contribution to the correlations Eq.~(\ref{DISD1})
may take any value in the interval $[-1,+1]$, independent of the values taken by other contributions,
yielding the trivial bound
\begin{eqnarray}
|C_{\Cac}\mp C_{\Cad}|+|C_{\Cbc}\pm C_{\Cbd}|&\le&4
\;.
\label{DISD2}
\end{eqnarray}

Without introducing a specific model for the process that generates the data,
we can derive a bound that is sharper than Eq.~(\ref{DISD2}) by making use of Eq.~(\ref{BASIC3}).
To this end, we first identify the contributions to $C_{\Cac}$, $C_{\Cad}$, $C_{\Cbc}$ and $C_{\Cbd}$
which, after suitable reshuffling of the terms, can be brought in the form $xz\mp xw+yz\pm yw$
to which Eq.~(\ref{BASIC3}) applies.

Unfortunately, writing down the idea of suitable reshuffling in mathematical notation requires a cumbersome notation,
possibly giving the wrong impression that the proof that follows is complicated.
The reader who is not interested in the technicalities of the proof should nevertheless
read the next paragraph to understand what is meant by ``quadruples'' and can then jump to
the final result Eq.~(\ref{DISD7}).

We introduce permutations $P(.)$, ${\widehat P}(.)$, ${\widetilde P}(.)$, and ${P'}(.)$
of the first $N$ integers and rewrite Eq.~(\ref{DISD1}) as
\begin{eqnarray}
C_{\Cac}=\frac{1}{N}\sum_{n=1}^N A_{\Cac,{P}(n)}B_{\Cac,{P}(n)}\;,\;
C_{\Cad}=\frac{1}{N}\sum_{n=1}^N A_{\Cad,{\widehat P}(n)}B_{\Cad,{\widehat P}(n)}\;,\;
C_{\Cbc}=\frac{1}{N}\sum_{n=1}^N A_{\Cbc,{\widetilde P}(n)}B_{\Cbc,{\widetilde P}(n)}\;,\;
C_{\Cbd}=\frac{1}{N}\sum_{n=1}^N A_{\Cbd,P'(n)}B_{\Cbd,P'(n)}
\;.
\label{QUAD0}
\end{eqnarray}
Obviously, reordering the terms of the sums does not change the value
of the sums themselves.
Suppose that we can find permutations $P(.)$, ${\widehat P}(.)$, ${\widetilde P}(.)$, and $P'(.)$ such that
\begin{eqnarray}
x=A_{\Cac,P(1)}=A_{\Cad,{\widehat P}(1)}
\;,\;
y=A_{\Cbc,{\widetilde P}(1)}=A_{\Cbd,P'(1)}
\;,
\;
z=B_{\Cac,P(1)}=B_{\Cbc,{\widetilde P}(1)}
\;,\;
w=B_{\Cad,{\widehat P}(1)}=B_{\Cbd,P'(1)}
\;,
\label{QUAD1}
\end{eqnarray}
showing that the variables of the octuple
$(A_{\Cac,P(1)},A_{\Cad,{\widehat P}(1)},
A_{\Cbc,{\widetilde P}(1)},A_{\Cbd,P'(1)},
B_{\Cac,P(1)},B_{\Cbc,{\widetilde P}(1)},
B_{\Cad,{\widehat P}(1)},B_{\Cbd,P'(1)})$
form the quadruple $(x,y,z,w)$, defined by Eq.~(\ref{QUAD1}).
In other words, if we can find $P(.)$, ${\widehat P}(.)$, ${\widetilde P}(.)$, and $P'(.)$
such that Eq.~(\ref{QUAD1}) holds,
the original data in terms of octuples exhibits structure that allows
at least that one octuple to be reduced to a quadruple.

From Eq.~(\ref{QUAD1}), it is clear that by definition, the permutations always interchange pairs of
data $(A_{s,n},B_{s,n})$ within a particular data set $s=1,2,3,4$, that is the mapping
is of the kind $(A_{s,n},B_{s,n})\rightarrow (A_{s,n'},B_{s,n'})$,
replacing $n$ by $n'$ for both the $A$ and $B$ simultaneously, for the same value of $s$.
There is no reshuffling of the items within pairs, as would be the case in for instance
$(A_{s,n},B_{s,n})\rightarrow (A_{s,n'},B_{s,n''})$ with $n'\not=n''$.
This is important because if the data pairs have been selected through a time-coincidence (see Fig.~\ref{weihsexp})
or any other procedure,
application of these permutations does not affect the pairing of events within one particular data set.
In other words, the permutations would never mix up the time tags of data pairs.

Using the triangle inequality and Eq.~(\ref{BASIC3b}), we have
\begin{align}
\vert C_{\Cac}\mp C_{\Cad}\vert+\vert C_{\Cbc}\pm C_{\Cbd}\vert&\le
\left\vert
\frac{1}{N}\sum_{\substack{n=2}}^N \left(A_{\Cac,{P}(n)}B_{\Cac,{P}(n)}
\mp A_{\Cad,{\widehat P}(n)}B_{\Cad,{\widehat P}(n)}
\right)
\right\vert
+\left\vert
\frac{1}{N}\sum_{\substack{n=2}}^N \left(A_{\Cbc,{\widetilde P}(n)}B_{\Cbc,{\widetilde P}(n)}
\pm A_{\Cbd,P'(n)}B_{\Cbd,P'(n)}
\right)
\right\vert
\nonumber \\
&+\frac{1}{N}\vert xz \mp  xw\vert + \vert yz \pm yw\vert
\;,
\nonumber \\
&\le
\frac{2}{N}+
\left\vert
\frac{1}{N}\sum_{\substack{n=2}}^N \left(A_{\Cac,{P}(n)}B_{\Cac,{P}(n)}
\mp A_{\Cad,{\widehat P}(n)}B_{\Cad,{\widehat P}(n)}
\right)
\right\vert
+\left\vert
\frac{1}{N}\sum_{\substack{n=2}}^N \left(A_{\Cbc,{\widetilde P}(n)}B_{\Cbc,{\widetilde P}(n)}
\pm A_{\Cbd,P'(n)}B_{\Cbd,P'(n)}
\right)
\right\vert
\;.
\label{QUAD2}
\end{align}
From Eq.~(\ref{QUAD2}) the importance of identifying quadruples is clear. For every quadruple
which we can create by reshuffling data pairs, the contribution to the expression on the left hand side of
Eq.~(\ref{QUAD2}) is limited in magnitude by two, not by four.

Now assume that we can find
permutations $Q(.)$, ${\widehat Q}(.)$, ${\widetilde Q}(.)$, and $Q'(.)$ such that for $k=1,\ldots,K_{\mathrm{max}}$,
\begin{equation}
x_k=A_{\Cac,Q(k)}=A_{\Cad,{\widehat Q}(k)}
\;,\;
y_k=A_{\Cbc,{\widetilde Q}(k)}=A_{\Cbd,Q'(k)}
\;,
\;
z_k=B_{\Cac,Q(k)}=B_{\Cbc,{\widetilde Q}(k)}
\;,\;
w_k=B_{\Cad,{\widehat Q}(k)}=B_{\Cbd,Q'(k)}
\;,
\label{QUAD3}
\end{equation}
where $K_{\mathrm{max}}\le N$ denotes the largest integer for
which we can find these four permutations, that is
$K_{\mathrm{max}}$ is the maximum number of pairs in each data set that form quadruples.
If it is not possible to find any such quadruple, we have $K_{\mathrm{max}}=0$ by definition.

The choice represented by Eq.~(\ref{QUAD3}) is motivated by the EPRB experiment, see Fig.~\ref{eprbidea}.
In general, other choices to define quadruples
are possible and may yield different values of
the maximum fraction of quadruples.
However, if
$\max(|C_{\Cac}- C_{\Cad}|+|C_{\Cbc}+ C_{\Cbd}|,
|C_{\Cac}+ C_{\Cad}|+|C_{\Cbc}- C_{\Cbd})=4-2\QUAD$ (possibly up to some statistical fluctuations),
as is the case in some of the numerical examples discussed below, we may be confident that the choice
Eq.~(\ref{QUAD3}) yields the largest value of $\QUAD$ that can be obtained by reshuffling of the data.

Repeating the steps the yielded Eq.~(\ref{QUAD2}) $K_{\mathrm{max}}$ times, we find
\begin{align}
\vert C_{\Cac}\mp C_{\Cad}\vert +\vert C_{\Cbc}\pm C_{\Cbd}\vert&\le
\frac{2K_{\mathrm{max}}}{N}+
\left\vert\frac{1}{N}
\sum_{n=K_{\mathrm{max}}+1}^N
\left(A_{\Cac,{Q}(n)}B_{\Cac,{Q}(n)}
\mp A_{\Cad,{\widehat Q}(n)}B_{\Cad,{\widehat Q}(n)}
\right)
\right\vert
\nonumber \\
&\hbox to 3 cm{}+
\left\vert\frac{1}{N}
\sum_{n=K_{\mathrm{max}}+1}^N
\left(A_{\Cbc,{\widetilde Q}(n)}B_{\Cbc,{\widetilde Q}(n)}\pm A_{\Cbd,Q'(n)}B_{\Cbd,Q'(n)}
\right)
\right\vert
\nonumber \\
&\le
\frac{2K_{\mathrm{max}}}{N}+
\frac{1}{N}
\sum_{n=K_{\mathrm{max}}+1}^N
\left(
\left\vert A_{\Cac,{Q}(n)}B_{\Cac,{Q}(n)}\right\vert
+\left\vert A_{\Cad,{\widehat Q}(n)}B_{\Cad,{\widehat Q}(n)}\right\vert
+\left\vert A_{\Cbc,{\widetilde Q}(n)}B_{\Cbc,{\widetilde Q}(n)}\right\vert
+ \left\vert A_{\Cbd,Q'(n)}B_{\Cbd,Q'(n)}\right\vert
\right)
\nonumber \\
&\le
\frac{2K_{\mathrm{max}}}{N}+\frac{4(N-K_{\mathrm{max}})}{N}= 4 - 2\QUAD
\;,
\label{QUAD4}
\end{align}
where
$0\le\QUAD = K_{\mathrm{max}}/{N}\le1$
is the ratio of the maximum number of quadruples $K_{\mathrm{max}}$ to the number of pairs $N$,
a measure for the ``hidden'' structure in the collection of octuples.

In summary, we have proven that independent of the origin of the four sets of discrete data
Eq.~(\ref{DISD0}), the correlations computed from these data sets must satisfy the ``model-free'' inequality
\begin{center}
\framebox{
\parbox[t]{0.8\hsize}{%
\begin{eqnarray}
\left|C_{\Cac}- C_{\Cad}\right| +\left|C_{\Cbc} + C_{\Cbd}\right|
\le 4-2\QUAD
\quad,\quad
\left|C_{\Cac}+ C_{\Cad}\right| +\left|C_{\Cbc} - C_{\Cbd}\right|
\le 4-2\QUAD
\;,
\label{DISD7}
\label{FUNDA}
\end{eqnarray}
}}
\end{center}
where $\QUAD = K_{\mathrm{max}}/{N}$
is the fraction of maximum number of quadruples one can find by reshuffling the
original data set of octuples.
Appealing to the triangle inequality once more, it follows from Eq.~(\ref{DISD7}) that
\begin{center}
\framebox{
\parbox[t]{0.6\hsize}{%
\begin{eqnarray}
S_{\mathrm{CHSH}}=\max_{(i,j,k,l)\in\bm\pi_4}
\left|C_{i}-C_{j}+C_{k}+C_{l}\right|
&\le& 4-2\QUAD
\;,
\label{CHSHz}
\end{eqnarray}
}}
\end{center}
where $\bm\pi_n$ denotes the set of all permutations of $(1,\ldots,n)$
and $S_{\mathrm{CHSH}}$ denotes the Bell-CHSH function.

The same procedure can be used to derive inequalities for the correlations computed from three instead of four data sets.
Alternatively, following Bell~\cite{BELL93}, we can obtain these inequalities by replacing $C_\Cbd$ in Eq.~(\ref{FUNDA}) by one and we have
\begin{eqnarray}
\left|C_{1}+ C_{2}\right|
\le 3-2\QUAD+ C_{3}
\quad,\quad
\left|C_{1}- C_{2}\right|
\le 3-2\QUAD- C_{3}
\;.
\label{BELLz}
\end{eqnarray}
From Eq.~(\ref{TRIPLE}) it then follows that
$\left|C_{i}\pm C_{j}\right|\le 3-2\QUAD\pm C_{k}$ for all $(i,j,k)\in\bm\pi_3$, are the appropriate, model-free ``Bell inequalities'' for discrete data.

Inequalities Eqs.~(\ref{DISD7}--(\ref{BELLz}) cannot be violated by data of a (real or thought) EPRB experiment,
unless the mathematical apparatus that we use is inconsistent (a possibility which we do not consider).
For example, assume that an EPRB laboratory experiment yields data
for which $S_{\mathrm{CHSH}}>2$,
that is the data violate the Bell-CHSH inequality.
Then we can use Eq.~(\ref{DISD7}) to find that $\QUAD \le 2-S_{\mathrm{CHSH}}/2$, implying
that the number of quadruples in the sets Eq.~(\ref{DISD0}) does not exceed $(2-S_{\mathrm{CHSH}}/2)N$.
On the basis of the experimental data only, this is all one can say with certainty.

It should be noted that we have not proven that all the octuples in the sets Eq.~(\ref{DATA0})
can be reshuffled to form quadruples if the correlations satisfy $\left|C_{\Cac}\mp C_{\Cad}|+|C_{\Cbc}\pm C_{\Cbd}\right|<2$.
In fact, it is easy to construct simple counterexamples.
For instance, if
${\cal D}_{\Cac}=\{(+1,-1),(+1,+1)\}$,
${\cal D}_{\Cad}=\{(-1,-1),(-1,+1)\}$,
${\cal D}_{\Cbc}=\{(-1,-1),(+1,-1)\}$, and
${\cal D}_{\Cbd}=\{(+1,+1),(-1,-1)\}$,
then $C_{\Cac}=C_{\Cad}=C_{\Cbc}=0$ and $C_{\Cbd}=1$, yet it is impossible
to reshuffle the data such that they can be extracted from two quadruples.
If we assume that the averages of $A_1$ and $A_2$, $A_3$ and $A_4$,
$B_1$ and $B_3$, $B_2$ and $B_4$ are the same and
$\left|C_{\Cac}\mp C_{\Cad}|+|C_{\Cbc}\pm C_{\Cbd}\right|<2$, our numerical experiments suggest that almost all of them can be reshuffled to create quadruples (see examples below). This observation in the realm of data can be understood in terms of Fine's theorem~\cite{FINE82}, see also Section~\ref{PROB}.

We emphasize that inequalities Eqs.~(\ref{DISD7}--(\ref{BELLz}) do not depend on how the data was generated and/or processed and
hold in general, independent of (any model for) the process that generates the data.

\subsection{The Eberhard inequality for discrete data}\label{CHEB}

The EPRB experiments reported in Refs. [34, 35] have only one detector in each of the two stations (see Fig. 2). To account for the photons that have been detected and would have been detected by the missing detectors and also to account
for undetected photons, the events are classified in two groups.
Adopting the notation used in Refs. [34, 35], the events that are
recorded by the detector and all other events are given the label ``+'' and ``0'', respectively. If desired, the averages and the correlations can
be calculated as usual (see Eq.~(\ref{CORR})) by assigning the values $x,y=+1$
to the former and $x,y=-1$ to the latter class of events. Of course, the correlation obtained from these experiments may be different from those used to compute the correlations appearing in inequalities Eqs.~(\ref{DISD7})--(\ref{BELLz}).

In the EPRB experiments reported in Refs. [34, 35], the number of interest is the combination of counts~\cite{EBER93}
\begin{eqnarray}
\mathrm{EBER}_{\mathrm{data}}&=&N_{\Cac}^{++}-N_{\Cad}^{+0}-N_{\Cbc}^{0+}-N_{\Cbd}^{++}
\;.
\label{CHEB0}
\end{eqnarray}
The rationale for considering $\mathrm{EBER}_{\mathrm{data}}$ is that if the data is generated in the form of quadruples or for instance, by a CM of a Bell-type (counterfactual definite) model, we have $\mathrm{EBER}_{\mathrm{data}}\le0$.

We derive the appropriate, model-free upper bound to $\mathrm{EBER}_{\mathrm{data}}$ by proceeding in a manner analogous to the one used to derive inequalities Eqs.~(\ref{DISD7})--(\ref{BELLz}).
First, we introduce two-valued, integer variables $X_{s,n}$ and $Y_{s,n}$
to represent the detection of an event at stations 1 and 2, respectively.
As before, the subscript $s=1,2,3,4$ refers to the data sets obtained
under the conditions $(\ba,\bc)$, $(\ba,\bd)$, $(\bb,\bc)$, and $(\bb,\bd)$,
respectively.
If station 1(2) reports a ``+'' event for the $n$-th pair, we set $X_{s,n}=1$ ($Y_{s,n}=1$). Otherwise, we set $X_{s,n}=0$ ($Y_{s,n}=0$).
In terms of these variables we have
\begin{eqnarray}
\mathrm{EBER}_{\mathrm{data}}&=&
\sum_{n=1}^N \left[
X_{1,n}Y_{1,n} -X_{2,n}(1-Y_{2,n})-(1-X_{3,n})Y_{3,n}-X_{4,n}Y_{4,n}
\right]
\;.
\label{CHEB1}
\end{eqnarray}

If $X_{1,n}=X_{2,n}$, $X_{3,n}=X_{4,n}$, $Y_{1,n}=Y_{3,n}$ and $Y_{2,n}=Y_{4,n}$,
as it would be if these data were generated in the form of a quadruple or by a CM of an LHVM, simply enumerating all sixteen possibilities shows that
$-1\le X_{1,n}Y_{1,n} -X_{2,n}(1-Y_{2,n})-(1-X_{3,n})Y_{3,n}-X_{4,n}Y_{4,n}\le0$.
In general, we have $-3\le X_{1,i}Y_{1,i'} -X_{2,j}(1-Y_{2,j'})-(1-X_{3,k})Y_{3,k'}-X_{4,l}Y_{4,l'}\le1$.

Next, as before, we assume that there exist
permutations $Q(.)$, ${\widehat Q}(.)$, ${\widetilde Q}(.)$, and $Q'(.)$
such that for all $k=1,\ldots,K_{\mathrm{max}}$,
\begin{eqnarray}
X_{\Cac,Q(k)}=X_{\Cad,{\widehat Q}(k)}
\;,\;
X_{\Cbc,{\widetilde Q}(k)}=X_{\Cbd,Q'(k)}
\;,
\;
Y_{\Cac,Q(k)}=Y_{\Cbc,{\widetilde Q}(k)}
\;,\;
Y_{\Cad,{\widehat Q}(k)}=Y_{\Cbd,Q'(k)}
\;.
\label{CHEB2}
\end{eqnarray}
If we can find these four permutations, we have identified $K_{\mathrm{max}}$ pairs in each data set that can be represented by $K_{\mathrm{max}}$ quadruples.
If it is not possible to find any such quadruple, we have $K_{\mathrm{max}}=0$ by definition.
Note that the permutations $Q(.)$, ${\widehat Q}(.)$, ${\widetilde Q}(.)$, and $Q'(.)$
and the value of $K_{\mathrm{max}}$ are not necessarily the same as in case where
the involved data has been obtained by analyzing the data of EPRB experiments with two detectors per station.

Writing Eq.~(\ref{CHEB1}) as
\begin{eqnarray}
\mathrm{EBER}_{\mathrm{data}}&=&
\sum_{k=1}^{K_{\mathrm{max}}} \left[
X_{1,Q(k)}Y_{1,Q(k)} -X_{2,{\widehat Q}}(1-Y_{2,{\widehat Q}}(k))-(1-X_{3,{\widetilde Q}(k)})Y_{3,{\widehat Q}(k)}-X_{4,Q'(k)}Y_{4,Q'(k)}
\right]
\nonumber\\
&&+\sum_{k=K_{\mathrm{max}}+1}^{N} \left[
X_{1,Q(k)}Y_{1,Q(k)} -X_{2,{\widehat Q}}(1-Y_{2,{\widehat Q}}(k))-(1-X_{3,{\widetilde Q}(k)})Y_{3,{\widehat Q}(k)}-X_{4,Q'(k)}Y_{4,Q'(k)}
\right]
\;,
\label{CHEB3}
\end{eqnarray}
and using the respective lower and upper bounds for each of the terms in the sums we
obtain
\begin{eqnarray}
-3(N-K_{\mathrm{max}})-K_{\mathrm{max}}\le \mathrm{EBER}_{\mathrm{data}}\le
N-K_{\mathrm{max}}
\;,
\label{CHEB4}
\end{eqnarray}
or, expressed in terms of the fraction of quadruples,
\begin{eqnarray}
-3+2\QUAD\le \frac{\mathrm{EBER}_{\mathrm{data}}}{N}\le
1-\QUAD
\;,
\label{CHEB5}
\end{eqnarray}
where the value of $\QUAD$ is not necessarily the same as the value of
$\QUAD$ obtained by analyzing the data of EPRB experiments with two detectors per station.
This is because the data of the $X$'s and $Y$'s
obtained by performing EPRB experiments with one detector per station
are not the same as the data of the $A$'s and $B$'s obtained by performing EPRB experiments with two detectors per station.
If {\bf all} the data pairs that contribute to $N_{\Cac}^{++}$, $N_{\Cad}^{+0}$, $N_{\Cbc}^{0+}$ and $N_{\Cbd}^{++}$ can be reshuffled to form quadruples, we have $\QUAD=1$ and
Eq.~(\ref{CHEB5}) becomes the CH inequality $\mathrm{EBER}_{\mathrm{data}}/N \le0$~\cite{CLAU74} for discrete data.

As an illustration, we take the data reported in the Supplemental material of Ref.~\cite{GIUS15}.
The valid trials for the four settings are
$N_{\Cac} = 875683790$,
$N_{\Cad} = 875518074$,
$N_{\Cbc} = 875882007$, and
$N_{\Cbd} = 875700279$
such that the total number of counts
$N_{\mathrm{tot}}=N_{\Cac}+N_{\Cad}+N_{\Cbc}+N_{\Cbd}=3 502 784 150$~\cite{GIUS15}.
After post-processing by adjusting voltage thresholds, the corresponding photon counts are
$N_{\Cac}^{++}=141 439$, $N_{\Cad}^{+0}=67 941$, $N_{\Cbc}^{0+}= 58 742$ and $N_{\Cbd}^{++}= 8 392$~\cite{GIUS15}.

To estimate $\QUAD$ from the data provided in Ref.~\cite{GIUS15}, we have to be able to truncate three of the four data sets such that they have the same number of pairs $N$.
In principle, this requires processing the four full sequences of individual events.
Fortunately, in view of the large number of events, this is not necessary if we proceed as follows.
We define
$N=(N_{\Cac}+N_{\Cad}+N_{\Cbc}+N_{\Cbd})/4$
and compute $\widehat{N}_{\Cac}^{++}=\nint{N N_{\Cac}^{++}/N_{1}}=141 441
=N_{\Cac}^{++}+2$ where $\nint{x}$ denotes the nearest integer to $x$.
Similarly, we obtain
$\widehat{N}_{\Cad}^{+0}=\nint{NN_{\Cad}^{+0}/N_2}=67955=N_{\Cad}^{+0}+14$,
$\widehat{N}_{\Cbc}^{0+}=\nint{NN_{\Cbc}^{0+}/N_3}=58730=N_{\Cbc}^{0+}-12$, and
$\widehat{N}_{\Cbd}^{++}=\nint{NN_{\Cbd}^{++}/N_4}=8392=N_{\Cbd}^{++}$.
Clearly, the errors made by using the procedure of estimates $N$,
$\widehat{N}_{\Cac}^{++}$ etc. is negligible. Therefore, to a very good approximation, we have
\begin{eqnarray}
\frac{\mathrm{EBER}_{\mathrm{data}}}{N}&\approx&
\frac{\widehat{N}_{\Cac}^{++}-\widehat{N}_{\Cad}^{+0}-\widehat{N}_{\Cbc}^{0+}-\widehat{N}_{\Cbd}^{++}}{N}=7.27\times10^{-6}
\;,
\label{CHEB6}
\end{eqnarray}
the same as the value of $J$ reported in Ref.~\cite{GIUS15}.
From Eq.~(\ref{CHEB5}) it then follows that $\QUAD\lesssim0.99999273$.

In Ref.~\cite{GIUS15}, the tiny number of $7.27\times10^{-6}$, an order of magnitude smaller than the expected statistical
error of $1/\sqrt{N}\approx 3\times 10^{-5}$, is taken as strong evidence that the data cannot
be described by an LHVM of the Bell-type~\cite{GIUS15,SHAL15}.
Moreover, the data was obtained by adjusting voltage thresholds~\cite{GIUS15,SHAL15},
a process that is not accounted for in the LHVM that is being rejected but is essential to create data such that
$\widehat{N}_{\Cac}^{++}-\widehat{N}_{\Cad}^{+0}-\widehat{N}_{\Cbc}^{0+}-\widehat{N}_{\Cbd}^{++}>0$.
Clearly, for the reasons explained in Section~\ref{INEQDATA}, the logic that leads to this conclusion needs to be revised.

On the basis of the experimental data,
the correct conclusion one can draw from $\QUAD\lesssim0.99999273$ is that a very small fraction of all the
selected pairs of photon events cannot be rearranged to create quadruples.

\subsection{The Clauser-Horn inequality for discrete data}

Given the same data sets, bounds on $\mathrm{CH}_{\mathrm{data}}$, also expressing structure in the data,
can be derived as follows. Consider the expression (or similar expressions with permutations of the subscripts (1,2,3,4))
\begin{align}
\mathrm{CH}_{\mathrm{data}}(x,y)=&f_1(x,y)-f_2(x,y)+f_3(x,y)+f_4(x,y)
\nonumber \\
&- \frac{1}{2}\sum_{z=-1,1}\left(
f_1(x,z)-f_2(x,z)
+f_3(x,z)+f_4(x,z)
+f_1(z,y)-f_2(z,y)
+f_3(z,y)+f_4(z,y)
\right)
\;,
\label{CHE0}
\end{align}
where $x,y=\pm1$ and the frequencies $f_s(x,y)$ for $s=1,2,3,4$ are given by Eq.~(\ref{DATA1}).
Recall that $x,y=1(-1)$ correspond to events of type ``+''(``0'').
Expressing these frequencies in terms of their moments, that is, using Eq.~(\ref{DATA2d}) yields
\begin{eqnarray}
\mathrm{CH}_{\mathrm{data}}(x,y)&=&
-\frac{1}{2}+\frac{xy}{4}\left(E^{(12)}_1-E^{(12)}_2+E^{(12)}_3+E^{(12)}_4\right)
\;.
\label{CHE1}
\end{eqnarray}
Adopting the notation for the correlations adopted in this, we have
\begin{eqnarray}
\mathrm{CH}_{\mathrm{data}}(x,y)&=&-\frac{1}{2}+\frac{xy}{4}\left(C_1-C_2+C_3+C_4\right)
\;,
\label{CHE2}
\end{eqnarray}
and using Eq.~(\ref{CHSHz}), we obtain
\begin{eqnarray}
-1-\frac{1-\Delta}{2}\le \mathrm{CH}_{\mathrm{data}}(x,y) \le \frac{1-\Delta}{2}
\;.
\label{CHE3}
\end{eqnarray}

The relation between $\mathrm{CH}_{\mathrm{data}}(x,y)$ and the CH
inequality becomes clear if we assume that
$E^{(1)}_1=E^{(1)}_2$,
$E^{(2)}_1=E^{(2)}_3$,
$E^{(1)}_3=E^{(1)}_4$, and
$E^{(2)}_2=E^{(2)}_4$.
This assumption complies with the idea of a description in terms of a joint distribution for EPRB experiments with four different pairs of settings, see Section~\ref{PROB}, or in terms of Bell's model Eq.~(\ref{IN0}).
With this assumption, Eq.~(\ref{CHE3}) reduces to
\begin{eqnarray}
\mathrm{CH}_{\mathrm{data}}(x,y) &=&
f_1(x,y)-f_2(x,y)+f_3(x,y)+f_4(x,y)-\sum_{z=-1,1}\left(
f_3(x,z)
+f_3(z,y)
\right)
\;.
\label{CHE4}
\end{eqnarray}
Assuming that all the data originates from a set of quadruples we have $\Delta=1$,
identifying $(1)=(\ba,\bc)$, $(2)=(\ba,\bd)$, $(3)=(\bb,\bc)$, and $(4)=(\bb,\bd)$ as before,
and adopting the notation used by CH, Eq.~(\ref{CHE3}) becomes
\begin{eqnarray}
-1\le P_{xy}(\ba,\bc)-P_{xy}(\ba,\bd)+P_{xy}(\bb,\bc)+P_{xy}(\bb,\bd)-P_x(\bb)-P_y(\bc) \le 0
\;,
\label{CHE5}
\end{eqnarray}
which, for every pair $(x,y)$, is the CH inequality in its usual form~\cite{CLAU74}.
Note that the assumption made to obtain Eq.~(\ref{CHE5})
implies that $E^{(1)}_1$, $E^{(1)}_3$, $E^{(2)}_1$, and $E^{(2)}_2$ only depend on
$\ba$, $\bb$, $\bc$, and $\bd$, respectively, justifying writing the marginals with respect to the variables
$x$ and $y$ as $P_x(\bb)$ and $P_y(\bc)$, respectively.

Using $P_{x=+}(\bb)=\sum_{u=+,-}P_{+u}(\bb,\bd)$ and
$P_{y=+}(\bc)=\sum_{u=+,-}P_{u+}(\bb,\bc)$,
Eq.~(\ref{CHE5}) becomes
\begin{eqnarray}
-1\le P_{++}(\ba,\bc)-P_{++}(\ba,\bd)-P_{-+}(\bb,\bc)-P_{+-}(\bb,\bd)\le 0
\;,
\label{CHE6}
\end{eqnarray}
a violation of the right-hand side of Eq.~(\ref{CHE6}) by {\bf data} being interpreted
as the break-down of local realism~\cite{GIUS15,SHAL15}.

In conclusion, as in the case of the Bell-CHSH inequality, also the CH inequalities Eqs.~(\ref{CHE5}) and~(\ref{CHE6})
are of little use to draw conclusions from the analysis of discrete data originating from (numerical) experiments, because for such data, the appropriate inequality is Eq.~(\ref{CHEB5}) with
$\mathrm{CH}_{\mathrm{data}}$ given by Eq.~(\ref{CHEB0}),
not inequality Eq.~\ref{CHE6} which has been derived within the context of a MM which assumes counterfactual definiteness.
Being model-free mathematical facts, inequalities Eqs.~(\ref{CHEB5}) and~\ref{CHE3} cannot be violated.

\subsection{Lower bounds to the fraction of quadruples}
It is easy to derive lower bounds to the fraction of quadruples $\QUAD$
if all the $A$'s and $B$'s that appear in Eq.~(\ref{DISD0}) take values $\pm1$ only.
The most naive method to compute such a lower bound
ignores the possibility of reshuffling the data such that the correlations remain the same and simply counts the  number of times the four conditions
\begin{eqnarray}
A_{\Cac,k}=A_{\Cad,k}
\;,\;
A_{\Cbc,k}=A_{\Cbd,k}
\;,
\;
B_{\Cac,k}=B_{\Cbc,k}
\;,\;
B_{\Cad,k}=B_{\Cbd,k}
\;.
\label{QUAD33}
\end{eqnarray}
are satisfied.
We denote the faction of quadruples thus obtained by $\widehat{\QUAD}$.
Obviously, calculating $\widehat{\QUAD}$ is easy and computationally inexpensive.

With similar computational effort, we can compute another lower bound as follows.
First we note that
\begin{eqnarray}
C_{\Cac}=\frac{1}{N}\left(n_{++}^{(\Cac)}+n_{--}^{(\Cac)}-n_{+-}^{(\Cac)}-n_{-+}^{(\Cac)}\right)
\;,
\label{DISD10}
\end{eqnarray}
where $n_{++}^{(\Cac)}$, $n_{--}^{(\Cac)}$, $n_{+-}^{(\Cac)}$, and $n_{-+}^{(\Cac)}$
are the numbers of times
$A_{\Cac,n}=+1$ and $B_{\Cac,n}=+1$,
$A_{\Cac,n}=-1$ and $B_{\Cac,n}=-1$,
$A_{\Cac,n}=+1$ and $B_{\Cac,n}=-1$, and
$A_{\Cac,n}=-1$ and $B_{\Cac,n}=+1$,
respectively.
Expressions similar to Eq.~(\ref{DISD10}) hold for $C_{\Cad}$, $C_{\Cbc}$, and $C_{\Cbd}$.
The number of quadruples that can be formed with all
$A$'s and $B$'s equal to $+1$ is given by
$N_{++}=\min\left(n_{++}^{(\Cac)},n_{++}^{(\Cad)},n_{++}^{(\Cbc)},n_{++}^{(\Cbd)}\right)$.
Denoting
$N_{--}=\min\left(n_{--}^{(\Cac)},n_{--}^{(\Cad)},n_{--}^{(\Cbc)},n_{--}^{(\Cbd)}\right)$,
$N_{+-}=\min\left(n_{+-}^{(\Cac)},n_{+-}^{(\Cad)},n_{+-}^{(\Cbc)},n_{+-}^{(\Cbd)}\right)$,
and
$N_{-+}=\min\left(n_{-+}^{(\Cac)},n_{-+}^{(\Cad)},n_{-+}^{(\Cbc)},n_{-+}^{(\Cbd)}\right)$,
it follows immediately that the fraction of quadruples that can be formed by reshuffling must be greater than or equal to
\begin{eqnarray}
\widetilde{\QUAD}=\frac{N_{++}+N_{--}+N_{+-}+N_{-+}}{N}
\;,
\label{DISD10a}
\end{eqnarray}
that is $\widetilde{\QUAD}$ is a lower bound to $\QUAD$ ($0\le\widetilde{\QUAD}\le\QUAD\le1$).
As $\widehat{\QUAD}$ has been obtained by imposing the condition                                             
Eq.~(\ref{QUAD33}) and $\widetilde{\QUAD}$ is given by Eq.~(\ref{DISD10a}),                                  
there is no order relation between $\widehat{\QUAD}$ and $\widetilde{\QUAD}$.                                
However, in many cases of interest (large data sets, etc.)                                                   
$\widehat{\QUAD}$ does not exceed $\widetilde{\QUAD}$.                                                       
In summary, we have                                                                                          
\begin{eqnarray}                                                                                             
S_{\mathrm{CHSH}}\le\left\vert C_{\Cac}\mp C_{\Cad}\right\vert+\left\vert C_{\Cbc}\pm C_{\Cbd}\right\vert\le 
4-2{\QUAD}\le4-2\max{(\widehat{\QUAD},\widetilde{\QUAD})}\le4                                                
\;,                                                                                                          
\label{DISD13}                                                                                               
\end{eqnarray}                                                                                               
where the upper bound of four is a mathematical triviality.                                                  

\subsection{Computing the maximum number of quadruples}
The technicalities of the proof of Eq.~(\ref{DISD7}), involving four permutations of $N$ numbers,
are of little use when we actually want to find all quadruples in the data sets Eq.~(\ref{DISD0}).
Indeed, enumerating all $(N!)^4$ possibilities by a computer quickly becomes prohibitive
as $N$ increases (for instance $(10!)^4\approx  173\times10^{24}$).
However, the proof of the model-free inequality Eq.~(\ref{DISD7}) only requires
the existence of a maximum number of quadruples, the actual value of this maximum
being irrelevant for the proof.

Nevertheless, it is instructive to write a computer program that
uses uniform pseudo-random numbers to generate the data sets Eq.~(\ref{DISD1})
and finds the number of quadruples.
By specifying an algorithm that generates the data, we have defined a CM.
At first sight, finding the value of $\QUAD$ itself may require ${\cal O}(N!^4)$ arithmetic operations.
Fortunately, the problem of determining the fraction of quadruples $\QUAD$ can be cast into an integer linear programming problem which, in the most relevant case for which the $A$'s and $B$'s take values $\pm1$ only, seems easy to solve by considering
the associated linear programming problem with real-valued unknowns.
In practice, we solve the latter by standard optimization techniques~\cite{PRES03}
and then check that the solution takes integer values only, which it always seems to do (an observation for which we have no proof). In this case, the solution of the linear programming problem is also the solution of the integer programming problem.
We have implemented the computer program in Mathematica\textsuperscript{\textregistered}.

\begin{table}[!htp]
\caption{
Lists of all possible combinations of the pairs of data which form quadruples, written
in a slightly different notation to emphasize the quadruple structure.
Given the data sets ${\cal D}_{\Cac}$, ${\cal D}_{\Cad}$, ${\cal D}_{\Cbc}$ and ${\cal D}_{\Cbd}$, 
the optimization task is to find the numbers $m_i\ge0$ for $i=0,\ldots,15$ that
maximize the number of quadruples.
}
\centering
\begin{tabular}{@{\extracolsep{1cm} } ccccc}
\noalign{\medskip}
\hline\hline\noalign{\smallskip}
& $(B_1,A_1)$  &$(A_2,B_2)$& $(B_4,A_4)$  & $(A_3,B_3)$ \\
\hline\noalign{\smallskip}
$m_0$&  $(+1,+1)$ & $(+1,+1)$ &$(+1,+1)$ & $(+1,+1)$\\
$m_1$&  $(-1,+1)$ & $(+1,+1)$ &$(+1,+1)$ & $(+1,-1)$\\
$m_2$&  $(+1,+1)$ & $(+1,+1)$ &$(+1,-1)$ & $(-1,+1)$\\
$m_3$&  $(-1,+1)$ & $(+1,+1)$ &$(+1,-1)$ & $(-1,-1)$\\
$m_4$&  $(+1,+1)$ & $(+1,-1)$ &$(-1,+1)$ & $(+1,+1)$\\
$m_5$&  $(-1,+1)$ & $(+1,-1)$ &$(-1,+1)$ & $(+1,-1)$\\
$m_6$&  $(+1,+1)$ & $(+1,-1)$ &$(-1,-1)$ & $(-1,+1)$\\
$m_7$&  $(-1,+1)$ & $(+1,-1)$ &$(-1,-1)$ & $(-1,-1)$\\
$m_8$&  $(+1,-1)$ & $(-1,+1)$ &$(+1,+1)$ & $(+1,+1)$\\
$m_9$&  $(-1,-1)$ & $(-1,+1)$ &$(+1,+1)$ & $(+1,+1)$\\
$m_{10}$& $(+1,-1)$ & $(-1,+1)$ &$(+1,-1)$ & $(-1,+1)$\\
$m_{11}$& $(-1,-1)$ & $(-1,+1)$ &$(+1,-1)$ & $(-1,-1)$\\
$m_{12}$& $(+1,-1)$ & $(-1,-1)$ &$(-1,+1)$ & $(+1,+1)$\\
$m_{13}$& $(-1,-1)$ & $(-1,-1)$ &$(-1,+1)$ & $(+1,-1)$\\
$m_{14}$& $(+1,-1)$ & $(-1,-1)$ &$(-1,-1)$ & $(-1,+1)$\\
$m_{15}$& $(-1,-1)$ & $(-1,-1)$ &$(-1,-1)$ & $(-1,-1)$\\
\hline\noalign{\smallskip}
\end{tabular}
\label{tab4}
\end{table}

The key step is to list all possible sixteen combinations of $A$'s and $B$'s that
form quadruples and to attach a variable to each of these combinations, as shown in Table~\ref{tab4}.
In terms of the $n_i$'s, the numbers of $(A,B)$ pairs
in each data set is given by the expressions listed in Table~\ref{tab5}.

\begin{table}[!htp]
\caption{
Total counts of different pairs $(A,B)$ belonging to the set of quadruples.
}
\centering
\begin{tabular}{@{\extracolsep{.2cm} } ccccc}
\noalign{\medskip}
\hline\hline\noalign{\smallskip}
$(A,B)$&$n_1(A_1,B_1)$&$n_2(A_2,B_2)$&$n_3(A_3,B_3)$&$n_4(A_4,B_4)$\\
\hline\noalign{\smallskip}
$(+1,+1)$&$m_0+m_2+m_4+m_6$          & $m_0+m_1+m_2+m_3$             & $m_0+m_4+m_8+m_{12}$    & $m_0+m_1+m_8+m_9$      \\
$(+1,-1)$&$m_1+m_3+m_5+m_7$          & $m_4+m_5+m_{6}+m_{7}$         & $m_1+m_5+m_{9}+m_{13}$  & $m_4+m_5+m_{12}+m_{13}$\\
$(-1,+1)$&$m_8+m_{10}+m_{12}+m_{14}$ & $m_8+m_9+m_{10}+m_{11}$       & $m_2+m_6+m_{10}+m_{14}$ & $m_2+m_3+m_{10}+m_{11}$\\
$(-1,-1)$&$m_9+m_{11}+m_{13}+m_{15}$ & $m_{12}+m_{13}+m_{14}+m_{15}$ & $m_3+m_7+m_{11}+m_{15}$ & $m_6+m_7+m_{14}+m_{15}$\\
\hline\noalign{\smallskip}
\end{tabular}
\label{tab5}
\end{table}

Next, we simply count the number of times a pair $(\pm1,\pm1)$ occurs in the data sets ${\cal D}_{\Cac}$, ${\cal D}_{\Cad}$, ${\cal D}_{\Cbc}$ and ${\cal D}_{\Cbd}$,
and denote the sixteen numbers thus obtained by $N_1(+1,+1),N_1(+1,-1),\ldots, N_4(+1,-1),N_4(-1,-1)$.
These numbers completely determine the values of the correlations
$C_{\Cac}$, $C_{\Cad}$, $C_{\Cbc}$, and $C_{\Cbd}$,
e.g., $C_{\Cac}=(N_1(+1,+1)-N_1(+1,-1)-N_1(-1,+1)+N_1(-1,-1))/N$.
The same sixteen numbers serve as input to the linear optimization problem.
Denoting the number of pairs $(A,B)$ in data set ${\cal D}_k$ that do
not belong to the set of quadruples by $u_k(A,B)\ge0$,
we must have
\begin{eqnarray}
N_k(A,B)=n_k(A,B)+u_k(A,B)
\;,\;
N=\sum_{x,y=\pm1} N_k(x,y)
\;,\;
U=\sum_{x,y=\pm1} u_k(x,y)
\;,
\label{QUA0}
\end{eqnarray}
for $k=1,2,3,4$ and all pairs $(A,B)=(\pm1,\pm1)$.
Note that $U$ cannot depend on $k$ because the number of pairs
which do not belong to the set of quadruples must be the same for all four data sets.

The final step is then to minimize the number of pairs which do not belong
to the set of quadruples, that is we solve the linear minimization problem
\begin{equation}
  \min\left(U=N-\sum_{i=0}^{15} m_i\right)
\end{equation}
in 32 unknowns (the $m_i$'s and $u_k(A,B)$'s),
subject to 33 inequality constraints ($m_i,u_k(\pm1,\pm1),U\ge0$ for all $i,k$) and 16 equality constraints (see Eq.~(\ref{QUA0})). For all cases that we have studied,
the linear programming solver yields integer-valued solutions only.

The results of several numerical experiments using $N=1000 000$ pairs per data set can be summarized as follows:
\begin{itemize}
\item
If the $A$'s and $B$'s are generated in the form of quadruples all taking random values $\pm1$,
the program returns $\QUAD=1$,  $|C_{\Cac}-C_{\Cad}|+|C_{\Cbc}+C_{\Cbd}|=0.00063$, and
$|C_{\Cac}+C_{\Cad}|+|C_{\Cbc}-C_{\Cbd}|=0.0028$
such that Eq.~(\ref{DISD7}) is satisfied.
\item
If all $A$'s and $B$'s take independent random values $\pm1$,
the $C$'s are approximately zero. We have $|C_{\Cac}\mp C_{\Cad}|+|C_{\Cbc}\pm C_{\Cbd}|\le4-2\QUAD=2(1+\epsilon)$
where $1\gg\epsilon\ge0$ reflects the statistical fluctuations in
$N_1(+1,+1)$, $N_1(+1,-1)$, ..., $N_4(+1,-1)$, $N_4(-1,-1)$.
This, perhaps counter intuitive, result may be understood by referring to Eq.~(\ref{DISD13}).
If $N$ is very large, we may (in the case at hand) expect that
$N_1(+1,+1)\approx N_1(+1,-1)\approx\ldots\approx N_4(+1,-1) \approx N_4(-1,-1)\approx {N/4}$.
From Eq.~(\ref{DISD10a}) it then follows that $\widetilde{\QUAD}=1-\epsilon$.
In our numerical experiment, $\epsilon=0.002$.
\item
If the pairs $(A_{\Cac,i},B_{\Cac,i})$,
$(A_{\Cad,j},B_{\Cad,j})$,
$(A_{\Cbc,k}B_{\Cbc,k})$, and
$(A_{\Cbd,l}B_{\Cbd,l})$
are generated randomly with frequencies
$(1-c_{\Cac}A_{\Cac,i}B_{\Cac,i})/4$,
$(1-c_{\Cad}A_{\Cad,j}B_{\Cad,j})/4$,
$(1-c_{\Cbc}A_{\Cbc,k}B_{\Cbc,k})/4$, and
$(1-c_{\Cbd}A_{\Cbd,l}B_{\Cbd,l})/4$, respectively,
the simulation mimics the case of the correlation of two spin-1/2 objects in the singlet state
if we choose $c_{\Cac}=-c_{\Cad}=c_{\Cbc}=c_{\Cbd}=1/\sqrt{2}$.
Recall that in this particular case, quantum theory yields
$\max\left(|C_{\Cac}\mp C_{\Cad}|+|C_{\Cbc}\pm C_{\Cbd}|\right)=2\sqrt{2}\approx2.83$~\cite{CIRE80}, see also Section~\ref{EXTE}.

Generating four times one million independent pairs, we obtain
$\widehat\QUAD\approx0.047$,
$\widetilde\QUAD\approx0.292$,
$\QUAD\approx0.585$,
$S_{\mathrm{CHSH}}=|C_{\Cac}-C_{\Cad}|+|C_{\Cbc}+C_{\Cbd}|\approx2.83$ and $4-2\QUAD\approx2.83$,
demonstrating that the value of the quantum-theoretical upper bound $2\sqrt{2}$
is reflected in the maximum fraction of quadruples that one can find by reshuffling the data.
\item
Same as in the previous case except that we use the probabilistic model of Section~\ref{TPM}
as the basis for the CM to generate pairs of data for the case $d=6$ for which $S_{\mathrm{CHSH}}\approx3.20>2\sqrt{2}\approx2.83$,
showing that these data cannot be described by a quantum-theoretical model of two spin-1/2 objects.
Choosing $c_{\Cac}=-c_{\Cad}=c_{\Cbc}=c_{\Cbd}=0.80$ and
generating four times one million independent pairs, we obtain
$S_{\mathrm{CHSH}}=|C_{\Cac}-C_{\Cad}|+|C_{\Cbc}+C_{\Cbd}|\approx3.20$ and $4-2\QUAD\approx3.20$.
\item
Same as in the previous case except that we consider the case $d=8$ for which $S_{\mathrm{CHSH}}\approx3.34>2\sqrt{2}\approx2.83$,
showing that these data cannot be described by a quantum-theoretical model of two spin-1/2 objects.
Choosing $c_{\Cac}=-c_{\Cad}=c_{\Cbc}=c_{\Cbd}=0.83$ and
generating four times one million independent pairs, we obtain
$S_{\mathrm{CHSH}}=|C_{\Cac}-C_{\Cad}|+|C_{\Cbc}+C_{\Cbd}|\approx3.34$ and $4-2\QUAD\approx3.34$.

\item
In~\ref{CHEB}, we have used
the data of the ``Significant-Loophole-Free Test of Bell’s Theorem with Entangled Photons'' experiment~\cite{GIUS15}
to estimate that the fraction of quadruples that can be created by reshuffling the data is $\QUAD\lesssim0.99999273$. Using the quantum state $(|HV\rangle+r|VH\rangle)/\sqrt{1+r^2}$ assumed to describe the ideal experiment~\cite{GIUS15},
we obtain $S_{\mathrm{CHSH}}\approx2.34>2$
and (see Eq.~(\ref{CHEB5}))
$(N_{++}(\ba,\bc)-N_{++}(\bb,\bd)-N_{-+}(\bb,\bc)-N_{+-}(\ba,\bd))/N\approx0.085 > 0$,
where $\ba$, $\bb$, $\bc$, and $\bd$ correspond to the angles $94.4^\circ$, $62.4^\circ$, $-6.5^\circ$, and $25.5^\circ$, respectively, and $r=-2.9$.
Generating four times one million independent pairs $(A_{\Cac,i},B_{\Cac,i})$,
$(A_{\Cad,j},B_{\Cad,j})$,
$(A_{\Cbc,k}B_{\Cbc,k})$, and
$(A_{\Cbd,l}B_{\Cbd,l})$
with frequencies corresponding to the quantum -theoretical probabilities, we obtain
$\widehat\QUAD\approx0.275$,
$\widetilde\QUAD\approx0.491$,
$\QUAD\approx0.829$, such that
$4-2\QUAD\approx2.34\approx S_{\mathrm{CHSH}}$.
As mentioned in~\ref{CHEB}, the value of $\QUAD\approx0.829$ obtained from the quantum-theoretical model does not necessarily relate to the value $\QUAD\lesssim0.99999273$ obtained by analyzing the experimental data.
\item
In the case of Bell's modified toy model (see~\ref{CBEL})
for which $C(\ba,\bc)=-(1/2)\cos(a-c)$ or
the probabilistic model of Section~\ref{TPM} with $d=0$ or $W>T_0$, we have $S_{\mathrm{CHSH}}=\sqrt{2}$.
Choosing $c_{\Cac}=-c_{\Cad}=c_{\Cbc}=c_{\Cbd}=1/2\sqrt{2}$ and
generating four times one million independent pairs, we obtain
$S_{\mathrm{CHSH}}=|C_{\Cac}-C_{\Cad}|+|C_{\Cbc}+C_{\Cbd}|\approx1.42$ and $4-2\QUAD\approx2.00$.
\end{itemize}
Except in the first case, the values of $\QUAD$  quoted in the other cases fluctuate a little if we repeat the $N=1000 000$ simulations with different random numbers.
Except for the first, second, and last case, the data suggest that inequality Eq.~(\ref{DISD7}) can be saturated.

\subsection{Illustration: Extended EPRB experiment}\label{EEPRB}

\begin{figure}[!htp]
\centering
\includegraphics[width=0.95\hsize]{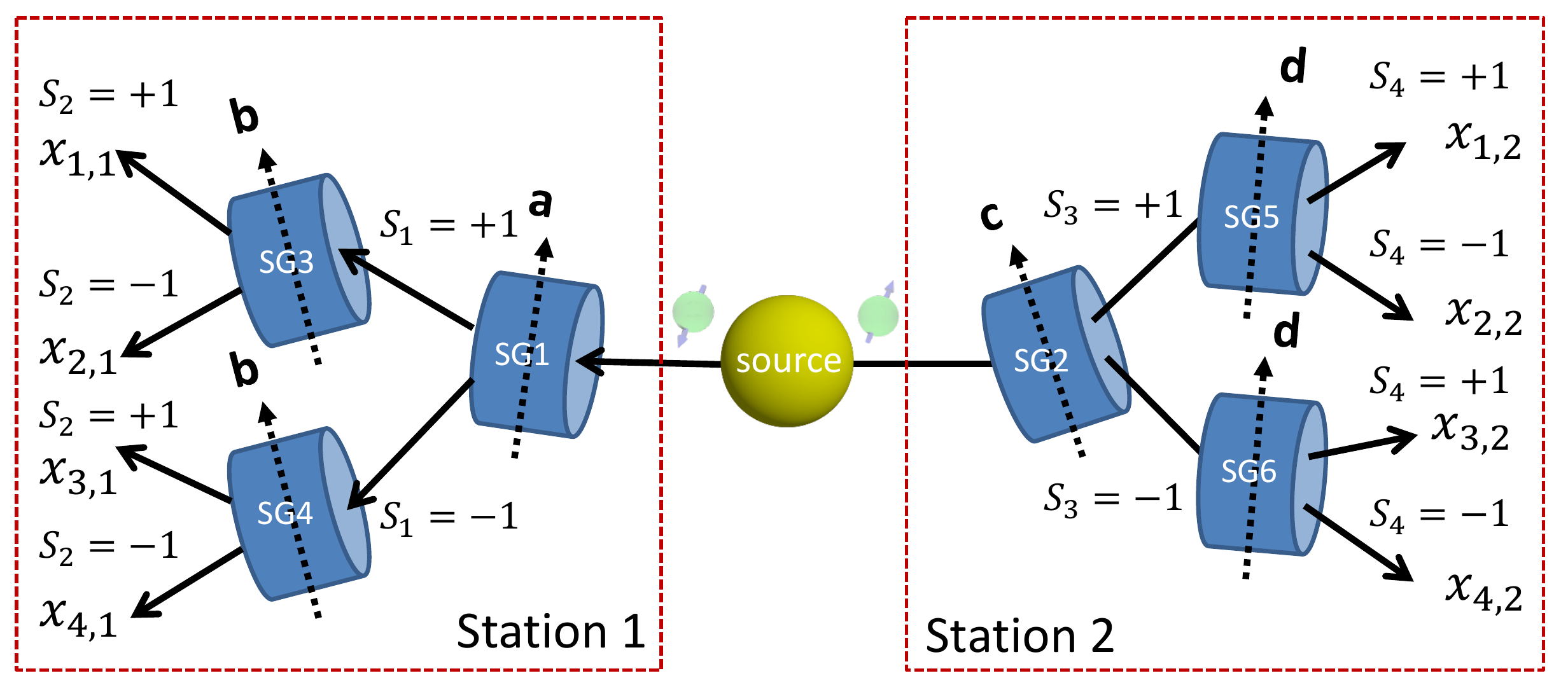}
\caption{
Layout of the extended Einstein-Podolsky-Rosen-Bohm thought experiment with spin-1/2 particles~\cite{SICA99,RAED20a}.
A source is emitting a pair of magnetic particles in two spatially separated directions, directed
towards observation station 1 and 2.
The observation station 1 (2) contains three identical Stern-Gerlach magnets
SG1, SG3, and SG4, (SG2, SG5, and SG6) with their uniform field component along
the directions $\ba$, $\bb$, and $\bb$ ($\bc$, $\bd$, and $\bd$), respectively.
Particles leaving SG3,...,SG6 are registered by identical, ideal detectors (not shown).
The binary variables $x_{i,j}=0,1$ for $i=1,2,3,4$ and $j=1,2$
indicate which of the four detectors at the left ($j=1$) and right ($j=2$) fire.
For each incoming particle, only one of the detectors in station 1 and
only one of the detectors in station 2 fires,
implying that for $j=1,2$, only one of $x_{1,j}$, $x_{2,j}$, $x_{3,j}$, and $x_{4,j}$ can be nonzero.
For each pair of particles emitted by the source, this experiment produces the quadruple
$(S_{1},S_{2},S_{3},S_{4})$.
}
\label{eeprb1}
\end{figure}

Figure~\ref{eeprb1} shows the layout of an
extended Einstein-Podolsky-Rosen-Bohm (EEPRB) experiment with spin-1/2 particles~\cite{SICA99,RAED20a}.
In this idealized experiment, all Stern-Gerlach magnets
perform selective (filtering) measurements~\cite{SCHW59,BALL03}.
Selective measurements allow us to attach an attribute with definite value
(e.g. the direction of the magnetic moment) to the particle.
For instance, assuming that SG1, SG3 and SG4 perform ideal selective measurements,
a particle leaving SG1 along path $S_1=+1$ ($S_1=-1$)
will always leave SG3 (SG4) along path $S_2=+1$ ($S_2=-1$) if $\bb=\ba$.
In this case, the value of this attribute (called spin in quantum theory) is given by $S_1$.
The same procedure is used to attach attributes to particles leaving the other Stern-Gerlach magnets.

As only one of $x_{1,1}$, $x_{2,1}$, $x_{3,1}$, and $x_{4,1}$
and only one of $x_{1,2}$, $x_{2,2}$, $x_{3,2}$, and $x_{4,2}$
can be nonzero (see also Fig.~\ref{eeprb1}), the four variables
\begin{eqnarray}
S_{1}&=&x_{1,1}+x_{2,1}-x_{3,1}-x_{4,1}\;,
\nonumber \\
S_{2}&=&x_{1,1}-x_{2,1}+x_{3,1}-x_{4,1}\;,
\nonumber \\
S_{3}&=&x_{1,2}+x_{2,2}-x_{3,2}-x_{4,2}\;,
\nonumber \\
S_{4}&=&x_{1,2}-x_{2,2}+x_{3,2}-x_{4,2}\;,
\label{EEPRB0}
\end{eqnarray}
can only take values $+1$ or $-1$.
Clearly, $S_1$ and $S_2$ ($S_3$ and $S_4$) encode, uniquely, manner,
the path that the left (right) going particle took.
The four variables Eq.~(\ref{EEPRB0}) form a quadruple $(S_1,S_2,S_3,S_4)$
which completely describes the outcome of the experiment for each pair of particles
emitted by the source.

Next, we attach a pair label $n$ to the $S$'s and compute correlations according to
\begin{eqnarray}
K_{ij}&=&\frac{1}{N}\sum_{n=1}^N S_{i,n}S_{j,n}
\;.
\label{EEPRB1}
\end{eqnarray}
Because the EEPRB experiment generates quadruples only, $\QUAD=1$.
Therefore, it follows from  Eq.~(\ref{FUNDA}) that
independent of the directions $\ba$, $\bb$, $\bc$ and $\bd$, we must have
\begin{eqnarray}
\left|K_{ik}-K_{il}+K_{jk}+K_{jl}\right|\le\left|K_{ik}-K_{il}\right|+\left|K_{jk}+K_{jl}\right|\le2
\;,\;(i,j,k,l)\in\bm\pi_4
\;.
\label{EEPRB2}
\end{eqnarray}

One run of the extended EPRB experiment yields enough data to compute all possible
correlations of the four $S$'s.
For instance, we have
$K_{13}=C_{\ba\bc}$,
$K_{14}=C_{\ba\bd}$,
$K_{23}=C_{\bb\bc}$, and
$K_{24}=C_{\bb\bd}$,
showing that one run of the extended EPRB experiment suffices
to compute all the correlations that would be obtained
by four runs of the EPRB experiment for the conditions
$(\ba,\bc)$,
$(\ba,\bd)$,
$(\bb,\bc)$, and
$(\bb,\bd)$.
Of course, the essential difference between these two experiments
is that the former {\bf always} generates quadruples $(S_{1,n},S_{2,n},S_{3,n},S_{4,n})$
whereas the latter not necessarily does.

In conclusion, if the four correlations that appear in the Bell-CHSH inequality
are obtained by performing the EEPRB experiment, we have
\begin{eqnarray}
S_{\mathrm{CHSH}}
\le2
\;,
\label{EEPRB3}
\end{eqnarray}
showing that these correlations can never violate the Bell-CHSH inequality~\cite{SICA99,RAED20a} even though all pair-wise correlations are given by the quantum-theoretical description in terms of the singlet state.

\section{Plausibility versus mathematical probability}\label{PLAUS}

Plausible reasoning is concerned with relating the truth of propositions
given the truth of other propositions~\cite{POLY54}.
The key concept is the plausibility, denoted by a real number $p(A|B)$, quantifying
that proposition $A$ is true conditional on proposition $B$ being true.
Logical inference is the mathematical framework, a set of rules, by which we perform calculations with plausibilities.
Logical inference allows us to reason in a logically consistent manner
which is both unambiguous and independent of the individual,
in particular if there are elements of uncertainty in the description.
For a detailed discussion of the foundations of
plausible reasoning, its relation to Boolean logic and the derivation of the
rules of logical inference, see Ref.~\cite{COX46,COX61,TRIB69,SMIT89,JAYN03}.
It can be shown that plausibilities
may be chosen to take values in the range $[0,1]$ and
obey the rules~\cite{COX46,COX61,TRIB69,SMIT89,JAYN03}
\begin{enumerate}[a.]
\item
$p(A|Z)+p({\bar A}|Z)=1$ where
${\bar A}$ denotes the negation of proposition $A$,
and $Z$ is a proposition assumed to be true.
\item
$p(AB|Z)=p(A|BZ)p(B|Z)=p(B|AZ)p(A|Z)$ where
the ``product'' $BZ$ denotes the logical product (conjunction) of the propositions $B$ and $Z$,
that is the proposition $BZ$ is true if both $B$ and $Z$ are true.
Defining a plausibility for a proposition
conditional on the conjunction of mutual exclusive propositions is regarded as nonsensical.
\item
$p(A{\bar A}|Z)=0$ and $p(A+{\bar A}|Z)=1$
where the ``sum'' $A+B$ denotes the logical sum (inclusive disjunction)
of the propositions $A$ and $B$,
that is the proposition $A+B$ is true if either $A$ or $B$ or both are true.
These two rules show that
Boolean algebra is contained in the algebra of plausibilities.
\end{enumerate}
The rules (a--c) are unique. Any
other rule which applies to plausibilities represented by real numbers
and is in conflict with rules (a--c)
will be at odds with common-sense reasoning and consistency~\cite{TRIB69,SMIT89,JAYN03}.

``Mathematical probability'' refers to the key concept in Kolmogorov's axiomatic framework of probability theory~\cite{KOLM56}.
Clearly, the rules (a--c) are identical to those of the calculus of probability theory~\cite{KOLM56,FELL68,GRIM95}.
However, logical inference does not involve concepts such as set theory, sample spaces, random variables,
probability measures, countable (or finite) additivity, etc., which all are essential to the mathematical
foundation of probability theory~\cite{KOLM56,FELL68,GRIM95}.
Perhaps most important is that in general, the logical inference approach
does not require (but can also deal with) a set of elementary events or propositions into which the propositions
under scrutiny can be resolved.

\section{Solution of the logical inference problem}\label{LIapp}

Expressing the requirements that the Fisher information Eq.~(\ref{LI1})
should be independent of $\theta$, positive and minimal~\cite{RAED14b} we have
\begin{eqnarray}
\frac{\partial I_\mathrm{F}(\theta)}{\partial \theta}
&=&\frac{2}{1-E_{12}^2(\theta)}\frac{\partial E_{12}(\theta)}{\partial \theta}
\left[
\frac{\partial^2 E_{12}(\theta)}{\partial \theta^2}
+\frac{E_{12}(\theta)}{1-E_{12}^2(\theta)}\left(\frac{\partial E_{12}(\theta)}{\partial \theta}\right)^2
\right]=0
\;.
\label{LI2}
\end{eqnarray}
The solution $\partial E_{12}(\theta)/\partial \theta=0$ can be discarded because
then $I_\mathrm{F}(\theta)=0$, corresponding to the uninteresting case
in which the correlation between the $x$ and $y$ does not change with $\theta$.
As $-1\le E_{12}(\theta)\le +1$ (otherwise Eq.~(\ref{LI0}) does not represent a plausibility),
we may substitute $E_{12}(\theta)=\cos g(\theta)$ in the expression in the right brackets and obtain 
\begin{eqnarray}
\frac{\partial^2 E_{12}(\theta)}{\partial \theta^2}
+\frac{E_{12}(\theta)}{1-E_{12}^2(\theta)}\left(\frac{\partial E_{12}(\theta)}{\partial \theta}\right)^2
=-\sin g(\theta)\frac{\partial^2 g(\theta)}{\partial \theta^2}=0
\;,
\label{LI2a}
\end{eqnarray}
of which the only nontrivial solution reads $g(\theta)=u\theta+\varphi$ where $u$ and $\varphi$ are
constants of integration.
Substituting $E_{12}(\theta)=\cos (u\theta+\varphi)$ in Eq.~(\ref{LI1}),
we obtain $I_\mathrm{F}(\theta)=u^2$, independent of $\theta$, as expected.
Moreover, for $n$ any integer, $\theta+2n\pi$ describes the same
experiment with $\ba\cdot\bc=\cos\theta$.
Therefore, we must have $E_{12}(\theta)=E_{12}(\theta+2n\pi)$, implying
that $u=n$ where $n$ is a positive integer
(we exclude $n=0$ because then $E_{12}(\theta)$ does not depend on $\theta$) and we have
\begin{eqnarray}
E_{12}(\theta)=\cos (n\theta+\varphi)
\;,\quad
I_\mathrm{F}(\theta)=n^2
\;,\quad
n=1,2,\ldots
\;.
\label{LI3}
\end{eqnarray}

\section{Bell's proof of his theorem}\label{BELLPROOF}

In the first proof of his theorem~\cite{BELL64}, Bell explicitly used
the assumption of perfect anticorrelation to derive an inequality which
is slightly different from Eq.~(\ref{BASIC4a}). The theorem then follows
from a contradiction derived by using this inequality.
Adopting Bell's notation (but omitting the bars), written in full detail Eq.~(\ref{BASIC4a}) reads
\begin{eqnarray}
\left\vert
\int A(\ba,\lambda)B(\bc,\lambda)\mu(\lambda)\,d\lambda
\pm
\int A(\ba,\lambda)B(\bd,\lambda)\mu(\lambda)\,d\lambda
\right\vert
\le 1\pm \int B(\bc,\lambda)B(\bd,\lambda)\mu(\lambda)\,d\lambda
\;.
\label{BABE1}
\end{eqnarray}

Let us temporarily allow for the idea that data and a model thereof live in the same universe.
If we were to insist that the functions $A(\ba,\lambda)$, $B(\bc,\lambda)$,
and $B(\bd,\lambda)$, living in the realm of the MM,
map one-to-one to the outcomes of the EPRB laboratory experiment, we face two problems.
In Eq.~(\ref{BABE1}) and also in Bell's original proof,
$A(\ba,\lambda),B(\bc,\lambda),B(\bd,\lambda)\in [-1,+1]$ and
as the outcomes take values $\pm1$ only, the mapping between the real numbers and the
discrete data does not exist.
Let us therefore consider the special case that
$A(\ba,\lambda),B(\bc,\lambda),B(\bd,\lambda)=\pm1$.
Then, the mapping exists, at least mathematically.

With the EPRB setup in mind (see Fig.~\ref{eprbidea}),
$A(\ba,\lambda)=\pm1$ represents data collected on, say the left side
whereas $B(\bc,\lambda)=\pm1$ and $B(\bd,\lambda)=\pm1$
would represent data collected on the opposite side.
Clearly, this creates an apparent conflict because in an EPRB laboratory experiment we cannot have one side collecting both
$B(\bc,\lambda)=\pm1$ and $B(\bd,\lambda)=\pm1$ simultaneously if $\bc\not=\bd$.
However, from the viewpoint of data collected in the EPRB laboratory experiment, there is no conflict at all.
Performing the experiment with settings $(\ba,\bc)$ yields data for $A(\ba,\lambda)$ and $B(\bc,\lambda)$.
Likewise, performing the experiment with settings $(\ba,\bd)$ {\sl and the same set of $\lambda$'s as used
in the first experiment} yields data for $A(\ba,\lambda)$ and $B(\bd,\lambda)$.
From these data we certainly can compute (approximations to) the three integrals in Eq.~(\ref{BABE1}),
even though we cannot perform an experiment to measure $B(\bc,\lambda)$ and $B(\bd,\lambda)$ simultaneously.
This apparent conflict illustrates once more that
empirical data and a model thereof do not live in the same universe.

Bell's proof does not suffer from the named conflict. Bell wrote~\cite{BELL64,BELL93}
\begin{align}
\int A(\ba,\lambda)B(\bc,\lambda)\mu(\lambda)\,d\lambda
-
\int A(\ba,\lambda)B(\bd,\lambda)\mu(\lambda)\,d\lambda
=&
\int A(\ba,\lambda)B(\bc,\lambda)\left(1+A(\bc,\lambda)B(\bd,\lambda)\right)\mu(\lambda)\,d\lambda
\nonumber \\
&
-\int A(\ba,\lambda)B(\bd,\lambda)\left(1+A(\bc,\lambda)B(\bc,\lambda)\right)\mu(\lambda)\,d\lambda
\;.
\label{BABE2}
\end{align}
Note that the integrands in Eq.~(\ref{BABE2}) contain products of terms which are not
accessible in an EPRB laboratory experiment. However, in the universe of MMs this is not an issue
as the functions that appear in Eq.~(\ref{BABE2}) are well-defined.
The last integral in Eq.~(\ref{BABE2}) vanishes by
Bell's assumption of perfect anticorrelation $A(\bc,\lambda)=-B(\bc,\lambda)$ for all $\bc$.
Using the triangle inequality and the assumption
that $|A(\ba,\lambda)|\le1$, $|B(\bc,\lambda)|\le1$, $|B(\bd,\lambda)|\le1$,
we obtain
\begin{eqnarray}
\left|\int A(\ba,\lambda)B(\bc,\lambda)\mu(\lambda)\,d\lambda
-
\int A(\ba,\lambda)B(\bd,\lambda)\mu(\lambda)\,d\lambda
\right|
&\le&
1+\int A(\bc,\lambda)B(\bd,\lambda)\mu(\lambda)\,d\lambda
\;,
\label{BABE3}
\end{eqnarray}
which is the inequality used by Bell to prove his theorem.
Equation~(\ref{BABE1}) is different from Eq.~(\ref{BABE3}) but, as Bell showed,
also leads to the conclusion that the expressions of the correlations
$\langle A(\ba)B(\bc)\rangle = \pm\ba\cdot\bc$,
$\langle A(\ba)B(\bd)\rangle = \pm\ba\cdot\bd$, and
$\langle A(\bc)B(\bc)\rangle = \pm\bc\cdot\bd$
is incompatible with Eq.~(\ref{BABE3}).
More generally, Bell's theorem states that the correlation
\begin{eqnarray}
C(\ba,\bc)=\int A(\ba,\lambda)B(\bc,\lambda)\mu(\lambda)\,d\lambda
\;,
\label{BABE4}
\end{eqnarray}
cannot arbitrarily closely approximate the function $-\ba\cdot\bc$
for all $\ba$ and $\bc$~\cite{BELL64,BELL93}.
Assuming that $C(\ba,\bc)=-\ba\cdot\bc$
and choosing, for instance, $\ba=(1,1,0)/\sqrt{2}$, $\bc=(1,0,0)$,
$\bd=(-1,0,0)$, Eq.~(\ref{BABE3}) becomes $\sqrt{2}\le0$, clearly a contradiction.
Bell's theorem is a restatement of the existence of contradictions, derived from Eq.~(\ref{BABE3}).

The assumption of perfect anticorrelation is necessary to arrive at Eq.~(\ref{BABE3})
but is not necessary to prove Bell's theorem, as Bell and CHSH (Clauser, Horn, Shimony and Holt)
showed by considering four instead of three functions~\cite{CLAU69,BELL71,BELL93}, also avoiding
the conflict of the kind mentioned earlier.
In the case of four functions, the proof of the theorem follows directly from Eq.~(\ref{BASIC4b}),
as we now show.

\subsection{Proof of the Bell-CHSH inequality}\label{BELLINQ}

Using the triangle inequality and
$|xy \pm xz|\le1\pm yz$ if $|x|\le1$, $|y|\le1$, and $|z|\le1$ (see Eq.~(\ref{BASIC3})) we have
\begin{align}
\left|{C}(\ba,\bc)-{C}(\ba,\bd)+{C}(\bb,\bc)+{C}(\bb,\bd)\right|
&\le
\left|{C}(\ba,\bc)-{C}(\ba,\bd)\right|+\left|{C}(\bb,\bc)+{C}(\bb,\bd)\right|
\nonumber \\
&\le
\int
\big(
\left|
{A}(\ba,\lambda){B}(\bc,\lambda)
-{A}(\ba,\lambda){B}(\bd,\lambda)
\right|
\nonumber \\
&\hbox to 3cm{}+
\left|
{A}(\bb,\lambda){B}(\bc,\lambda)
+{A}(\bb,\lambda){B}(\bd,\lambda)
\right|
\big)
\,\mu(\lambda)\,d\lambda
\nonumber\\
&\le
\int
\left(
1-{B}(\bc,\lambda){B}(\bd,\lambda) +1+{B}(\bc,\lambda){B}(\bd,\lambda)
\right)
\,\mu(\lambda)\,d\lambda
\nonumber\\
&\le 2\int
\,\mu(\lambda)\,d\lambda
=2
\;.
\label{BELLINQ1}
\end{align}
In the proof of Eq.~(\ref{BELLINQ1}), there is no conflict of the type
encountered in the proof of Eq.~(\ref{BABE1}).
If we assume that $C(\ba,\bc)=-\ba\cdot\bc$,
it is not difficult to find $\ba$'s, $\bb$'s, $\bc$'s
and $\bd$'s for which inequality Eq.~(\ref{BELLINQ1}) is violated.
Bell's theorem is a restatement of the existence of these violations.

We emphasize that inequalities Eqs.~(\ref{BABE3}) and~(\ref{BELLINQ1}) have been derived within
the context of the MM Eq.~(\ref{IN0}) in which the integration over
$\lambda$ is over the full domain of $\lambda$.
From the perspective of EPRB laboratory experiments, the application
of Eq.~(\ref{BELLINQ1}) can only be justified if the data produced
by the EPRB laboratory experiment comes in the form of quadruples,
as in the case of an extended EPRB experiment~\cite{RAED20a}.
For all EPRB experiments performed up to this day, there is no evidence that this is the case.

\subsection{Bell's theorem and separation of conditions}\label{BELLSEP}
Applied to separating conditions in the description of EPRB data (see section~(\ref{SOC}),
Bell's theorem tells us that if we are given data collected under
conditions $(c_1,c_2)$ with correlation $C(c_1,c_2)$ it may, depending on the values taken by the latter, be mathematically impossible
to find functions $f(c_1,\lambda)$ and $g(c_2,\lambda)$ such that
\begin{eqnarray}
C(c_1,c_2)=\int f(c_1,\lambda)g(c_2,\lambda)\mu(\lambda)\,d\lambda
\;,
\label{BASIC6}
\end{eqnarray}
if we impose the constraints  $|f(c_1,\lambda)|\le1$ and $|g(c_2,\lambda)|\le1$.
Given $C(c_1,c_2)$, if it is found that there do not exist scalar functions $f(c_1,\lambda)$ and $g(c_2,\lambda)$ satisfying the aforementioned constraints, and a density $\mu(\lambda)$ such that Eq.~(\ref{BASIC6}) holds, using matrices instead of scalar function seems like the first alternative to explore.

\section{Application of Bell's theorem to Stern-Gerlach experiments}\label{SOCapp}
\begin{figure}[!htp]
\centering
\includegraphics[width=0.90\hsize]{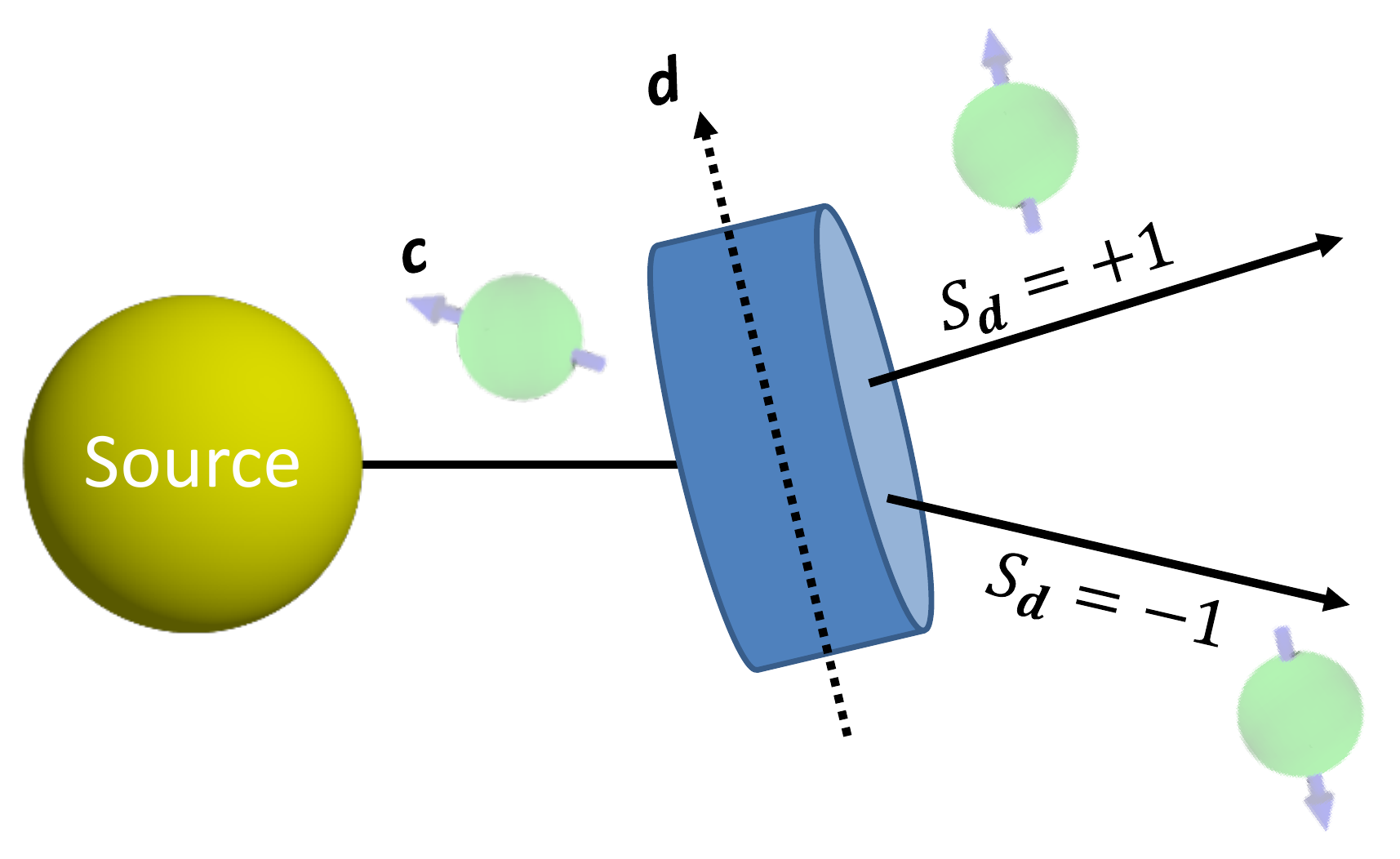}
\caption{(color online)
Conceptual representation of an experiment
with an idealized Stern-Gerlach magnet (cylinder). A source produces neutral particles carrying magnetic moments that are aligned
along the direction represented by the unit vector $\bc$.
The magnetic field gradient of the Stern-Gerlach magnet
with its uniform magnetic field component
along the directions of the unit vectors $\bd$ diverts each incoming
particle into one of two, spatially separated directions labeled by $S_\bd=+1$ and $S_\bd=-1$.
The discrete values of these labels describe the quantum state of the magnetic moment of the particle.
Particles leave the Stern-Gerlach magnet
with their magnetic moment along $\pm\bd$,
labeled by the spin quantum number $S_{\bd}=\pm1$.
Quantum theory predicts that the number of particles
with quantum numbers $S_\bd=\pm1$ is proportional
to $(1+ S_\bd\,\bc\cdot\bd)/2$.
}
\label{doublesg}
\end{figure}

Bell's theorem, see Section~\ref{BELL}, stripped from all relations to the EPRB experiment, tells us that
scalar functions $A(\bc,\lambda)$ and $B(\bd,\lambda)$ such that
\begin{eqnarray}
\bc\cdot\bd&=&\int {A}(\bc,\lambda){B}(\bd,\lambda)\,\mu(\lambda)\,d\lambda
\;,\;
|{A}(\bc,\lambda)|\le1
\;,\;
|{B}(\bd,\lambda)|\le1
\;,\;0\le\mu(\lambda)
\;,\;\int d\lambda \;\mu(\lambda)=1
\;,
\label{SOCapp0}
\end{eqnarray}
do not exist.
We apply Bell's theorem to the experiment sketched and explained in the caption of Fig.~\ref{doublesg}.
According to quantum theory, the number of particles leaving the Stern-Gerlach magnet along the direction labeled by $S_\bd=+1$ is proportional to $(1+S_\bd\,\bc\cdot\bd)/2$.
Consequently, the average value of $S_\bd$ is $\langle S_\bd\rangle=\sum_{S_\bd=\pm1}S_d(1+S_\bd\,\bc\cdot\bd)/2=\bc\cdot\bd$.

Thus, in the notation of Section~\ref{SOC}, Bell's theorem guarantees that there do not exist functions
$-1\le{\widetilde z}(\lambda,\bd)\le1$, $0\le{\widetilde f}(\lambda,\bc)\le1$
(with the symbol $\lambda$ denoting an arbitrary collection of variables)
and a measure $0\le\mu(\lambda)\le1$ such that
\begin{eqnarray}
\int {\widetilde f}(\lambda,\bc)\mu(\lambda)\,d\lambda=1
\quad,\quad
\int {\widetilde z}(\lambda,\bd){\widetilde f}(\lambda,\bc)\mu(\lambda)\,d\lambda
=\bc\cdot\bd
\;,
\label{SOC5b}
\end{eqnarray}
for all unit vectors $\bc$ and $\bd$.

From Eq.~(\ref{SOC5b}), it is clear that ``locality'' actually means ``separation'',
in terms of scalar functions in this case, which depend on distinct conditions ($\bc$ and $\bd$ in this case).
We emphasize that this no-go theorem is the result of applying `Bell's theorem to the Stern-Gerlach experiment, not to the EPRB experiment.

\section{Direct proof of a less general Bell theorem: I}\label{NOINEQ}

We simplify matters by replacing the three-dimensional unit vectors $\ba$, $\bb$
and the hidden variable $\lambda$
by two-dimensional unit vectors that are specified by the angles $a,b\in[0,2\pi]$
and a real variable $\phi\in[0,2\pi]$, respectively.
We give a simple, direct proof that for any differentiable, periodic, real-valued functions
$f(x)=f(x+2\pi)$ or $g(x+2\pi)$ having a finite number of zeros $K_f$ and $K_g$, respectively, the expression
\begin{eqnarray}
I(a,c)&=&\frac{1}{2\pi}\int_{0}^{2\pi} \sign\left[f(\phi-a)\right] \sign\left[g(\phi-c)\right]\,d\phi
\;,
\label{NOINEQ0}
\end{eqnarray}
cannot be equal to $\pm \cos(a-c)$ for all $a$ and $b$.
Making use of the periodicity, it is sufficient to demonstrate this fact by considering the function
\begin{eqnarray}
I(\theta)&=&\frac{1}{2\pi}\int_{0}^{2\pi} \sign\left[f(\phi-\theta)\right] \sign\left[g(\phi)\right]\,d\phi
\;,
\label{NOINEQ1}
\end{eqnarray}
where $\theta=a-c$.

Taking the derivative with respect to $\theta$ and using $d \sign(x)/dx=2\delta(x)$ we find
\begin{eqnarray}
\frac{\partial I(\theta)}{\partial \theta}
&=&\frac{1}{\pi}\int_{0}^{2\pi} \delta(f(\phi-\theta))
\frac{\partial f(\phi-\theta)}{\partial \theta}\sign\left[g(\phi)\right]\,d\phi
=-\frac{1}{\pi}\int \delta(f(\phi))
\frac{\partial f(\phi)}{\partial \phi}\sign\left[g(\phi+\theta)\right]\,d\phi
\;.
\label{NOINEQ2}
\end{eqnarray}
Performing the integral over $\phi$ by using the identity
\begin{eqnarray}
\delta(f(\phi))= \sum_{k=1}^{K_f}\frac{\delta(\phi-\phi_k)}{|\frac{\partial f(x)}{\partial x}\vert_{\phi_k}}
\;,
\label{NOINEQ4}
\end{eqnarray}
where $k=1,\ldots,K_f$ labels the zeros of $f(\phi)$, that is $f(\phi_k)=0$, we obtain
\begin{eqnarray}
\frac{\partial I(\theta)}{\partial \theta}
&=&-\frac{1}{\pi}\sum_{k=1}^{K_f}
\frac{1}{|\frac{\partial f(\phi_k)}{\partial \phi_k}\vert}
\frac{\partial f(\phi_k)}{\partial (\phi_k)}
\sign\left[g(\phi_k+\theta)\right]
=-\frac{1}{\pi}\sum_{k=1}^{K_f}
\sign\left[
\frac{\partial f(\phi_k)}{\partial (\phi_k)}
g(\phi_k+\theta)\right]
\nonumber \\
&\in&
\left\{-\frac{K_f}{\pi},-\frac{K_f-1}{\pi},\ldots,\frac{K_f-1}{\pi},\frac{K_f}{\pi}\right\}
\;.
\label{NOINEQ5}
\end{eqnarray}
Interchanging the roles of $f(\phi)$ and $g(\phi)$ leads to a similar expression.

If $I(\theta)=\pm\cos\theta$ then
${\partial I(\theta)}/{\partial \theta}=\mp\sin\theta\in[-1,+1]$.
But Eq.~(\ref{NOINEQ5}) shows that within this range, ${\partial I(\theta)}/{\partial \theta}$
obtained from Eq.~(\ref{NOINEQ1}) can at most take the values $\pm1/\pi, \pm2/\pi,\pm3/\pi$
which clearly does not arbitrarily closely approximate $\mp\sin\theta$ for all $\theta$.
For instance, $f(\phi)=g(\phi)=\cos\phi$ (see~\ref{CBEL}) has two zeros and
${\partial I(\theta)}/{\partial \theta}=-(2/\pi)\sign[\sin\theta]=\mp2/\pi\not=\mp\sin\theta$ for almost
all $\theta$.

\section{Direct proof of a less general Bell theorem: II}\label{NOINEQII}

In~\ref{NOINEQ}, the $\pm1$'s are obtained from the periodic functions $\sign[f(x)]$
and  $\sign[g(x)]$.
Then, simply because $\cos^2(x)\not=\sign[f(x)]$, it is obviously impossible to reproduce Malus' law.
We can recover Malus' law if we consider periodic functions $0\le f(x)=f(x+2\pi)\le 1$ and define
\begin{eqnarray}
A(a,\phi)=\int_0^1 \sign[f(\phi-a)-r]\,dr\quad,\quad -1\le A(a,\phi)\le 1
\;.
\label{NOINEQ6}
\end{eqnarray}
Indeed, if we choose $f(x)=\cos^2(x)$ and use $f(x)$ to define a CM that generates $+1$ ($-1$) events if $\cos^2(x)>r$ ($\cos^2(x)\le r$), these events appear with a frequency given by Malus' law.

Without invoking Bell-type inequalities, we now prove that for any well-defined, real-valued periodic function $0\le f(x)=f(x+2\pi)\le 1$
\begin{eqnarray}
I(a,c)=\frac{1}{2\pi}\int_0^1 \int_0^1
\int_{0}^{2\pi} \sign\left[{f}(\phi-a)-r\right] \sign\left[f(\phi-c)-r'\right]\,d\phi \,dr\,dr'
\;,
\label{NOINEQ7}
\end{eqnarray}
cannot be equal to $\pm \cos k(a-c)$ for all $a$ and $c$ and $k=1,2,\ldots$.

Substituting $\phi\to\phi+c$ into Eq.~(\ref{NOINEQ7}),
and using the periodicity of $f(x)$, we obtain
\begin{eqnarray}
I(\theta)=
\frac{1}{2\pi}\int_0^1 \int_0^1
\int_{0}^{2\pi} \sign\left[{f}(\phi-\theta)-r\right] \sign\left[f(\phi)-r'\right]\,d\phi \,dr\,dr'
\;,
\label{NOINEQ8}
\end{eqnarray}
where $\theta=a-c$.
Calculating the second derivative of $I(\theta)$ with respect to $\theta$
and using $d \sign(x)/dx=2\delta(x)$ twice we find
\begin{eqnarray}
\frac{\partial^2 I(\theta)}{\partial \theta^2}
=-\frac{2}{\pi}\int_{0}^{2\pi}
\frac{\partial f(\phi)}{\partial \phi}
\frac{\partial f(\phi+\theta)}{\partial \theta}\,d\phi
\;.
\label{NOINEQ9}
\end{eqnarray}
Substituting the Fourier series
$f(\phi)=\sum_{n=-\infty}^{+\infty} f_n e^{in\phi}$ in Eq.~(\ref{NOINEQ9}),
performing the integral over $\phi$ and using $f_{-n}=f^\ast_n$ (because $f(x)$ is real-valued) yields
\begin{eqnarray}
\frac{\partial^2 I(\theta)}{\partial \theta^2}=
-8\sum_{n>0} n^2|f_n|^2 \cos n\theta
\;.
\label{NOINEQ10}
\end{eqnarray}
Equation~(\ref{NOINEQ10}) is equal to $\partial^2_\theta \cos k\theta=-k^2\cos k\theta$
if $|f_k|^2=1/8$  and $|f_n|^2=0$ for all $n\not= k$.
Writing $f_k=|f_k|e^{i\psi}$ with $\psi$ a real number, we have $f(\theta)=f_0+(1/\sqrt{2})\cos k(\theta+\psi)$.
By assumption $0\le f(\theta)$, implying that we must have $f_0 \ge 1/\sqrt{2}$.
Then $f(\theta)\ge [1+\cos k(\theta+\psi)]/\sqrt{2}$.
For $\theta=-\psi$, we have
$f(\theta) \ge \sqrt{2}> 1$, contradicting the assumption that $f(\theta)\le 1$.
Furthermore, Eq.~(\ref{NOINEQ10}) can never be equal to $\partial^2_\theta [-\cos k\theta]=+k^2\cos k\theta$
because $-8|f_k|^2$ can never be equal to one.
This completes the proof.

Consider the case in which we require
$I(\theta)=(1/2)\cos\theta$ instead of
$I(\theta)=\cos\theta$.
Equation~(\ref{NOINEQ10}) is equal to $(1/2)\partial^2_\theta [\cos\theta]$ $=-(1/2)\cos\theta$
if $|f_1|^2=1/16$  and $|f_n|^2=0$ for all $n>1$.
It follows that we must have $0\le f_0\pm1/2\le1$ or $f_0=1/2$
such that $f(\theta)=[1+\cos(\theta+\psi)]/2$.
Thus, the model Eq.~(\ref{NOINEQ6}) can produce a correlation
\begin{eqnarray}
\frac{1}{2}\cos(a-c)=\int A(a,\phi)A(c,\phi)\,d\phi
\;,
\label{NOINEQ11}
\end{eqnarray}
but not a correlation with the factor $1/2$ removed.

\section{Local hidden variable models: discrete data}\label{ERGO}

In this, we derive inequalities for LHVMs
in the case of discrete data.
Adopting the notation used in Eq.~(\ref{IN2}), application of the triangle inequality
and Eq.~(\ref{BASIC3}) yields
\begin{align}
|C(\ba,\bc)-C(\ba,\bd)+C(\bb,\bc)+C(\bb,\bd)|&\le
|C(\ba,\bc)-C(\ba,\bd)|+|C(\bb,\bc)+C(\bb,\bd)|
\nonumber \\
&\le
\sum_{i=1}^P
\big(
\left\vert
 {A}(\ba,\lambda_i){B}(\bc,\lambda_i)
-{A}(\ba,\lambda_i){B}(\bd,\lambda_i)\right\vert
\nonumber \\
&\hbox to 3cm{}+
\left\vert
{A}(\bb,\lambda_i){B}(\bc,\lambda_i)
+{A}(\bb,\lambda_i){B}(\bd,\lambda_i)\right\vert
\big)
\mu(V_i)
\nonumber \\
&\le 2
\;,
\label{IN2z}
\end{align}
showing that the Bell-CHSH inequality also holds for LHVMs defined by the finite sum Eq.~(\ref{IN2}).

The derivation of model-free inequality Eq.~(\ref{DISD7}) did not rely on any assumption
about the data other than that the data are discrete, taking values in the interval $[-1,+1]$.
In contrast, LHVMs assume that there is a rule
that specifies how the value of a data item depends on hidden variables, collectively denoted by
$\lambda$~\cite{BELL64,BELL93}.
In the following, we change the notation somewhat to make it is easier to recognize
the relation with the model-free derivation of inequality Eq.~(\ref{DISD7}).

Imagine repeating the EPRB experiment four times
with specific combinations $(\ba,\bc)$, $(\ba,\bd)$, $(\bb,\bc)$, and $(\bb,\bd)$.
We denote the outcomes for condition $\bx\in\{\ba,\bb,\bc,\bd\}$ by $A(\bx,\lambda)=\pm1$
and $B(\bx,\lambda)=\pm1$ where $\lambda$ plays the role of the hidden variable,
taking values in the domain $\Lambda$.
In LHVMs, the actual value of $A(\bx,\lambda)$ or $B(\bx,\lambda)$
depends on both $\bx$ and on $\lambda$ but, for the purpose of this section,
there is no need to specify this dependence in more detail.
Thus, instead of e.g., $A_{\Cac,n}$ in~\ref{DISD}, in LHVMs we have $A(\ba,\lambda_n)$.
Whereas the $n$ in $A_{\Cac,n}$ is simply a label for the $n$th data item recorded under condition $\Cac$,
$A(\bx,\lambda_n)$ or $B(\bx,\lambda_n)$ are assumed to be known mathematical functions for all $\bx$ and $\lambda_n$,
$\lambda_n$ being the value of $\lambda$ for the $n$-th pair.

In four independent but equally long runs of length $N$, the EPRB experiment
produces the discrete data $A(\ba,\lambda_n)$, etc., for $n=1,\ldots,N$.
With these data, we compute the correlations
\begin{eqnarray}
C(\ba,\bc)&=&\frac{1}{N}\sum_{n=1}^N A(\ba,{\lambda_n})   B(\bc,{\lambda_n})\;,\;
C(\ba,\bd)=\frac{1}{N}\sum_{n=1}^N A(\ba,{\lambda'_n})  B(\bd,{\lambda'_n})\;,\;
\nonumber \\
C(\bb,\bc)&=&\frac{1}{N}\sum_{n=1}^N A(\bb,{\lambda''_n}) B(\bc,{\lambda''_n})\;,\;
C(\bb,\bd)=\frac{1}{N}\sum_{n=1}^N A(\bb,{\lambda'''_n})B(\bd,{\lambda'''_n})
\;.
\label{ERGO1}
\end{eqnarray}
The single, double and triple primes indicate that, in principle, the $\lambda$'s in each of the
independent four runs may be different.

Without any specification of the domain the $\lambda$'s and without
any knowledge about the process that generates them, each
of the correlations in Eq.~(\ref{ERGO1}) can take the value $\pm1$ independent
of the values taken by the others, yielding the trivial bounds
\begin{eqnarray}
|C(\ba,\bc)-C(\ba,\bd)+C(\bb,\bc)+C(\bb,\bd)|&\le&
|C(\ba,\bc)-C(\ba,\bd)|+|C(\bb,\bc)+C(\bb,\bd)|
\le4
\;.
\label{ERGO1a}
\end{eqnarray}
Assume that, for whatever reason,
$\{\lambda_n\,|\,n=1,\ldots,N\}$=$\{\lambda'_n\,|\,n=1,\ldots,N\}$=$\{\lambda''_n\,|\,n=1,\ldots,N\}$=
$\{\lambda'''_n\,|\,n=1,\ldots,N\}$.
Then, it is possible to rearrange the terms in the sums in Eq.~(\ref{ERGO1}) such that for
each $n$, the $A$'s and $B$'s appearing in Eq.~(\ref{ERGO1}) form a quadruple, that is $\QUAD=1$.
From Eq.~(\ref{FUNDA}) it then follows that
\begin{eqnarray}
|C(\ba,\bc)-C(\ba,\bd)+C(\bb,\bc)+C(\bb,\bd)|&\le&
|C(\ba,\bc)-C(\ba,\bd)|+|C(\bb,\bc)+C(\bb,\bd)|
\le2
\;,
\label{ERGO1b}
\end{eqnarray}
which takes the form of the Bell-CHSH inequality Eq.~(\ref{BELLINQ1}).

We now ask ourselves if it is possible to ``interpolate'' between the case of total absence
of knowledge about the $\lambda$'s, yielding inequality Eq.~(\ref{ERGO1a}),
and the special case that led to the Bell-CHSH inequality Eq.~(\ref{ERGO1b}).
To derive a useful inequality, we do not need to specify the domain of the $\lambda$'s,
(they may represent e.g., different animals)
but we assume that the number of different $\lambda$'s is finite.
In symbols $\lambda\in \Lambda =\{\lambda_1,\ldots,\lambda_K\}$.
At this stage, we make no assumption about which $\lambda$'s
of the set $\Lambda$ appear in a particular run.

As the number of different $\lambda$'s is assumed to be finite,
the sums in Eq.~(\ref{ERGO1}) may contain terms such that
$\lambda_n=\lambda'_{n'}=\lambda''_{n''}=\lambda'''_{n'''}$.
We denote the largest set of quadruples for which the latter condition is satisfied by
$Q=\{(n,n',n'',n''')\,|\lambda_n=\lambda'_{n'}=\lambda''_{n''}=\lambda'''_{n'''}\,;\,n,n',n'',n'''\in\{1,\ldots,N\}\}$,
whereby it is implicitly understood that different quadruples $(n,n',n'',n''')$
of $Q$ differ in all four elements $n$, $n'$, $n''$ and $n'''$.
Note that the number of elements in the set $Q$ is a lower bound to the maximum number of quadruples one can find by considering the values of the $A$'s and $B$'s instead of the values of the $\lambda$'s.

At this point, the problem of deriving an upper bound to $\left|C(\ba,\bc)-C(\ba,\bd)|+|C(\bb,\bc)+C(\bb,\bd)\right|$
is identical to the one solved in~\ref{DISD},
Therefore, we have
\begin{eqnarray}
|C(\ba,\bc)-C(\ba,\bd)+C(\bb,\bc)+C(\bb,\bd)|&\le&
|C(\ba,\bc)-C(\ba,\bd)|+|C(\bb,\bc)+C(\bb,\bd)|
\le4-2\QUAD'
\;,
\label{ERGO8d}
\end{eqnarray}
where in this case, $0\le \QUAD'\le\QUAD\le1$ quantifies the fraction of times the same $\lambda$'s
appear simultaneously in the four data sets used to compute $C(\ba,\bc)$, $C(\ba,\bd)$, $C(\bb,\bc)$, and $C(\bb,\bd)$.

\subsection{Illustration I}
As a concrete realization of an LHVM,
assume that there is a fixed rule $R$, an ``equation of motion'' that, given the current
value of $\lambda$, yields the value of the next $\lambda$. For instance
\begin{eqnarray}
\lambda_{n+1}=R(\lambda_{n})\quad,\quad n=1,2,\ldots
\;.
\label{ERGO0}
\end{eqnarray}
We further require that the process $\lambda_1\to\lambda_2\to \ldots$ is
periodic with period $K$. Symbolically, $R^K(\lambda)=\lambda$ for any $\lambda\in\Lambda$.

If $K\le N$ we may write $N=mK +r$ where $m\ge1$ and $0\le r<K$.
In other words, the number of quadruples with the same $\lambda$'s is at least equal to $mK$ and $4-2\QUAD'=2+2 (1-\QUAD')\le2+2 (N-mK)/N = 2+2r/N$.
Therefore, from Eq.~(\ref{ERGO8d})
\begin{eqnarray}
|C(\ba,\bc)-C(\ba,\bd)+C(\bb,\bc)+C(\bb,\bd)|&\le&
|C(\ba,\bc)-C(\ba,\bd)|+|C(\bb,\bc)+C(\bb,\bd)|
\le 2+\frac{2r}{N}
\;.
\label{ERGO0a}
\end{eqnarray}
Obviously, if $K\to N$, then $m\to 1$ and $r/N\to0$
or if $K$ is independent of $N$ and $N\to\infty$,
Eq.~(\ref{ERGO0a}) reduces to the Bell-CHSH inequality
$\left|C(\ba,\bc)-C(\ba,\bd)+C(\bb,\bc)+C(\bb,\bd)\right| \le 2$.
If $K> N$, we have to determine $\QUAD'$ by computation and use Eq.~(\ref{ERGO8d}).

\subsection{Illustration II}
We consider four independent but equally long runs of length $N$
and assume a from an experimental viewpoint, more ``realistic'' scenario in which the order in which the
$K$ different values of $\lambda\in \Lambda =\{\lambda_1,\ldots,\lambda_K\}$
appear is unpredictable.
As the number $K$ of different $\lambda$'s is finite,
the correlations Eq.~(\ref{ERGO1}) can be written as
\begin{subequations}
\label{ERGO5}
\begin{eqnarray}
C(\ba,\bc)
&=&\frac{1}{N}\sum_{k=1}^K n_k^{(\Cac)} A(\ba,{\lambda_k})B(\bc,{\lambda_k})
\;,\;
C(\ba,\bd)
=\frac{1}{N}\sum_{k=1}^K n_k^{(\Cad)} A(\ba,{\lambda_k})B(\bd,{\lambda_k})
\;,\;
\\
C(\bb,\bc)
&=&\frac{1}{N}\sum_{k=1}^K n_k^{(\Cbc)} A(\bb,{\lambda_k})B(\bc,{\lambda_k})
\;,\;
C(\bb,\bd)
=\frac{1}{N}\sum_{k=1}^K n_k^{(\Cbd)} A(\bb,{\lambda_k})B(\bd,{\lambda_k})
\;,
\end{eqnarray}
\end{subequations}
where $0\le n_k^{(\Cac)}\le N$, constrained by $\sum_{k=1}^K {n}^{(\Cac)}_k=N$, is the number of times $\lambda_k$ appears in the
sum of the $A(\ba,{\lambda_n})B(\bc,{\lambda_n})$ terms,
and the same for $n_k^{(\Cad)}$ etc.

For each value of $k$ in Eq.~(\ref{ERGO5}), we introduce the symbol $N_k=\min(n_k^{(\Cac)},n_k^{(\Cad)},n_k^{(\Cbc)},n_k^{(\Cbd)})$
and we have
\begin{align}
|C(\ba,\bc)-C(\ba,\bd)+C(\bb,\bc)+C(\bb,\bd)|&\le
|C(\ba,\bc)-C(\ba,\bd)|+|C(\bb,\bc)+C(\bb,\bd)|
\nonumber \\
&\le
\frac{1}{N}\sum_{k=1}^K N_k
\bigg|A(\ba,{\lambda_k})\big[B(\bc,{\lambda_k})-B(\bd,{\lambda_k})\big]\bigg|
+\bigg|A(\bb,{\lambda_k})\big[B(\bc,{\lambda_k})+B(\bd,{\lambda_k})\big]\bigg|
\nonumber \\
&+\frac{1}{N}\sum_{k=1}^K \bigg|
 (n_k^{(\Cac)}-N_k)A(\ba,{\lambda_k})B(\bc,{\lambda_k})
-(n_k^{(\Cad)}-N_k)A(\ba,{\lambda_k})B(\bd,{\lambda_k}) \bigg|
\nonumber \\
&\hbox to 1.5cm{}
+\bigg|(n_k^{(\Cbc)}-N_k)A(\bb,{\lambda_k})B(\bc,{\lambda_k})
+(n_k^{(\Cbd)}-N_k)A(\bb,{\lambda_k})B(\bd,{\lambda_k})
\bigg|
\nonumber \\
&\le
2 \frac{1}{N}\sum_{k=1}^K N_k +\frac{1}{N}\sum_{k=1}^K \left(
 |n_k^{(\Cac)}-N_k| +|n_k^{(\Cad)}-N_k|+|n_k^{(\Cbc)}-N_k|+|n_k^{(\Cbd)}-N_k|\right)
\nonumber \\
&=
2\frac{1}{N}\sum_{k=1}^K N_k +\frac{1}{N}\sum_{k=1}^K \left(n_k^{(\Cac)}-N_k
+n_k^{(\Cad)}-N_k+n_k^{(\Cbc)}-N_k+n_k^{(\Cbd)}-N_k\right)
\nonumber \\
&=
2\frac{1}{N}\sum_{k=1}^K N_k +4\left(1-\frac{1}{N}\sum_{k=1}^K N_k\right)=4-2\QUAD''
\;,
\label{ERGO6}
\end{align}
which has the same form as Eq.~(\ref{DISD7}),
except that in this particular case
$0\le\QUAD''
=N^{-1}\sum_{k=1}^K \min(n_k^{(\Cac)},n_k^{(\Cad)},n_k^{(\Cbc)},n_k^{(\Cbd)})\le\QUAD\le1$.

Let us consider the simple model
for which the probability to select $k\in\{1,\ldots,K\}$ is $1/K$.
Then, for sufficiently large $N$, $n_k^{(\Cac)}\approx n_k^{(\Cad)}\approx n_k^{(\Cbc)} \approx n_k^{(\Cbd)}\approx N/K$
implying that $\QUAD''\lesssim1$.
In other words,
$|C(\ba,\bc)-C(\ba,\bd)|+|C(\bb,\bc)+C(\bb,\bd)|\le 2 +\epsilon$
where $\epsilon$ is a small number that reflects the statistical fluctuations in
$n_k^{(\Cac)}$, $n_k^{(\Cad)}$, $n_k^{(\Cbc)}$, and $n_k^{(\Cbd)}$.
Note that if $K\gg N$, $\QUAD''\approx 0$ and Eq.~(\ref{ERGO6}) reduces to Eq.~(\ref{ERGO1a}).

\section{Proof of Lemma I}\label{LEMMA}
{\bf Lemma I:} Given four, pair-wise compatible, nonnegative, normalized bivariates
$f(x_1,x_3|\ba,\bc)$, $f(x_1,x_4|\ba,\bd)$, $f(x_2,x_3|\bb,\bc)$, and $f(x_2,x_4|\bb,\bc)$
with moments $K_1$, $K_2$, $K_3$, $K_4$, $K_{13}$, $K_{14}$, $K_{23}$, and $K_{24}$
satisfying the Bell-CHSH inequalities $|K_{13}\mp K_{14}|+|K_{23}\pm K_{24}|\le2$,
there exists a number $\alpha$ satisfying
\begin{eqnarray}
-1\le -1 + \max(\vert K_{13} + K_{14}\vert,\vert K_{23} + K_{24}\vert,
\vert K_{3} + K_{4}\vert)
\le \alpha \le 1 - \max(\vert K_{13} - K_{14} \vert,
\vert K_{23} - K_{24} \vert,\vert K_{3} - K_{4} \vert)\le 1
\;.
\label{CHSH10}
\end{eqnarray}
Proof: From Eq.~(\ref{CHSH10}), it follows that in order to prove the existence of (a range of) $\alpha$,
the following inequalities must hold
\begin{subequations}
\label{LEMMA0}
\begin{eqnarray}
-1+\vert K_{13} + K_{14}\vert &\le& 1- \vert K_{13} - K_{14}\vert \;,\label{LEM1}\\
-1+\vert K_{13} + K_{14}\vert &\le& 1- \vert K_{23} - K_{24}\vert \;,\label{LEM2}\\
-1+\vert K_{13} + K_{14}\vert &\le& 1- \vert K_{3} - K_{4}\vert   \;,\label{LEM3}\\
-1+\vert K_{23} + K_{24}\vert &\le& 1- \vert K_{13} - K_{14}\vert \;,\label{LEM4}\\
-1+\vert K_{23} + K_{24}\vert &\le& 1- \vert K_{23} - K_{24}\vert \;,\label{LEM5}\\
-1+\vert K_{23} + K_{24}\vert &\le& 1- \vert K_{3} - K_{4}\vert   \;,\label{LEM6}\\
-1+\vert K_{3} + K_{4}\vert   &\le& 1- \vert K_{13} - K_{14}\vert \;,\label{LEM7}\\
-1+\vert K_{3} + K_{4}\vert   &\le& 1- \vert K_{23} - K_{24}\vert \;,\label{LEM8}\\
-1+\vert K_{3} + K_{4}\vert   &\le& 1- \vert K_{3} - K_{4}\vert   \;.\label{LEM9}
\end{eqnarray}
\end{subequations}
Recall that all the moments that appear in Eq.~(\ref{LEMMA0}) do not exceed one on absolute value.
Then Eqs.~(\ref{LEM1}),~(\ref{LEM5}), and~(\ref{LEM9}) follow directly from the basic
inequality $|x+y|+|x-y|\le 1-xy +1 +xy\le2$ (see ~\ref{BASIC}).
Equation~(\ref{LEM3}) follows from
$|K_3-K_4|=|K_3\pm K_1 -(K_4\pm K_1)|\le|K_3\pm K_1|+|K_4\pm K_1|\le 2\pm (K_{13}+K_{14}) \le 2 - |K_{13}+K_{14}|$.
Replacing subscript 1 by 2, we prove Eq.~(\ref{LEM6}).
In the same way, we can prove Eqs.~(\ref{LEM7}) and~(\ref{LEM8}).
Finally, the assumption that Bell-CHSH inequalites Eq.~(\ref{CHSH4z}) hold
is just rewriting Eqs.~(\ref{LEM2}) and~(\ref{LEM4}).

Lemma I shows that we may use any choice of $\alpha$ bounded as in Eq.~(\ref{CHSH10})
to assign a value to $K_{34}$.
By construction and appeal to Eq.~(\ref{TRIPLE}), $K_1$, $K_3$, $K_4$, $K_{13}$, $K_{14}$, and $K_{34}$ can be shown to satisfy all the inequalities Eqs.~(\ref{THREE3a})--(\ref{THREE3c}) and so do
$K_2$, $K_3$, $K_4$, $K_{23}$, $K_{24}$, and $K_{34}$ (with subscript 1 replaced by 2).

\section{Transpiled circuits of the EPRB quantum computer experiment}\label{APPQC}

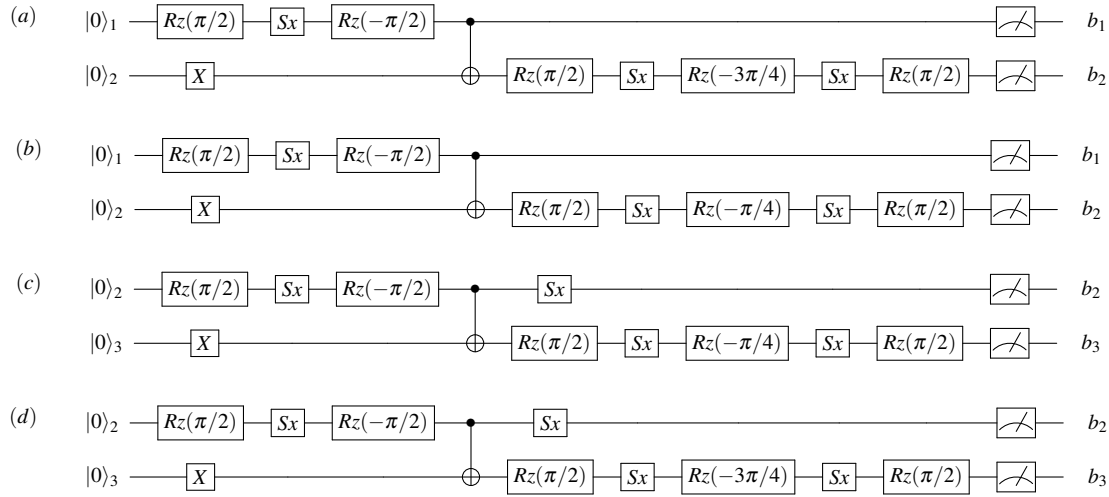
\begin{figure}[!htp]
{\footnotesize 
  \[
  \begin{array}{ccc}
   (a)\hbox to 0.5cm{}&\mbox{\Qcircuit @C=1.3em @R=.95em {
      &\lstick{\ket{0}_1}&\gate{Rz(\pi/2)} &\gate{Sx} &\gate{Rz(-\pi/2)}&\ctrl{1}&\qw              &\qw      &\qw               &\qw      &\qw             &\meter&\qw& &\lstick{b_1}\\
      &\lstick{\ket{0}_2}&\gate{X}         &\qw       &\qw              &\targ   &\gate{Rz(\pi/2)} &\gate{Sx}&\gate{Rz(-3\pi/4)}&\gate{Sx}&\gate{Rz(\pi/2)}&\meter&\qw& &\lstick{b_2}\\
    }
    }
  \end{array}
  \]
  }
\bigskip
{\footnotesize 
  \[
  \begin{array}{ccc}
   (b)\hbox to 0.5cm{}&\mbox{\Qcircuit @C=1.3em @R=.95em {
      &\lstick{\ket{0}_1}&\gate{Rz(\pi/2)} &\gate{Sx} &\gate{Rz(-\pi/2)}&\ctrl{1}&\qw              &\qw      &\qw               &\qw      &\qw             &\meter&\qw& &\lstick{b_1}\\
      &\lstick{\ket{0}_2}&\gate{X}         &\qw       &\qw              &\targ   &\gate{Rz(\pi/2)} &\gate{Sx}&\gate{Rz(-\pi/4)}&\gate{Sx}&\gate{Rz(\pi/2)}&\meter&\qw& &\lstick{b_2}\\
    }
    }
  \end{array}
  \]
  }
\bigskip
{\footnotesize 
  \[
  \begin{array}{ccc}
   (c)\hbox to 0.5cm{}&\mbox{\Qcircuit @C=1.3em @R=.95em {
      &\lstick{\ket{0}_2}&\gate{Rz(\pi/2)} &\gate{Sx} &\gate{Rz(-\pi/2)}&\ctrl{1}&\gate{Sx}        &\qw      &\qw               &\qw      &\qw            &\meter&\qw& &\lstick{b_2}\\
      &\lstick{\ket{0}_3}&\gate{X}         &\qw       &\qw              &\targ   &\gate{Rz(\pi/2)} &\gate{Sx}&\gate{Rz(-\pi/4)}&\gate{Sx}&\gate{Rz(\pi/2)}&\meter&\qw& &\lstick{b_3}\\
    }
    }
  \end{array}
  \]
  }
\bigskip
{\footnotesize 
  \[
  \begin{array}{ccc}
   (d)\hbox to 0.5cm{}&\mbox{\Qcircuit @C=1.3em @R=.95em {
      &\lstick{\ket{0}_2}&\gate{Rz(\pi/2)} &\gate{Sx} &\gate{Rz(-\pi/2)}&\ctrl{1}&\gate{Sx}        &\qw      &\qw               &\qw      &\qw             &\meter&\qw& &\lstick{b_2}\\
      &\lstick{\ket{0}_3}&\gate{X}         &\qw       &\qw              &\targ   &\gate{Rz(\pi/2)} &\gate{Sx}&\gate{Rz(-3\pi/4)}&\gate{Sx}&\gate{Rz(\pi/2)}&\meter&\qw& &\lstick{b_3}\\
    }
    }
  \end{array}
  \]
  }
  \caption{Transpiled versions of the circuits used to generate the raw data used to compute
  the data shown in Table~\ref{tab1}.
  (a): $(a,b)=(45,0)$;
  (b): $(a,b)=(135,0)$;
  (c): $(a,b)=(45,90)$;
  (d): $(a,b)=(135,90)$.
  \label{CIRCfig3}
  }
\end{figure}

Figure~\ref{CIRCfig3} shows the transpiled circuits used to compute the four contributions
to the Bell-CHSH function $S_{\mathrm{CHSH}}$, see Table~\ref{tab1}.

\section{How to obtain the correlation $C(\ba,\bc)=-\ba\cdot\bc$}\label{RECO}

As Bell's theorem is mathematically sound, there is no way to obtain the correlation $C(\ba,\bc)=-\ba\cdot\bc$
if one sticks to
the conditions under which the theorem has been proven~\cite{BELL64}.
Thus, one way to obtain the correlation while retaining the factorized form of Eq.~(\ref{IN0})
is to discard one or more of the conditions under which the theorem has been proven.
Another alternative is to forget altogether about Eq.~(\ref{IN0})
because as explained in Section~\ref{TPM}, Eq.~(\ref{IN0}) may be too primitive to capture the way the experimental data is collected and processed.

In this section, we keep the mathematics as simple as possible by focusing
on models for EPRB experiments with polarized light.
Then $\ba=(\cos a,\sin a,0)$, $\bc=(\cos c,\sin c,0)$, Malus' law reads
$P(x|a,\phi)=[1+x\cos2(\phi-a)]/2$, and the quantum-theoretical result of the correlation
of two photons with their polarizations in the singlet state
reads
\begin{eqnarray}
C(\ba,\bc)=-\cos2(a-c)
\;,
\label{RECO0}
\end{eqnarray}
the extra factor two stemming from the fact
that we are not considering spin-1/2 objects but photon polarizations, see Section~\ref{INSTEAD}.
Repeating the calculations with 3D vectors is a little more tedious but straightforward.

An NQM model for the EPRB experiment is considered to be physically relevant
if (1) in the case of photons, the model also complies with Malus' law,
or (2) in the case of spin-1/2 objects, the model yields $P(x|\ba,\bS)=(1+x\,\ba\cdot\bS)/2$.
We first review NQMs that fail to
and then present NQMs that succeed to yield the correlation $C(\ba,\bc)=-\cos2(a-c)$.

\subsection{Bell's toy model}\label{CBEL}

In Bell's toy model, the correlation is given by Eq.~(\ref{IN0}) with
\begin{eqnarray}
A({\ba,\lambda}) &=& -B({\ba,\lambda})= \sign\left[\cos2(\lambda-a)\right]
=
\left\{
\begin{array}{ccc}
+1 & \hbox{if}& 0\le 2(\lambda-a)<{\pi}/{2}\\
\\
-1 & \hbox{if}& {\pi}/{2}\le 2(\lambda-a)<{3\pi}/{2}\\
\\
+1 & \hbox{if}& {3\pi}/{2}\le 2(\lambda-a)<2\pi
\end{array}
\right.
\;,
\label{CBEL0}
\end{eqnarray}
where $0\le2(\lambda-a)<2\pi$ and $\lambda$ denotes the polarization of the light beam.
Model Eq.~(\ref{CBEL0}) is in blatant contradiction with Malus' law which predicts a sinusoidal dependence as a function of $\lambda-a$ but has the virtue that the correlation Eq.~(\ref{IN0}) has some interesting features.

With the explicit form Eq.~(\ref{CBEL0}) and $\mu(\lambda)=1/2\pi$,
the integral in Eq.~(\ref{IN0}) can be carried out analytically, yielding~\cite{BELL64}
\begin{eqnarray}
{C}(\ba,\bc)&=&\int {A}(\ba,\lambda){B}(\bc,\lambda)\mu(\lambda)\, d\lambda
= -1 + \frac{2}{\pi}\arccos\left(\cos2(a-c)\right)
\;.
\label{CBEL1}
\end{eqnarray}
For $-\pi/2\le(a-c)\le\pi/2$, the correlation ${C}(\ba,\bc)=-1+4|a-c|/\pi$.
Furthermore, for any $a=c$, we have ${C}(\ba,\bc)=-1$ implying that is there is perfect anticorrelation,
independent of the choice of $a=c$.

In Eq.~(\ref{CBEL1}), the integration over $\lambda$ with weight $\mu(\lambda)$ can, but does not
have to, be interpreted as the integration over the realizations of the random variable $\lambda$
with probability density $\mu(\lambda)$.
If we adopt this view, then the individual values of $A({\ba,\lambda})$
will be random, either $+1$ or $-1$.
Nevertheless, in Bell's model, the $\pm1$ events observed at the two
sides are completely correlated if $a=b$.
Thus, in this probabilistic version of Bell's toy model, if $a=c$ (arbitrary)
knowing the value of say $A({\ba,\lambda})$,
we can predict with certainty that the value of $B({\bc,\lambda})$
will be the opposite of $A({\ba,\lambda})$, even though the values of the $A$'s are random themselves.
It is this feature of the correlation $-\ba\cdot\bc$ which is commonly
referred to in popular accounts of the EPRB experiment.

\subsection{Bell's modified toy model: Malus' law}\label{CBELM}

As mentioned in~\ref{CBEL}, the model defined by Eq.~(\ref{CBEL0})
does not comply with Malus' law.
However, fixing this only requires the simple modification
\begin{eqnarray}
A({\ba,\phi,r})= \sign\left[1+\cos2(\phi-a) - 2r\right]
\;,\quad
B({\bc,\phi,r'})=-\sign\left[1+\cos2(\phi-c) - 2r'\right]
\;,
\label{CBEL0a}
\end{eqnarray}
where $0\le\phi<2\pi$ denotes the polarization of the light beam and $0\le r,r'\le 1$ are uniform random variables.
From Eq.~(\ref{CBEL0a}), it follows immediately that the probability density
to find $A({\ba,\phi,r})=+1$ is given by $\cos^2(\phi-a)$, in agreement with Malus' law.

With the explicit form Eq.~(\ref{CBEL0a}) and $\mu(\phi)=1/2\pi$,
the integral in Eq.~(\ref{IN0}) can be carried out analytically, yielding
\begin{eqnarray}
{C}(\ba,\bc)&=&\frac{1}{2\pi}\int_0^{2\pi} d\phi \;\int_0^{1} dr \;\int_0^{1} dr' {A}(\ba,\phi,r){B}(\bc,\phi,r')
= -\frac{1}{2}\cos2(a-c)
\;.
\label{CBEL1a}
\end{eqnarray}
It is important to note that, as already indicated by Eq.~(\ref{CBEL0a}),
there are two different random variables $r$ and $r'$ at play.
Thus, in Bell's language one might be tempted to write $\lambda=\{\phi,r,r'\}$.
However, then it is difficult to imagine how the station measuring $A({\ba,\phi,r})$ can know
that it should use the $r$-part of $\lambda$ whereas the station measuring $B({\ba,\phi,r'})$ should use the $r'$-part of $\lambda$.
We might try to avoid this conflict by assuming that $r=r'$ but then ${C}(\ba,\bc)=-1$
for all $\ba$ and $\bc$, which is unacceptable.

Thus, it is not obvious that Bell's $\lambda$ is sufficiently general
to include the simple variant Eq.~(\ref{CBEL0a}) of Bell's toy model.
However, Bell also showed that his theorem holds true if we replace
$A(\ba,\lambda)$ and $B(\bc,\lambda)$ by $\overline{A}(\bc,\lambda)$ and $\overline{B}(\bc,\lambda)$
obtained by averaging with respect to ``distributions of instrument variables''~\cite{BELL71,BELL93}.
In Eq.~(\ref{CBEL0a}), $r$ and $r'$ play the role of these instrument variables.
In summary, model Eq.~(\ref{CBEL0a}) satisfies the conditions for proving Bell's theorem,
as confirmed by Eq.~(\ref{CBEL1a}).

The factor $1/2$ that appears in front of the cosine in Eq.~(\ref{CBEL1a})
is a common feature of factorable models (such as the one discussed in~\ref{MAXW})
that comply with Malus' law.
In essence, recovering the quantum-theoretical results
$C(\ba,\bc)=-\cos2(a-c)$ amounts to constructing models that change $1/2$ into $1$.


\subsection{Classical electrodynamics}\label{MAXW}

According to empirical evidence, the intensity of light passing through a polarizer is given by Malus' law
\begin{eqnarray}
I(x|\ba,\phi,I_0(t))&=&I_0(t)\frac{1 + x\, \cos2(\phi-a)}{2}
\;,
\label{MAXW1a}
\end{eqnarray}
where $x=\pm1$ labels the directions of the outgoing light,
$\phi$ and $a$ represent the polarization of the incoming light and orientation
of the polarizer, respectively.
The total intensity of light impinging on the polarizer is $I_0(t)$
which is assumed to fluctuate with time.
In this section, $\langle f(t) \rangle=T^{-1}\int_0^T f(t)\,dt$ denotes
the time average of a function $f(t)$ over the time of observation $T$.

As usual, it is expedient to work with dimensionless variables.
To this end, we divide both sides of Eq.~(\ref{MAXW1a}) by the time-averaged intensity and obtain
\begin{eqnarray}
\frac{I(x|\ba,\phi,r(t))}{\langle I_0(t)\rangle}&=&
\frac{I_0(t)}{\langle I_0(t)\rangle}\frac{1 + x\, \cos2(\phi-a)}{2}
=r(t)\frac{1 + x\, \cos2(\phi-a)}{2}
\;,
\label{MAXW1h}
\end{eqnarray}
where $r(t)={I_0(t)}/{\langle I_0\rangle}\ge0$ is a dimensionless random variable with time average
$\langle r(t)\rangle=1$.
By construction we have
\begin{eqnarray}
\frac{\langle I(x|\ba,\phi,r(t))\rangle}{\langle I_0\rangle}&=&\frac{1 + x\, \cos2(\phi-a)}{2}
\;,
\label{MAXW1b}
\end{eqnarray}
which is the dimensionless form of Malus' law, as expected.

For simplicity, we assume that the left and right going light beams in the
EPRB setup have exactly the same intensity at any time and that
their polarizations differ by $\phi_0$, which is fixed in time.
Then, the expression of the time-averaged correlation of the two normalized intensities reads
\begin{eqnarray}
\frac{\langle I(x,y|\ba,\bc,\phi,\phi_0)\rangle}{\langle I_0\rangle^2} &=&
\langle r^2(t)\rangle \frac{1+x\,\cos2(\phi-a)}{2}\frac{1+y\,\cos2(\phi-c+\phi_0)}{2}
\;.
\label{MAXW1c}
\end{eqnarray}
Next, we imagine that we collect data for Eq.~(\ref{MAXW1c}) by repeating the experiments for many values of the polarization $0\le\phi<\pi$.
Integrating over all polarizations with uniform density $1/\pi$,
the correlated intensity is found to be
\begin{eqnarray}
I(x,y|\ba,\bc,\phi_0)&=&
\frac{1}{\pi}\int_0^{\pi}\frac{\langle I(x,y|\ba,\bc,\phi,\phi_0) \rangle}{\langle I_0\rangle^2}\, d\phi
=\langle r^2(t)\rangle\frac{2+x\,y\,\cos2(a-c+\phi_0)}{8}
\;.
\label{MAXW1d}
\end{eqnarray}
Thus, this classical, Maxwell-theory model yields for the correlation
\begin{eqnarray}
C(\ba,\bc)&=&\sum_{x,y=\pm1} xy I(x,y|\ba,\bc,\phi_0)
=\frac{\langle r^2(t)\rangle}{2}\cos2(a-c+\phi_0)
\;.
\label{MAXW1f}
\end{eqnarray}

A very simple model for the fluctuating intensity can be constructed as follows.
We start from
\begin{eqnarray}
\frac{1}{T}\int_0^T r^k(t)\,dt &\approx&\frac{1}{N}\sum_{n=0}^{N-1} r^k(n T/N)=\langle r^k \rangle_N
\;.
\label{MAXW1fa}
\end{eqnarray}
and assume that $r(n T/N)$ is a random variable
with the probability density $p(r(n T/N))=\exp(-r(n T/N))$ for each $n=0,\ldots,N-1$.
It is easy to check numerically that for each realization of the set of variables $\{r(0),r( T/N),\ldots,r((N-1) T/N)\}$,
we have $\langle r(t)\rangle_N\to1$ and $\langle r^2(t)\rangle_N\to2$ as $N\to\infty$.
Of course, the same result is obtained by replacing $r^k(n T/N)$ by its average.
Thus, this simple model yields $\langle r^2(t)\rangle=2$ and therefore the correlation Eq.~(\ref{MAXW1f})
becomes
\begin{eqnarray}
C(\ba,\bc)&=&\sum_{x,y=\pm1} xy I(x,y|\ba,\bc,\phi_0)=\cos2(a-c+\phi_0)
\;.
\label{MAXW1g}
\end{eqnarray}
If $\phi_0=\pi/2$, Eq.~(\ref{MAXW1g}) agrees with
the desired quantum-theoretical expression $-\cos2(a-c)$ for the correlation of two photons
with their polarizations described by the singlet state.

A key difference between the classical wave mechanical model and
quantum theory of spin-1/2 objects is that in the latter,
the results of quantum measurements, an abstract theoretical concept, are discrete,
either $+1$ or $-1$ whereas in the former,
the intensity can, in principle, take all possible non-negative real values.
Of course, the light intensity measured in real experiments takes discrete values (the resolution
of any measurement device being finite) but the fact remains that these discrete values
are not bound to the interval $[0,1]$.

In conclusion, the local realistic, classical wave mechanical model
based on Eqs.~(\ref{MAXW1a}) and~(\ref{MAXW1c})
\begin{enumerate}
\item
complies with Malus' law and can yield the desired correlation $C(\ba,\bc)=-\cos2(a-c)$
for two light beams with opposite but otherwise random polarization,
\item
does not satisfy the conditions necessary to prove Bell's theorem
because $0\le r(t)$ can exceed one,
implying that the conditions $A(x|\ba,\lambda)\le1$  and
$B(x|\bc,\lambda)\le1$ in Eq.~(\ref{IN0}) are not satisfied,
\end{enumerate}

\subsection{A system of two classical spins}\label{MOME}
We start from the representation
$\bS_j=S_j(\cos\phi_j\sin\theta_j,\sin\phi_j\sin\theta_j,\cos\theta_j)^{\mathrm{T}}$ for $j=1,2$,
of the classical spins and introduce the unit vectors
$\ba_j=(\cos\alpha_j\sin\beta_j,\sin\alpha_j\sin\beta_j,\cos\beta_j)^{\mathrm{T}}$ for $j=1,2$
to specify two directions.
We assume that the length of the spins $S_j\ge0$ is distributed according to a yet unspecified
(probability) density $p(\bS_1,\bS_2)$.

We consider a pair of spins that is perfectly anticorrelated, that is $\bS_1=-\bS_2$,
implying that $S_1=S_2=S$, $\theta_1=\theta_2+\pi=\theta$, $\phi_1=\phi_2=\phi$.
We assume that the vectors representing pairs of the same length $S$
uniformly cover the 3D sphere of radius $S$.
With these assumptions, the (probability) density for all pairs reads
$p(\bS_1,\bS_2)=p(\bS_1)\delta(\bS_1+\bS_2)=S^2\mu(S)\sin\theta\;\delta(\bS_1+\bS_2)$,
where the function $\mu(S)\ge0$ is to be determined later.

Expressing the requirement that the density $p(\bS_1,\bS_2)$ is normalized to one yields
\begin{eqnarray}
\frac{1}{4\pi}\int_0^{\infty} \int_{0}^{\pi} \int_0^{2\pi} S^2\,\mu(S)\,\sin\theta\,dS\,d\theta\,d\phi
&=&\int_0^{\infty} S^2\,\mu(S)\,dS =1
\;,
\label{MOME0a}
\end{eqnarray}
which is a first constraint on candidates for the nonnegative function $\mu(S)$.
For the single-spin averages we have
\begin{eqnarray}
\langle \ba_i\cdot\bS_i \rangle &=&
\frac{1}{4\pi}\int_0^{\infty} \int_{0}^{\pi} \int_0^{2\pi} S^3\,\mu(S)\,\sin\theta\,dS\,d\theta\,d\phi
\left[\cos(\phi-\alpha_i)\sin\theta\sin\beta_i+\cos\theta\cos\beta_i\right]=0\;,\;i=1,2
\;,
\label{MOME0b}
\end{eqnarray}
independent of the choice of $\mu(S)$.
For the correlation, we have
\begin{eqnarray}
\langle \ba_1\cdot\bS_1\; \ba_2\cdot\bS_2 \rangle &=& -\langle \ba_1\cdot\bS_1\; \ba_2\cdot\bS_1 \rangle
\nonumber \\
&=&-\frac{1}{4\pi}\int_0^{\infty} \int_{0}^{\pi} \int_0^{2\pi} S^2\,\mu(S)\,\sin\theta\,dS\,d\theta\,d\phi
\bigg\{S^2\left[\cos(\phi-\alpha_1)\sin\theta\sin\beta_1+\cos\theta\cos\beta_1\right]
\nonumber \\
&&\hbox to 6cm{}\left[\cos(\phi-\alpha_2)\sin\theta\sin\beta_2+\cos\theta\cos\beta_2\right]\bigg\}
\nonumber \\
&=&-\frac{1}{4\pi}\int_0^{\infty} \int_{0}^{\pi} \int_0^{2\pi} S^4\,\mu(S)\,dS\,d\theta\,d\phi
\bigg\{\frac{1}{2}\big[\cos(2\phi-\alpha_1-\alpha_2)+\cos(\alpha_1-\alpha_2)\big]\sin^3\theta\sin\beta_1\sin\beta_2
\nonumber \\
&&\hbox to 6cm{}+\sin\theta\cos^2\theta\cos\beta_1\cos\beta_2\bigg\}
\nonumber \\
&=&-\frac{1}{2}\int_0^{\infty} S^4\,\mu(S)\,dS \int_{0}^{\pi} \,d\theta
\bigg[\frac{1}{2}\cos(\alpha_1-\alpha_2)\sin^3\theta\sin\beta_1\sin\beta_2+\sin\theta\cos^2\theta\cos\beta_1\cos\beta_2\bigg]
\;.
\label{MOME0}
\end{eqnarray}
Using
$\int_{0}^{\pi}\,\sin^3\theta\,d\theta = {4}/{3}$ and
$\int_{0}^{\pi}\,\sin\theta\cos^2\theta\,d\theta = {2}/{3}$
we find
\begin{eqnarray}
\langle \ba_1\cdot\bS_1 \;\ba_2\cdot\bS_2 \rangle &=&
-\frac{1}{3}\int_0^{\infty} S^4\,\mu(S)\,dS
\bigg[\cos(\alpha_1-\alpha_2)\sin\beta_1\sin\beta_2+\cos\beta_1\cos\beta_2\bigg]
=-\frac{\ba_1\cdot\ba_2}{3}\int_0^{\infty} S^4\,\mu(S)\,dS
\;.
\label{MOME1}
\end{eqnarray}
We recover the quantum-theoretical result
$\widehat{E}_{12}(\ba_1,\ba_2)=\langle \ba_1\cdot\bm\sigma_1 \;\ba_2\cdot\bm\sigma_2 \rangle =-\ba_1\cdot\ba_2$ if
we choose the density $\mu(S)$ such that Eq.~(\ref{MOME0a}) holds and that
\begin{eqnarray}
\int_0^{\infty} S^4\,\mu(S)\,dS=3
\;,
\label{MOME2}
\end{eqnarray}
which is always possible.
For instance, if we choose $\mu(S)=4 \exp(-2 S)$,
we have $\int_0^{\infty} S^2\,\mu(S)\,dS=1$ and
$\int_0^{\infty} S^4\,\mu(S)\,dS=3$.
There are many other solutions to Eqs.~(\ref{MOME0a}) and~(\ref{MOME2}),
for instance $\mu(S)=(1/3)\delta(S-1) + (1/6)\delta(S-2)$, a very simple one.

As in the classical electrodynamics model discussed in~\ref{MAXW},
the result Eq.~(\ref{MOME1}) together with $\int_0^{\infty} S^4\,\mu(S)\,dS=3$ does
not contradict Bell's theorem because the latter
involves functions $|{A}(\ba,\lambda)|\le1$ and $|{B}(\bb,\lambda)|\le1$
whereas Eq.~(\ref{MOME1}) is obtained by computing the correlation between two 3D vectors of lengths exceeding one.

To mimic a product state,
we assume that $p(\bS_1,\bS_2)=\delta(\bS_1-\bM_1)\delta(\bS_2-\bM_2)$
where $\bM_1$ and $\bM_2$ are 3D vectors which are considered to be fixed.
Then we have
\begin{subequations}
\label{MOME4}
\begin{eqnarray}
\langle \ba_1\cdot\bS_1 \rangle &=& \ba_1\cdot\bM_1
\;,
\\
\langle \ba_2\cdot\bS_2 \rangle &=& \ba_2\cdot\bM_2
\;,
\\
\langle \ba_1\cdot\bS_1 \;\ba_2\cdot\bS_2 \rangle &=& \ba_1\cdot\bM_1 \;\ba_2\cdot\bM_2
\;,
\end{eqnarray}
\end{subequations}
in full agreement with the quantum-theoretical result Eq.~(\ref{QTDE2z}) below.

In summary: we can recover the averages and the correlation $-\ba_1\cdot\ba_2$ of the singlet and product state
by replacing the ``quantum spins'' by 3D vectors of variable length and making an appropriate
choice of the density $p(\bS_1,\bS_2)$.

\section{Standard quantum theory of the EPRB experiment}\label{QTDE}

We briefly review the standard quantum-theoretical description of the EPRB experiment.
Recall that a basic premise of such a description is that the state of the quantum system, represented by the density matrix $\bm\rho$,
does not depend on the kind of measurements that are carried out~\cite{BALL03,RAED19b}.

\subsection{Singlet state}
If the statistics of repeated experiments with pairs of spin-1/2 objects
are captured by the singlet state defined by the density matrix
\begin{eqnarray}
\bm\rho&=&
\frac{1-\bm\sigma_1\cdot\bm\sigma_2}{4}=
\left( \frac{|{\uparrow}{\downarrow}\rangle-|{\downarrow}{\uparrow}\rangle}{\sqrt{2}}\right)
\left( \frac{\langle{\uparrow}{\downarrow}|-\langle{\downarrow}{\uparrow}|}{\sqrt{2}}\right)
\;,
\label{QTDE1}
\end{eqnarray}
the probability for observing the outcomes $x,y=\pm1$ when the first spin is measured along
the direction $\ba$ and the second one is measured along the direction $\bc$ is given by
\begin{eqnarray}
P(x,y|{\ba},{\bc})&=&
\hbox{\bf Tr\;}\bm\rho\,\frac{1+x\,\bm\sigma_1\cdot\ba}{2}\frac{1+y\,\bm\sigma_2\cdot\bc}{2}
=\frac{1 - x\,y\,\ba\cdot\bc}{4}
\;,
\label{QTDE3}
\end{eqnarray}
from which it immediately follows that
\begin{eqnarray}
\widehat{E}_1(\ba,\bc)=\sum_{x,y=\pm1} x\,P(x,y|{\ba},{\bc})=0
\;,\;
\widehat{E}_2(\ba,\bc)=\sum_{x,y=\pm1} y\,P(x,y|{\ba},{\bc})=0
\;,\;
\widehat{E}_{12}(\ba,\bc)=\sum_{x,y=\pm1} x\,y\,P(x,y|{\ba},{\bc})=-\ba\cdot\bc
\;,
\label{QTDE4}
\end{eqnarray}
in agreement with Eq.~(\ref{SOC11}).

\subsection{Product state}\label{PRODUCTSTATE}
If the density matrix of a two-particle system
can be written as $\bm\rho=\bm\rho_1\otimes\bm\rho_2$,
the system is said to be described by a product state~\cite{BALL03}.
Here $\bm\rho_i$ is the density matrix of particle $i=1,2$.
For two spin-1/2 objects described by a product state
we have $\widehat{E}_{12}({\bf a},{\bf b})=\widehat{E}_1({\bf a},{\bf b})\widehat{E}_2({\bf a},{\bf b})$
and the correlation
$\widehat{E}_{12}({\bf a},{\bf b})- \widehat{E}_{1}({\bf a},{\bf b})\widehat{E}_{2}({\bf a},{\bf b})=0$.
Therefore the quantum state $\bm\rho=\bm\rho_1\otimes\bm\rho_2$ is called uncorrelated.

For two spin-1/2 objects, the product state takes the generic form
\begin{eqnarray}
\bm\rho&=&\frac{1+\bm\sigma_1\cdot\bM_1}{2}\frac{1+\bm\sigma_2\cdot\bM_2}{2}
\;,
\label{QTDE1a}
\end{eqnarray}
where $\bM_1$ and $\bM_2$ are 3D vectors with a length less than or equal to one.
If $||\bM_1||=||\bM_2||=1$, the product state Eq.~(\ref{QTDE1a}) is said to be pure,
otherwise it is said to be mixed~\cite{BALL03}.

From Eq.~(\ref{QTDE1a}) it follows that
\begin{eqnarray}
P(x,y|{\ba},{\bc})&=&
\hbox{\bf Tr\;}\bm\rho\,\frac{1+x\,\bm\sigma_1\cdot\ba}{2}\frac{1+y\,\bm\sigma_2\cdot\bc}{2}
=\frac{1 + x\,\ba\cdot\bM_1}{2}\frac{1 + y\,\bc\cdot\bM_2}{2}
\;,
\label{QTDE3a}
\end{eqnarray}
and
\begin{eqnarray}
\widehat{E}_1(\ba,\bc)=
\langle \bm\sigma_1\cdot\ba\rangle=\ba\cdot\bM_1
\;,\;
\widehat{E}_2(\ba,\bc)=
\langle \bm\sigma_2\cdot\bc\rangle=\bc\cdot\bM_2
\;,\;
\widehat{E}_{12}(\ba,\bc)=
\langle \bm\sigma_1\cdot\ba\; \bm\sigma_2\cdot\bc \rangle=\ba\cdot\bM_1\,\bc\cdot\bM_2
\;.
\label{QTDE2z}
\end{eqnarray}

\subsection{Factorability and independence}\label{FACT}

In general, the probability of two dichotomic variables taking values $+1$ and $-1$ can be written as
\begin{eqnarray}
P(x,y|Z)&=&\frac{1 + x\,E_1(Z)+y\,E_2(Z)+x\,y\,E_{12}(Z)}{4}
=\frac{1 + x\,E_1(Z)}{2}\frac{1 + y\,E_2(Z)}{2}+x\,y\,\frac{E_{12}(Z)-E_1(Z)\,E_2(Z)}{4}
\nonumber\\
&=&P(x|Z)P(y|Z)+x\,y\,\frac{E_{12}(Z)-E_1(Z)\,E_2(Z)}{4}
\;,
\label{QTDE5}
\end{eqnarray}
where $Z$ stands for a collection of conditions.

By definition, if $P(x,y|Z)=P(x|Z)P(y|Z)$
the variables $x$ and $y$ are (logically/statistically) independent~\cite{GRIM01}.
If the variables $x$ and $y$ are (logically/statistically) independent
it immediately follows from Eq.~(\ref{QTDE5}) that their correlation $E_{12}(Z)-E_1(Z)\,E_2(Z)=0$.

In general, zero correlation does not imply independence~\cite{GRIM01}. However,
from Eq.~(\ref{QTDE5}) it also follows that $P(x,y|Z)=P(x|Z)P(y|Z)$
if the correlation $E_{12}(Z)-E_1(Z)\,E_2(Z)=0$.

In summary, the dichotomic variables $x=\pm1$ and $y\pm1$ are (logically/statistically) independent {\bf if and only if}
their correlation $E_{12}(Z)-E_1(Z)\,E_2(Z)=0$. This is a special property of a probability distribution of dichotomic variables.

In contrast, in quantum theory we can have zero correlation even though the density matrix does not factorize.
For example, the density matrix Eq.~(\ref{QTDE1}) cannot be factorized ($\bm\rho\not=\bm\rho_1\otimes\bm\rho_2$)
yet the correlation
$\widehat{E}_{12}(\ba,\bc)- \widehat{E}_{1}({\ba},{\bc})\widehat{E}_{2}({\ba},{\bc})=\widehat{E}_{12}(\ba,\bc)=-\ba\cdot\bc$
is zero if the vectors $\ba$ and $\bc$ are orthogonal.

\subsection{Extension of Bell's theorem to quantum-theoretical models}\label{EXQT}

In this subsection, we generalize Bell's theorem to the realm of quantum-theoretical models.
As nothing is gained by limiting the discussion to spin-1/2 systems
we prove the theorem in full generality.
We consider a composite quantum system that consists of two identical subsystems $i=1,2$.
The state of subsystem $i$ is represented by the density matrix $\bm\rho_i(\lambda)$.
The variable $\lambda$ is an element of a set that does not need to be defined in detail and plays exactly
the same role as in Bell's theorem.
We define a matrix $\bm\rho$ of the composite system  by
\begin{eqnarray}
\bm\rho&=&
\int \bm\rho_1(\lambda) \bm\rho_2(\lambda)\mu(\lambda) d\lambda
,
\label{EXQT0}
\end{eqnarray}
where $\mu(\lambda)$ is a probability density, that is a nonnegative
function, which satisfies $\int \mu(\lambda)\,d\lambda =1$  (compare with Eq.~(\ref{IN0})).
Using the properties of the trace $\mathbf{Tr\;}$,
$\mathbf{Tr\;} \bm\rho_1(\lambda)\bm\rho_2(\lambda)=
\left[\mathbf{Tr\;}_1 \bm\rho_1(\lambda)\right]\left[\mathbf{Tr\;}_2\bm\rho_2(\lambda)\right]=1$
and the fact that a sum of nonnegative definite
matrices with positive weights is a nonnegative definite matrix~\cite{HORN13},
it follows that Eq.~(\ref{EXQT0}) is a proper density matrix for the system
consisting of subsystems one and two.
Density matrices $\bm\rho$ of the form Eq.~(\ref{EXQT0}) are called separable.
A product state is a special case of a separable state.
In general, a separable state, being a sum (integral) of uncorrelated states, is correlated.

To avoid mathematical technicalities, in the following
we only consider quantum systems represented by finite-dimensional Hilbert spaces.
Consider two observables of each subsystem,
represented by the matrices $\bA_1$, $\bB_1$, $\bC_2$ and $\bD_2$, respectively.
The entries of these matrices are assumed to have been rescaled
such that all the eigenvalues of these four matrices lie in the interval $[-1,1]$
(the equivalent of the conditions on $A(\ba,\lambda)$ and $B(\bb,\lambda)$ in Eq.~(\ref{IN0})).
From the definition of the quantum-theoretical expectation
$\langle \bX_i \rangle_{\lambda} = \mathbf{Tr\;}_i \bm\rho_i(\lambda)\bX_i$,
it follows that for any $\lambda$,
$|\langle \bA_1 \rangle_{\lambda}|\le 1$,
$|\langle \bB_2 \rangle_{\lambda}|\le 1$,
$|\langle \bC_1 \rangle_{\lambda}|\le 1$, and
$|\langle \bD_2 \rangle_{\lambda}|\le 1$.

The correlation between observables of the two subsystems is defined by
\begin{equation}
Q(\bA_1,\bC_2)=\langle \bA_1\bC_2 \rangle=\mathbf{Tr\;}\bm\rho  \bA_1\bC_2 =
\int \left[\mathbf{Tr\;}\bm\rho_1(\lambda)\bA_1\right]
\left[\mathbf{Tr\;}\bm\rho_2(\lambda)\bC_2\right]
\mu(\lambda) d\lambda
=\int \langle \bA_1 \rangle_\lambda\langle \bC_2 \rangle_\lambda\mu(\lambda) d\lambda
\;.
\label{EXQT1}
\end{equation}
The correlations $Q(\bA_1,\bD_2)$, $Q(\bB_1,\bC_2)$ and $Q(\bB_1,\bD_2)$ are defined similarly.

Suppose that the two-spin system is described by the singlet state. Then we have
$Q(\bA_1,\bC_2)=\widehat{E}_{12}(\ba,\bc)=-\ba\cdot\bc$,
$Q(\bA_1,\bD_2)=\widehat{E}_{12}(\ba,\bd)=-\ba\cdot\bd$,
$Q(\bB_1,\bC_2)=\widehat{E}_{12}(\bb,\bc)=-\bb\cdot\bc$,
and $Q(\bB_1,\bD_2)=\widehat{E}_{12}(\bb,\bd)=-\bb\cdot\bd$.
Making use of the Cauchy-Schwarz inequality and recalling
that $\ba$, $\bb$, $\bc$ and $\bd$ are unit vectors, we find
\begin{align}
\left|
\widehat{E}_{12}(\ba,\bc)-
\widehat{E}_{12}(\ba,\bd)+
\widehat{E}_{12}(\bb,\bc)+
\widehat{E}_{12}(\bb,\bd)
\right|^2
&=\left| \ba\cdot(\bc-\bd)+ \bb\cdot(\bc+\bd) \right|^2
\nonumber \\
&\le
\left| \ba\cdot(\bc-\bd)\right|^2+ \left| \bb\cdot(\bc+\bd) \right|^2
+2\left| \ba\cdot(\bc-\bd)\right|\,\left| \bb\cdot(\bc+\bd) \right|
\nonumber \\
&\le
\Vert\ba\Vert^2\,\Vert\bc-\bd\Vert^2
+\Vert\bb\Vert^2\,\Vert\bc+\bd\Vert^2
+2\sqrt{\Vert\ba\Vert^2\,\Vert\bc-\bd\Vert^2\,\Vert\bb\Vert^2\,\Vert\bc+\bd\Vert^2}
\nonumber \\
&\le8
\;,
\label{EXQT4}
\end{align}
where we also used
$\Vert\bc-\bd\Vert^2+\Vert\bc+\bd\Vert^2=\bc^2+\bd^2=2$ and
$\Vert\bc-\bd\Vert^2\,\Vert\bc+\bd\Vert^2=(\bc^2+\bd^2)^2-4(\bc\cdot\bd^2)^2\le4$.
Therefore, for the singlet state we have~\cite{CIRE80}
\begin{eqnarray}
\left|
\widehat{E}_{12}(\ba,\bc)-
\widehat{E}_{12}(\ba,\bd)+
\widehat{E}_{12}(\bb,\bc)+
\widehat{E}_{12}(\bb,\bd)
\right|
&\le&2\sqrt{2}
\;.
\label{EXQT5}
\end{eqnarray}

There exists a choice for $\ba$, $\bb$, $\bc$ and $\bd$
for which equality in Eq.~(\ref{EXQT5}) can be reached.
To show this it is sufficient to consider four vectors that lie in the $x$-$y$ plane.
Let us write $\ba=(\cos a,\sin a, 0)$, etc. For the singlet state, we have
$\widehat{E}_{12}(\ba,\bc)=-\cos(a-c)$, $\widehat{E}_{12}(\ba,\bd)=-\cos(a-d)$,
$\widehat{E}_{12}(\bb,\bc)=-\cos(b-c)$, and $\widehat{E}_{12}(\bb,\bd)=-\cos(b-d)$.
Then take $a=0$, $b=\pi/2$, $c=\pi/4$ and $d=3\pi/4$ to find
$\widehat{E}_{12}(\ba,\bc)-\widehat{E}_{12}(\ba,\bd)+\widehat{E}_{12}(\bb,\bc)
+\widehat{E}_{12}(\bb,\bd)=-2\sqrt{2}$.

On the other hand, using the triangle inequality and Eq.~(\ref{BASIC3}) we obtain
\begin{align}
\vert Q(\bA_1,\bC_2)- Q(\bA_1,\bD_2)+&Q(\bB_1,\bC_2)- Q(\bB_1,\bD_2)\vert
\nonumber \\
&=\left|\int \big[
 \langle\bA_1 \rangle_\lambda\langle \bC_2 \rangle_\lambda
-\langle\bA_1 \rangle_\lambda\langle \bD_2 \rangle_\lambda
+
\langle\bB_1 \rangle_\lambda\langle \bC_2 \rangle_\lambda
+\langle\bB_1 \rangle_\lambda\langle \bD_2 \rangle_\lambda
\big]
\,\mu(\lambda) d\lambda
\right|
\nonumber \\
&\le\int \big|
\langle\bA_1 \rangle_\lambda\langle \bC_2 \rangle_\lambda
-\langle\bA_1 \rangle_\lambda\langle \bD_2 \rangle_\lambda
+
\langle\bB_1 \rangle_\lambda\langle \bC_2 \rangle_\lambda
+\langle\bB_1 \rangle_\lambda\langle \bD_2 \rangle_\lambda
\big|
\,\mu(\lambda) d\lambda
\nonumber \\
&\le 2\int\mu(\lambda) d\lambda=2
\;.
\label{EXQT3}
\end{align}

In summary, the correlation $\widehat{E}_{12}(\ba,\bb)=\langle \bm\sigma_1\cdot\ba\; \bm\sigma_2\cdot\bb \rangle$
of two spin-1/2 objects in the singlet state satisfies the bound Eq.~(\ref{EXQT5})~\cite{CIRE80}
but may violate the bound  Eq.~(\ref{EXQT3}).
The only conclusion one can draw from violation of the bound  Eq.~(\ref{EXQT3}) is that
there does not exist a separable density matrix that yields
$\widehat{E}_{12}(\ba,\bc)=-\ba\cdot\bc$ for all
$\ba$ and $\bc$.
From a violation Eq.~(\ref{EXQT3}), it would be a logical fallacy to draw any other conclusion
than the one just mentioned simply because the derivation of Eq.~(\ref{EXQT3}) pertains to quantum theory only.

\section{Basic inequalities}\label{BASIC}

For any pair of real numbers $u$ and $v$, the triangle inequality
\begin{eqnarray}
|u+v|\le |u|+|v|
\;,
\label{BASIC0a}
\end{eqnarray}
and the identity
\begin{eqnarray}
(u \pm v)^2+(1-u^2)(1-v^2)=(1\pm uv)^2
\;,
\label{BASIC0}
\end{eqnarray}
hold.

In this section, the symbols $w$, $x$, $y$, and $z$ represent real numbers in the range $[-1,1]$.
Applying Eq.~(\ref{BASIC0}) with $u=x$ and $v=y$, the second term in Eq.~(\ref{BASIC0}) is nonnegative such that
\begin{eqnarray}
(x \pm y)^2\le (1\pm xy)^2
\;,
\label{BASIC1}
\end{eqnarray}
or, equivalently,
\begin{eqnarray}
|x \pm y|\le 1\pm xy
\;.
\label{BASIC2}
\end{eqnarray}
Using Eq.~(\ref{BASIC0a}) and Eq.~(\ref{BASIC2}) we obtain
\begin{subequations}
\label{BASIC3}
\begin{eqnarray}
|xy \pm xz| &=&|x||y\pm z|\le1\pm yz
\;,
\label{BASIC3x}
\\
|xz - xw + yz + yw| &\le&
|xz - xw| + |yz + yw|\le |x||z-w| + |y||z + w|
\le 1 - zw + 1 + zw = 2
\;.
\label{BASIC3b}
\end{eqnarray}
\end{subequations}
The variables appearing in Eqs.~(\ref{BASIC3x}) and~(\ref{BASIC3b}) form the triple $(x,y,z)$
and the quadruple $(x,y,z,w)$, respectively.
The triple/quadruple structure is essential to prove Eq.~(\ref{BASIC3}).
For instance, an expression of the form  $|xz - xw| + |yz + yw'|$ can be larger than 2 (e.g., $(x,y,z,w,w')=(1,1,1,-1,1)$ yields $|xz - xw| + |yz + yw'| = 4$).

Next, we prove that for any triple of real numbers $a$, $b$, and $c$,
\begin{eqnarray}
|a\pm b|\le 1 \pm c \iff |a\pm c|\le 1 \pm b \iff |b\pm c|\le 1 \pm a
\;.
\label{TRIPLE}
\end{eqnarray}
Written more explicitly, the inequalities $|a\pm b|\le 1 \pm c$  read $-1\mp c \le a\pm b\le 1 \pm c$ from which
\begin{eqnarray}
-1+c &\le& a-b \le 1-c\; \Rightarrow\; a+c\le 1+b\;\;\hbox{and}\;\;
-1+b \le a-c\;,
\nonumber \\
-1-c &\le& a+b \le 1+c \;\Rightarrow\; -1-b \le a+c\;\;\hbox{and}\;\;
a-c\le 1-b\;,
\end{eqnarray}
or, written more compactly, $|a\pm c|\le 1 \pm b$. In the same manner, we can prove that $|b\pm c|\le 1 \pm a$.

Attaching subscripts to the $x$'s, $y$'s, etc.,  and denoting the correlation of $x$'s and $y$'s by
\begin{eqnarray}
\langle xy\rangle=\frac{1}{N}\sum_{i=1}^N x_i y_i
\;,
\label{BASIC3a}
\end{eqnarray}
etc., repeated use of Eqs.~(\ref{BASIC0a}) and~(\ref{BASIC3}) yields
\begin{subequations}
\label{BASIC4}
\begin{eqnarray}
| \langle xy \rangle\pm \langle xz\rangle| &\le&1\pm \langle yz\rangle
\label{BASIC4a}
\;,\\
|\langle xz \rangle- \langle xw \rangle+ \langle yz\rangle + \langle yw\rangle| &\le&
|\langle xz \rangle- \langle xw \rangle|+ |\langle yz\rangle + \langle yw\rangle| \le2
\;,
\label{BASIC4b}
\end{eqnarray}
\end{subequations}
In exactly the same manner, one proves that
$|\langle xz \rangle+\langle xw \rangle|+ |\langle yw\rangle - \langle yz\rangle| \le2$,
$|\langle xz \rangle+\langle xw \rangle|+ |\langle yz\rangle - \langle yw\rangle| \le2$,
and
$|\langle xw \rangle+\langle yz\rangle |+ |\langle yw\rangle -\langle xz \rangle | \le2$.

\subsection{Application: discrete data}\label{BASICa}

Inequalities Eqs.~(\ref{BASIC4}) can be used to detect inconsistencies between the data and their correlations~\cite{BO1862}.
Suppose that we are given a set of discrete data
${Q}_3=\{(x_i, y_i, z_i)\,|\,i=1,\ldots, n\;;\; x_i=\pm1, y_i=\pm1, z_i=\pm1\}$,
consisting of triples $(x_i, y_i, z_i)$.
Also suppose that $\langle xy \rangle=0.7$ and $\langle xz\rangle = -0.7$.
Then inequality Eq.~(\ref{BASIC4a}) puts a constraint on the values that
$\langle yz\rangle$ may take, namely $ \langle yz\rangle \le 1-1.4=-0.4$.

Conversely, assume that we are given three numbers $\alpha=0.7$,
$\beta=-0.7$ and e.g., $\gamma=0.4$.
Does there exist a set ${Q}_3$ of triples of discrete data
such that $\langle xy \rangle=\alpha$ and $\langle xz\rangle =\beta$ and $\langle yz\rangle =\gamma$\ ?
The answer is {\bf no} for if there was, the value of $\gamma$ would be in conflict with the constraint
$ \langle yz\rangle \le -0.4$.

For a collection of two-valued quadruples
${Q}_4=\{(x_i, y_i, z_i, w_i)\,|\,i=1,\ldots, n\;;\; x_i=\pm1, y_i=\pm1, z_i=\pm1, w_i=\pm1\}$
we can, in addition to Eq.~(\ref{BASIC4a}), use Eqs.~(\ref{BASIC4})
to find constraints on the pairwise correlations.

In summary, if the averages $\langle xy \rangle$, $\langle xz \rangle$, and $\langle yz \rangle$
violate at least one of the inequalities Eq.~(\ref{BASIC4a})
these averages cannot have been computed from the data set consisting of triples.
Similarly, if the averages $\langle xz \rangle$, $\langle xw \rangle$, $\langle yz \rangle$, and $\langle yw \rangle$
violate at least one of the inequalities Eqs.~(\ref{BASIC4})
these averages cannot have been computed from the data set consisting of quadruples.
To the best of our knowledge, the inequality Eq.~(\ref{BASIC4a}) was (in a different but equivalent form)
first given by Boole, who called it a condition of possible experience~\cite{BO1862}.

\subsection{Application: real-valued functions}

If we define the correlation of two functions $x(\lambda)$ and $y(\lambda)$ by
\begin{eqnarray}
\langle xy\rangle=\int x(\lambda) y(\lambda)\mu(\lambda)\,d\lambda,
\quad,\quad
\mu(\lambda)\ge0
\quad,\quad
\int \mu(\lambda)\,d\lambda=1
\;,
\label{BASIC5}
\end{eqnarray}
then the inequalities Eqs.~(\ref{BASIC4}) hold as long as
$|x(\lambda)|\le1$, $|y(\lambda)|\le1$, $|z(\lambda)|\le1$, and $|w(\lambda)|\le1$.

Suppose that we are given three functions $x(\lambda)$, $y(\lambda)$
and $z(\lambda)$, satisfying $|x(\lambda)|\le1$, $|y(\lambda)|\le1$ and $|z(\lambda)|\le1$,
and for which $\langle xy \rangle=1/\sqrt{2}$ and $\langle xz\rangle = -1/\sqrt{2}$.
Then inequality Eq.~(\ref{BASIC4a}) forces $\langle yz\rangle$ to be in the range $\langle yz\rangle \le 1-\sqrt{2}$.

Next assume that we have three unit vectors $\ba$, $\bb$, $\bc$,
and three functions $x(\ba,\lambda)$, $y(\bb,\lambda)$, and $z(\bc,\lambda)$.
Then inequality Eq.~(\ref{BASIC4a}) rules out that there exist
functions $x(\ba,\lambda)$, $y(\bb,\lambda)$, $z(\bc,\lambda)$,
satisfying $|x(\ba,\lambda)|\le1$, $|y(\bb,\lambda)|\le1$, $|z(\bc,\lambda)|\le1$
such that
$\langle x(\ba)y(\bb)\rangle = \ba\cdot\bb$,
$\langle x(\ba)z(\bc)\rangle = \ba\cdot\bc$, and
$\langle y(\bb)z(\bc)\rangle = \bb\cdot\bc$.
Indeed, if we take $\ba=(1,0,0)$, $\bb=(1,1,0)/\sqrt{2}$,
$\bc=(-1,1,0)/\sqrt{2}$, and use
$| \langle xy \rangle\pm \langle xz\rangle| \le1\pm \langle yz\rangle$,
we obtain $\sqrt{2}\le 1$ which contradicts elementary arithmetic.
In essence, a similar argument was used by Bell to prove his theorem (see Section~\ref{BELL}).

\bibliographystyle{elsarticle-num}
\bibliography{/D/papers/all24}
\end{document}